\documentclass{elsarticle}
\usepackage{placeins}
\usepackage{amsmath,amssymb}
\usepackage{color}
\usepackage{fullpage}
\usepackage[english]{babel}
\usepackage{graphicx}
\usepackage{xcolor}
\usepackage[unicode]{hyperref}
\usepackage{float}
\usepackage{enumitem}
\usepackage{appendix}
\usepackage{algorithm}
\usepackage{algpseudocode}
\usepackage{caption}
\usepackage{subcaption}
\usepackage{xspace}
\usepackage{upgreek}
\usepackage{mathtools}
\usepackage{setspace}
\usepackage{tikz}
\usepackage{siunitx}
\usetikzlibrary{arrows.meta}
\usepackage{multirow}
\usepackage{booktabs}
\newcommand{\rom}{\mathrm{rom}}
\newcommand{\fom}{\mathrm{fom}}

\newcommand{\eg}{\emph{e.g.}\xspace}
\newcommand{\ie}{\emph{i.e.}\xspace}
\newcommand{\online}{\emph{online}\xspace}
\newcommand{\offline}{\emph{offline}\xspace}
\renewcommand{\vec}[1]{\boldsymbol{#1}}
\newcommand{\ten}[1]{\boldsymbol{#1}}
\newcommand{\mat}[1]{\mathsf{#1}}
\newcommand{\params}{\boldsymbol{\mu}}
\newcommand{\newparams}{\boldsymbol{\mu}^*}
\newcommand{\wi}{\mathrm{Wi}}

\begin{document}

\begin{frontmatter}

\title{Intrusive versus non-intrusive reduced-order modeling
of generalized Newtonian fluid flows}

\author[1]{Parajal Rai\corref{cor1}}
\ead{p.rai@tue.nl}
\author[1]{Michelle Spanjaards}
\author[1]{Patrick D. Anderson}
\author[1]{Ye Wang}
\author[1]{Nick O. Jaensson}
\cortext[cor1]{Corresponding author.}

\address[1]{Department of Mechanical Engineering, Eindhoven University of
Technology, 5600 MB Eindhoven, The Netherlands}

\begin{abstract}
This study compares three reduced-order modeling (ROM) approaches for flow
simulations of generalized Newtonian fluids described by the Carreau rheological
model. All three methods rely on \offline snapshot generation in the rheological
parameter space using the full-order model (FOM), followed by a proper orthogonal
decomposition (POD) of the snapshot matrix to obtain a reduced basis, but they differ in
how they reconstruct the solution for new parameter values in the \online phase.
The three ROM approaches examined are: (i) intrusive Galerkin projection onto the
reduced basis with full operator reassembly (ROM-FULL), (ii) intrusive hyper-reduced
Galerkin projection using the discrete empirical interpolation method with GappyPOD for the nonlinear term (ROM-DEIM), and (iii) a non-intrusive interpolation approach using radial basis
function interpolation (ROM-RBF).
We demonstrate these three ROM approaches on two benchmark flows: a lid-driven cavity
and a sphere settling in a closed container, spanning boundary-driven and
force-driven flows.
ROM-FULL achieves the highest accuracy but requires reassembling the full-order nonlinear operator
during the \online phase, whereas ROM-RBF is fully non-intrusive, and its accuracy is closely tied to data availability and deteriorates outside the training data range. ROM-DEIM offers a balance between
efficiency and accuracy, even when data are sparse. The results provide guidelines for selecting an appropriate
ROM strategy based on solver accessibility, computational efficiency, and desired
accuracy.
\end{abstract}


\end{frontmatter}

\section{Introduction}

Generalized Newtonian fluids arise in a broad range of natural and industrial
applications, including blood flow modeling
\cite{abbasian2020effects,quarteroni2016geometric}, food processing \cite{krishnan2013simulation}, and industrial mixing \cite{liu2006laminar}.
Unlike Newtonian fluids with constant viscosity, generalized Newtonian fluids
exhibit a viscosity that depends on the local shear rate. This shear rate dependence introduces strong nonlinearities into the governing equations and typically requires iterative solution schemes, such as Picard or Newton methods,
with repeated matrix assemblies and linear solves for each parameter value.
The resulting computational cost becomes particularly restrictive in
many-query settings, such as optimization \cite{lee2019simultaneous},
uncertainty quantification \cite{kim2019uncertainty,rinkens2023uncertainty},
parameter inference \cite{kontogiannis2025learning}, and model selection
\cite{Rinkens2026Bayesian}, especially in three-dimensional simulations.

Reduced-order models (ROMs) address this challenge by approximating
high-fidelity solutions in a low-dimensional subspace constructed from
representative solution snapshots. In the \offline phase, full-order model
(FOM) simulations are computed at selected parameter values, and a reduced basis
is commonly obtained using proper orthogonal decomposition (POD)
\cite{Holmes_Lumley_Berkooz_Rowley_2012,kunisch2002galerkin}. In the \online phase,
solutions at new parameter values are then obtained either by projecting the
governing equations onto the reduced basis, leading to intrusive ROMs, or by
interpolating the reduced coefficients directly from data, leading to
non-intrusive ROMs. While simplified low-fidelity models, such as coarser
discretizations or reduced-physics formulations form another important class
of reduced models \cite{peherstorfer2018survey}, the present work focuses on
POD-based intrusive and non-intrusive ROM strategies.

Intrusive ROMs preserve the structure of the governing equations by projecting
them onto the reduced space. For problems with affine parameter dependence, this
enables an efficient \offline-\online decomposition and can reduce the \online
cost by orders of magnitude compared with the FOM
\cite{hesthaven2016certified,quarteroni2014reduced}. For generalized Newtonian
fluids, however, the shear-dependent viscosity must still be evaluated during
the \online phase, so the cost may remain tied to the full-order discretization.
Hyper-reduction techniques are therefore needed to obtain a truly reduced
\online complexity in projection-based ROMs. Examples include
interpolation-based methods such as the empirical interpolation method (EIM)
\cite{barrault2004empirical} and the discrete empirical interpolation method
(DEIM) \cite{chaturantabut2010nonlinear}, as well as sampling and
quadrature-based approaches, such as the energy-conserving sampling and weighting
(ECSW) method \cite{farhat2015structure} and the empirical cubature method
\cite{hernandez2017dimensional}.

Non-intrusive ROMs, in contrast, treat the high-fidelity solver as a black box.
They learn a direct mapping from input parameters to reduced coefficients using
snapshot data and therefore do not require access to the governing equations or
solver internals during the \online phase. Common approaches include radial
basis function (RBF) interpolation \cite{xiao2015non}, artificial neural
networks (ANNs) \cite{hesthaven2018non,wang2019non}, and Gaussian process
regression (GPR) \cite{guo2018reduced,cicci2023uncertainty}.

In fluid mechanics, ROM research has focused primarily on Newtonian flows, for
which extensive reviews and comparative studies are available
\cite{balajewicz2012stabilization,hajisharifi2023non,xiao2019domain,stabile2018finite}.
Several works have directly compared intrusive and non-intrusive ROM strategies
in this setting; see, for example,
\cite{benner2015survey,czech2022data,padula2024brief}. Less attention has been
devoted to non-Newtonian flows. For viscoelastic fluids, both intrusive and
non-intrusive ROMs have been investigated
\cite{chetry2023comparing,oishi2024nonlinear,amor2026reduced}. For generalized
Newtonian fluids, however, ROM studies remain more limited, and existing work
has often considered unbounded power-law viscosity models
\cite{reyes2023reduced}. Such models can exhibit numerical difficulties as the
shear rate approaches zero, where the viscosity may diverge. A systematic
comparison of intrusive, hyper-reduced, and non-intrusive ROM strategies for
bounded-viscosity models, such as the Carreau law, is therefore still largely
missing.

The present work addresses this gap by comparing three ROM strategies:
(i) intrusive Galerkin projection with full operator reassembly (ROM-FULL),
(ii) intrusive hyper-reduced Galerkin projection using DEIM for the nonlinear
viscosity term (ROM-DEIM), and (iii) non-intrusive RBF interpolation of the
reduced coefficients (ROM-RBF). The methods are applied to two benchmark
problems governed by a generalized Newtonian constitutive model: a lid-driven
cavity with prescribed velocity and a settling sphere in a closed cylindrical
container with prescribed force. For each benchmark, we assess the accuracy,
robustness, and computational behavior of the three ROMs under interpolation
and extrapolation in the rheological parameter space. We also examine the
effect of training-set size by systematically increasing the number of snapshot samples.

The remainder of this paper is organized as follows.
Section~\ref{sec:fom} presents the governing equations and the full-order finite
element formulation. Section~\ref{sec:rom} describes the three reduced-order
modeling approaches. Section~\ref{sec:results} discusses the numerical results
for the lid-driven cavity and settling-sphere benchmarks.
Section~\ref{sec:conclusion} provides the conclusions.

\section{Full-order model}\label{sec:fom}
Modeling generalized Newtonian flows requires solving nonlinear partial differential
equations due to shear-rate-dependent viscosity. In this section, we summarize the
governing equations and the corresponding finite element weak formulation.

\subsection{Governing equations}

Let $\Omega \subset \mathbb{R}^d$ (with $d=2$ or $3$) be the spatial domain.
Assuming steady creeping flow, the momentum and continuity equations are:
\begin{align}
    -\nabla \cdot \ten{\sigma} &= \vec{f}, \quad \text{in }\Omega, \label{eq:momentum}\\
    \nabla \cdot \vec{u} &= 0, \quad \text{in }\Omega,
\end{align}
where $\vec{u}$ is the velocity field, $\vec{f}$ is a body-force vector, and $\ten{\sigma}$ is the Cauchy stress
tensor, which is decomposed as
\begin{equation}
    \ten{\sigma} = -p\,\ten{I} + \ten{\tau},
    \label{eq:stress-split}
\end{equation}
with $p$ the pressure, $\ten{\tau}$ the extra (deviatoric) stress tensor, and
$\ten{I}$ the identity tensor. 
These equations are closed with the boundary conditions:
\begin{align}
    \vec{u} &= \vec{u}_\mathrm{D} \quad \text{on } \Gamma_\mathrm{D}, \\
    \ten{\sigma}\cdot \vec{n} &= \vec{t}_\mathrm{N} \quad
    \text{on } \Gamma_\mathrm{N},
\end{align}
where $\Gamma_\mathrm{D}$ and $\Gamma_\mathrm{N}$ are the Dirichlet
(velocity $\vec{u}_\mathrm{D}$ prescribed) and Neumann (traction $\vec{t}_\mathrm{N}$ prescribed) portions of the
boundary, respectively, and $\vec{n}$ is the outwardly directed unit normal
on the boundary.

For generalized Newtonian fluids, the viscosity
depends on the local shear rate. The extra stress is defined by
\begin{equation}
    \ten{\tau} = 2\,\eta(\dot{\gamma})\,\ten{D}(\vec{u}),
\end{equation}
where the rate-of-deformation tensor is the symmetric part of the velocity gradient:
\begin{equation}
    \ten{D}(\vec{u}) = \tfrac{1}{2}\bigl(\nabla \vec{u} + (\nabla \vec{u})^\top \bigr),
\end{equation}
and the scalar shear rate $\dot{\gamma}$ is
\begin{equation} \label{eq:shear_rate}
    \dot{\gamma} = \sqrt{2\,\ten{D}(\vec{u}):\ten{D}(\vec{u})}\,.
\end{equation}
Several constitutive models exist for generalized Newtonian fluids, \eg, Carreau, Carreau-Yasuda, Cross, and power-law \cite{bird1987dynamics}.
Although the power-law model has a low number of adjustable parameters and
has been used in previous ROM work for generalized Newtonian fluids
\cite{reyes2023reduced}, in this work, we use the Carreau model, which provides
a smooth, bounded transition between the low-shear (Newtonian plateau) and
high-shear viscosity limits and avoids the diverging viscosities at low shear
rates observed with the power-law model. The Carreau viscosity law is:
\begin{equation} \label{eq:carreau}
    \eta(\dot{\gamma}) = \eta_\infty + (\eta_0 - \eta_\infty)\,\Big[1 +
    (\lambda\,\dot{\gamma})^2\Big]^{\frac{n-1}{2}},
\end{equation}
where $\eta_0$ and $\eta_\infty$ are the zero-shear and infinite-shear viscosities,
$\lambda$ is a characteristic time scale, and $n$ is the power-law index characterizing shear-thinning ($n<1$) or shear-thickening ($n>1$) behavior. For $n=1$, the model
reduces to a Newtonian fluid with viscosity $\eta_0$.

\subsection{Weak formulation}

The generalized Newtonian flow problem is formulated in a variational framework.
For brevity, we present the formulation for the case of homogeneous Dirichlet
boundary conditions; non-homogeneous Dirichlet boundary conditions, such as a
prescribed lid velocity, are incorporated through the standard lift and imposed
strongly in the discrete spaces \cite{brenner2008mathematical, boffi2013mixed}.
Moreover, the settling-sphere problem requires special treatment, which is
elaborated in \ref{sec:force_constraint}. For the numerical approximation,
we follow the usual procedure of discretizing the weak formulation using finite element spaces $\mathcal{V}_h \subset H^1_0(\Omega)^d$ for the velocity and
$\mathcal{Q}_h \subset L^2(\Omega)$ for the pressure.
The discrete weak form is then given by: find a velocity-pressure pair $(\vec{u}_h,p_h)\in \mathcal{V}_h \times \mathcal{Q}_h$ such that
\begin{align}
    \int_\Omega 2\,\eta(\dot{\gamma}_h)\,\ten{D}(\vec{u}_h):\ten{D}(\vec{v}_h) \,\mathrm{d}V
    - \int_\Omega p_h\,(\nabla\cdot\vec{v}_h)\,\mathrm{d}V
    &= \int_\Omega \vec{f}\cdot\vec{v}_h\,\mathrm{d}V
    + \int_{\Gamma_\mathrm{N}} \vec{t}_\mathrm{N}\cdot\vec{v}_h\,\mathrm{d}A, 
    \label{weak_form1} \\
    \int_\Omega q_h\,(\nabla\cdot\vec{u}_h)\,\mathrm{d}V &= 0, \label{weak_form2}
\end{align}
for all $(\vec{v}_h,q_h)$ in the same function spaces. Here $\eta(\dot{\gamma}_h)$
denotes the numerical approximation of shear-dependent Carreau viscosity.

The system given by Eqs.~\eqref{weak_form1} and \eqref{weak_form2} can, after multiplying the continuity
equation by $-1$, be written as a nonlinear system of equations
\cite{john2016finite,larson2013finite}:
\begin{equation}
    \begin{bmatrix} \mat{A} & -\mat{L}^\top \\ - \mat{L} & \mat{0} \end{bmatrix}
\begin{bmatrix} \mat{u} \\ \mat{p} \end{bmatrix}
=
\begin{bmatrix} \mat{b} \\ \mat{0} \end{bmatrix}
\end{equation}
where $\mat{A} \in\mathbb{R}^{N_u\times N_u}$ is the diffusion matrix,
$\mat{L} \in\mathbb{R}^{N_p\times N_u}$ is the discrete divergence matrix,
$\mat{u} \in\mathbb{R}^{N_u}$ are the velocity unknowns,
$\mat{p} \in\mathbb{R}^{N_p}$ are the pressure unknowns, and
$\mat{b} \in\mathbb{R}^{N_u}$ is the load vector (\eg, including forces and
traction boundary conditions). For brevity, $N_u$ denotes the \emph{total} number
of velocity degrees of freedom (\ie, the sum of the number of velocity degrees of
freedom in each spatial direction), and $N_p$ denotes the number of pressure degrees of
freedom.

The discrete problem is nonlinear because the viscosity $\eta(\dot{\gamma}_h)$
depends on the unknown velocity field. To handle this nonlinearity in the FOM, we employ a
hybrid iterative solution strategy: the computation begins with Picard iterations
and, after 20 Picard steps, the solver is switched to the Newton-Raphson method
to accelerate convergence near the solution. The choice of 20 Picard steps is
motivated by robustness: for weakly non-Newtonian behavior, a smaller number of
steps (\eg, 5 to 10) is typically sufficient, but employing 20 Picard steps ensures convergence in the strongly nonlinear regimes considered in this work. Iterations are continued
until the difference between successive solutions falls below the prescribed
tolerances, namely, a maximum point-wise change of $10^{-13}$ for the velocity and
$10^{-7}$ for the pressure, where the latter is more affected by the loss of accuracy in floating-point operations. The converged solutions serve as snapshot data
for constructing the reduced-order models described in the following sections.

Throughout this work, we adopt the Taylor-Hood element pair, with quadratic basis
functions for the velocity and linear basis functions for the pressure. This
element combination satisfies the Ladyzhenskaya-Babu\v{s}ka-Brezzi (LBB)
stability condition \cite{boffi2013mixed,ern2004theory}.

\section{Reduced-order models}\label{sec:rom}\label{sec:svd}

Reduced-order modeling (ROM) reduces the computational cost of full-order
simulations by replacing the large system matrix arising from the full-order
discretization with a much smaller reduced system matrix defined on a
low-dimensional subspace obtained through projection or interpolation. In this section, we describe the three ROM strategies investigated in this work.

The construction of all ROM approaches presented here begins with the computation
of a set of $M$ high-fidelity solutions, referred to as snapshots, using parameters
within the domain of interest. The snapshots yield the nodal values of the
velocity, which are stored in column vectors $\mat{u}_i \in \mathbb{R}^{N_u}$ for
$i=1, \ldots, M$ and are subsequently assembled into a snapshot matrix
\begin{equation}
\mat{X} = [\mat{u}_1, \mat{u}_2, \ldots, \mat{u}_M] \in \mathbb{R}^{N_u \times M}.
\end{equation}
A singular value decomposition (SVD) of $\mat{X}$ yields the proper
orthogonal decomposition (POD) basis, which, after truncation to the leading $r$
modes, forms a reduced basis $\mathcal{V}_r$, given by
\begin{equation} \label{eq:reduced_space}
    \mathcal{V}_r \coloneqq \mathrm{span}\{\vec{\xi}_1,\ldots,\vec{\xi}_r\}
    \subset \mathcal{V}_h,
\end{equation}
which is, by construction, a subspace of $\mathcal{V}_h$, since each $\vec{\xi}_i$
is obtained from snapshots that themselves lie in $\mathcal{V}_h$.
Note that we use Taylor-Hood finite elements and that the velocity snapshots
are weakly divergence-free; thus, the reduced basis $\mathcal{V}_r$ only
describes weakly divergence-free functions.

The reduced basis $\mathcal{V}_r$ is represented by the reduced basis matrix
$\mat{\Phi} \in \mathbb{R}^{N_u\times r}$, whose columns contain the nodal values
of the vector-valued reduced basis functions $\vec{\xi}_i(\vec{x})$ 
and $i=1,\ldots,r$, expressed in the FEM basis. The value of $r$ is selected such
that all basis functions with normalized singular values above a prescribed
threshold $\epsilon_\mathrm{pod}$ are retained. The procedure to obtain the
reduced basis matrix $\mat{\Phi}$ is summarized in Algorithm~\ref{alg:SVD}.

\begin{algorithm}[H]
\caption{POD basis construction for the velocity field}
\label{alg:SVD}
\begin{algorithmic}[1]
\State \textbf{Input:} Snapshot matrix
$\mat{X} = [\mat{u}_1, \mat{u}_2, \ldots, \mat{u}_M]
\in \mathbb{R}^{N_u \times M}$, POD tolerance $\epsilon_\mathrm{pod}$
\State Compute the (economy) singular value decomposition:
\begin{equation}
\mat{X} = \mat{U}\,\mat{\Sigma}\,\mat{V}^{\top},
\end{equation}
where
$\mat{U} \in \mathbb{R}^{N_u \times M}$ contains the left singular vectors,
$\mat{V} \in \mathbb{R}^{M \times M}$ contains the right singular vectors,
and
$\mat{\Sigma} =
\mathrm{diag}(\sigma_1,\sigma_2,\ldots,\sigma_M) \in \mathbb{R}^{M \times M}$
with $\sigma_1 \geq \sigma_2 \geq \cdots \geq \sigma_M \geq 0$
contains the singular values.
\State Normalize the singular values:
\begin{equation}
\bar{\sigma}_i = \sigma_i / \sigma_1
\end{equation}
\State Determine the reduced dimension
\begin{equation}
r = \begin{cases}
\max\{\, i : \bar{\sigma}_i > \epsilon_\mathrm{pod} \,\} &
\text{if } \bar{\sigma}_M \le \epsilon_\mathrm{pod}, \\
M & \text{otherwise.}
\end{cases}
\end{equation}
\State Form the POD basis from the leading $r$ left singular vectors:
\begin{equation}
\mat{\Phi} = \mat{U}(:,1\!:\!r).
\end{equation}
\State \textbf{Output:} Reduced basis matrix $\mat{\Phi} \in \mathbb{R}^{N_u \times r}$.
\end{algorithmic}
\end{algorithm}

Given the reduced basis, the velocity field $\vec{u}_h$ can be approximated as
\begin{equation} \label{eq:reduced_u}
    \vec{u}_h(\vec{x}) \approx \vec{u}_\mathrm{rom}(\vec{x}) = \hat{u}_i\,\vec{\xi}_i(\vec{x}),
\end{equation}
where $\hat{u}_i \in \mathbb{R}$ are the corresponding reduced velocity coefficients, and repeated indices imply summation over $i=1,\ldots,r$ (Einstein summation), with $r$ 
the dimension of the reduced subspace.
This expression can be written in matrix notation, using the reduced basis
matrix, as
\begin{equation} \label{eq:u_rom}
\mat{u}_\mathrm{rom} = \mat{\Phi}\,\hat{\mat{u}},
\end{equation}
where $\mat{u}_\mathrm{rom} \in \mathbb{R}^{N_u}$ are the full-order nodal values
of the solution field reconstructed from the reduced basis, and
$\hat{\mat{u}} \in \mathbb{R}^r$ are the reduced POD coefficients.
We note that the reduced coefficients are functions of the parameter values,
collected in the vector $\params$ with components $\mu^{(i)}$ for 
$i=1\ldots N_\text{params}$, where $N_{\text{params}}$ denotes the number of parameters.
 Finding the mapping
$\params \mapsto \hat{\mat{u}}(\params)$ forms a crucial part of reduced-order
modeling.

Two main approaches are used to efficiently compute these reduced coefficients for
new parameter values. In projection-based (or intrusive) ROMs, the governing
equations of the full-order model are projected onto the reduced space spanned by
$\mat{\Phi}$, which results in a low-dimensional system for $\hat{\mat{u}}(\params)$.
However, for nonlinear problems, the evaluation of reduced operators may still
involve full-order operations, which motivates the use of hyper-reduction
techniques. In contrast, interpolation-based (or non-intrusive) ROMs treat the
full-order solver as a black box and learn the direct mapping from parameters to
reduced coefficients using the training data. As a consequence, no reduced
governing equations need to be assembled or solved during the \online phase.

We now describe the specific ROM implementations used: (i) intrusive Galerkin
projection onto the reduced basis with full operator reassembly (ROM-FULL),
(ii) intrusive hyper-reduced Galerkin projection using the discrete empirical
interpolation method for the nonlinear term (ROM-DEIM), and (iii) non-intrusive interpolation using radial basis function (ROM-RBF).

\subsection{ROM-FULL} \label{sec:rom-full}

A reduced system of equations is obtained by a Galerkin projection, \ie the test
functions $\vec{v}_h$ are chosen in the same space as the velocity $\vec{u}_h$
(Eq.~\eqref{eq:reduced_u}), yielding
\begin{equation} \label{eq:reduced_v}
    \vec{v}_h(\vec{x}) \approx \vec{v}_\mathrm{rom}(\vec{x}) =
    \hat{v}_i\,\vec{\xi}_i(\vec{x}),
\end{equation}
where $\hat{v}_i \in \mathbb{R}$ are the corresponding reduced velocity coefficients.
Substitution of these expressions for $\vec{u}_h$ and $\vec{v}_h$ in the weak form
Eq.~\eqref{weak_form1} yields the discrete system
\cite{hesthaven2016certified,rozza2022advanced}:
\begin{equation}
    \hat{\mat{A}}\,\hat{\mat{u}} = \hat{\mat{b}},
\end{equation}
which is an $r$-dimensional nonlinear system for $\hat{\mat{u}}$, where
$\hat{\mat{A}}=\mat{\Phi}^\top \mat{A} \mat{\Phi} \in \mathbb{R}^{r \times r}$
and $\hat{\mat{b}} = \mat{\Phi}^\top \mat{b} \in \mathbb{R}^r$. Note that the
full-order velocity system matrix depends on the velocity solution and the
problem parameters, \ie $\mat{A}=\mat{A}(\mat{u}_\mathrm{rom}, \params)$, and
the load vector can depend on the problem parameters, \ie $\mat{b}=\mat{b}(\params)$.
The pressure-coupling term is neglected under the assumption that the POD basis inherits the weak divergence-free property of the discrete snapshots. Alternatively, one could include supremizer modes
\cite{ballarin2015supremizer, Fonn2019} for pressure stability, but we omit that approach
here.

For \emph{affine} problems, the reduced matrix $\mat{\Phi}^\top \mat{A} \mat{\Phi}$
and the reduced load vector $\mat{\Phi}^\top \mat{b}$ need to be assembled
only once during the \offline phase and could be reused during the \online phase
\cite{hesthaven2016certified}. However, this approach is not possible for
generalized Newtonian fluids, where the full-order system matrix depends on the
velocity through the viscosity function. In the ROM-FULL approach, we
solve the reduced system with a Picard iteration, using the high-fidelity solver
as needed to assemble the operator $\mat{A}(\mat{u}_\mathrm{rom}, \params)$ (see
Algorithm~\ref{alg:reduced_iteration}).

\begin{algorithm}[H]
\caption{ROM-FULL: \online phase (Picard iteration)}
\label{alg:reduced_iteration}
\begin{algorithmic}[1]
\State \textbf{Input:} Reduced basis $\mat{\Phi}$, new parameter values $\newparams$,
initial guess $\hat{\mat{u}}^{(0)}$ (vector of zeros)
\State \textbf{repeat}
\State \quad Assemble the full-order operator evaluated at the reconstructed reduced state
$\mat{A}(\mat{\Phi}\,\hat{\mat{u}}^{(k)}, \newparams)$.
\State \quad Form the $r \times r$ reduced matrix
\begin{equation}
\hat{\mat{A}} = \mat{\Phi}^\top
\mat{A}(\mat{\Phi}\,\hat{\mat{u}}^{(k)},\,\newparams)\,\mat{\Phi}
\end{equation}
and reduced load $\hat{\mat{b}} = \mat{\Phi}^\top \mat{b}(\newparams)$.
\State \quad Solve
\begin{equation}
\hat{\mat{A}}\,\hat{\mat{u}}^{(k+1)} = \hat{\mat{b}}
\end{equation}
for the updated reduced solution.
\State \quad Compute the convergence indicator
\begin{equation}
\delta u^{(k)} =
\|\mat{\Phi}\,\hat{\mat{u}}^{(k+1)} - \mat{\Phi}\,\hat{\mat{u}}^{(k)}\|_2.
\end{equation}
\State \textbf{until} $\delta u^{(k)} < 10^{-6}$
\State \textbf{Output:} Converged reduced solution $\hat{\mat{u}}^{(k+1)}$.
\end{algorithmic}
\end{algorithm}

In problems with non-homogeneous Dirichlet boundary conditions, such as the
lid-driven cavity, POD modes constructed directly from full-order snapshots do
not satisfy the prescribed boundary data. Following
\cite{gunzburger2007reduced,reyes2023reduced}, we address this by introducing
a lifting function $\mat{u}_\mathrm{lift}$ that satisfies the Dirichlet boundary
conditions. In this work, $\mat{u}_\mathrm{lift}$ is taken as the mean of the
velocity snapshots,
\begin{equation}
    \mat{u}_\mathrm{lift} = \frac{1}{M}\sum_{i=1}^{M} \mat{u}_i,
\end{equation}
where $\mat{u}_\mathrm{lift}$ denotes the vector of nodal values of the full-order velocity
lifting function, which, by construction, satisfies the prescribed lid velocity
and provides a natural centering of the snapshot data.
The POD basis is then constructed from shifted velocity snapshots
$\mat{u} - \mat{u}_\mathrm{lift}$ that satisfy homogeneous boundary conditions,
and the lifting is added back during reconstruction. The resulting reduced system
reads:
\begin{equation}
    \mat{\Phi}^\top \mat{A}\,
    \mat{\Phi}\,\hat{\mat{u}} = \mat{\Phi}^\top \left(
    \mat{b}
    - \mat{A}\,
    \mat{u}_\mathrm{lift} \right),
\end{equation}
and the full-order velocity solution is then obtained from
\begin{equation}
    \mat{u}_\mathrm{rom} =
    \mat{\Phi}\,\hat{\mat{u}} + \mat{u}_\mathrm{lift}\,.
\end{equation}

We also apply snapshot centering to the settling-sphere benchmark.
However, the primary constraint is not a prescribed boundary velocity but a
constant force acting on the particle. In this case, the ROM must enforce a
force balance rather than non-homogeneous Dirichlet conditions. The corresponding formulation used to impose the constant-force constraint in the ROM-FULL and ROM-DEIM approaches is described in~\ref{sec:force_constraint}.

ROM-FULL serves as a reliable reference model: it preserves the projected FOM physics,
but its \online cost remains tied to the assembly of the full-order system matrix
at each iteration. Improvements in computation time are still anticipated when solving the linear system is the limiting step because the reduced solve itself
is cheap, with $r \ll N_u$.

\subsection{ROM-DEIM}

Although ROM-FULL reduces the size of the system to be solved, the assembly of
the nonlinear operator still depends on the high-fidelity solver. ROM-DEIM
mitigates this cost through hyper-reduction by approximating the nonlinear terms
at only a small set of carefully selected interpolation points.

In ROM-DEIM, the viscosity field is approximated using a separate reduced basis
$\zeta_k(\vec{x})$, $k = 1,\ldots,s$, constructed from viscosity snapshots
in the same way as the velocity basis (see Algorithm~\ref{alg:SVD}).
The viscosity field is then approximated as
\begin{equation}\label{eq:reduced_visc}
    \eta(\dot{\gamma}_h(\vec{x})) \approx \eta_\mathrm{rom}(\vec{x})
    = \hat{\eta}_k\,\zeta_k(\vec{x}),
\end{equation}
where $\hat{\eta}_k \in \mathbb{R}$ are the corresponding reduced viscosity coefficients. This expression
can be written in matrix notation using the reduced basis matrix for the
viscosity, denoted $\mat{\Psi}$, as
\begin{equation}
  \mat{\upeta}_\mathrm{rom} = \mat{\Psi}\,\hat{\mat{\upeta}},
\end{equation}
where the columns of the reduced basis matrix
$\mat{\Psi} \in \mathbb{R}^{N_{\eta}\times s}$ contain the nodal values of the
scalar-valued reduced basis functions $\zeta_k(\vec{x})$,
$k=1,\ldots,s$, expressed in the FEM basis,
$\mat{\upeta}_\mathrm{rom} \in \mathbb{R}^{N_{\eta}}$ are the nodal values of the
viscosity field reconstructed from the reduced basis, and
$\hat{\mat{\upeta}} \in \mathbb{R}^s$ are the reduced POD coefficients. Here, $N_\eta$ denotes the number of degrees of freedom associated with the scalar viscosity field.

Note that obtaining the viscosity from the FOM simulations is non-trivial
because the velocity gradient $\ten{G}_h = (\nabla\vec{u}_h)^\top$ is not globally
continuous across elements in a continuous Galerkin discretization, and directly
evaluating $\dot{\gamma}(\vec{x})$ at arbitrary points can introduce spurious
discontinuities. To avoid this, the discrete gradient is first projected onto a
continuous tensor-valued finite element space $\mathcal{T}_h \subset
[H^1(\Omega)]^{d \times d}$, where each component of $\ten{G}_h$ is represented
using the same continuous piecewise-quadratic Lagrange basis as the velocity
components. The projection is computed by solving the $L^2$-projection problem
\begin{equation}\label{eq:l2proj}
    \text{find } \ten{G}_h \in \mathcal{T}_h \;\text{ such that }\;
    \int_\Omega \ten{G}_h : \ten{V}_h\,\mathrm{d}V
    = \int_\Omega (\nabla\vec{u}_h)^\top : \ten{V}_h\,\mathrm{d}V,
    \quad \forall\,\ten{V}_h \in \mathcal{T}_h,
\end{equation}
yielding a smooth, continuous approximation $\ten{G}_h \approx (\nabla\vec{u}_h)^\top$.
The shear rate is then computed point-wise from $\ten{G}_h$ via
Eq.~\eqref{eq:shear_rate}, and the viscosity snapshot is obtained by evaluating
the Carreau model, Eq.~\eqref{eq:carreau}, at every node.

Substituting the expansions in Eqs.~\eqref{eq:reduced_u}, \eqref{eq:reduced_v}, and
\eqref{eq:reduced_visc} into the weak formulation \eqref{weak_form1} yields the first term
\begin{equation} \label{eq:rom_deim_solve}
\int_\Omega 2\,\eta(\dot{\gamma}_h)\,\ten{D}(\vec{u}_h):\ten{D}(\vec{v}_h) \,\mathrm{d}V
   \approx
   2\,\hat{\upeta}_k\, \hat{v}_j
   \underbrace{\left(\int_\Omega \zeta_k(\vec{x})\,\ten{D}(\vec{\xi}_i) :
   \ten{D}(\vec{\xi}_j)\,\mathrm{d}V \right)}_{T_{ijk}}\,\hat{u}_i ,
\end{equation}
where we recognize the components $T_{ijk}$ of the three-dimensional array
$\mat{T}$, which can be precomputed in the \offline phase. During the \online
phase, the viscosity-dependent reduced stiffness matrix is obtained by a single
contraction of $\mat{T}$ with the reduced viscosity coefficients,
\begin{equation}
  \bigl[\hat{\mat{A}}(\hat{\mat{\upeta}})\bigr]_{ij}
  \;=\; 2\,\hat{\upeta}_k\, T_{ijk},
\end{equation}
yielding an $r \times r$ matrix whose assembly requires no access to the
full-order discretization.

The remaining difficulty is that $\eta$ depends on the full-field shear rate
$\dot{\gamma}_h(\vec{x})$, which requires evaluating the current velocity
solution in the entire domain, an $\mathcal{O}(N_{\eta})$ operation. Empirical
interpolation methods \cite{barrault2004empirical, hesthaven2016certified}
address this by selecting a small number of spatial points at which to sample
the local value of the viscosity and construct an approximation of the viscosity field from those samples. Here, we use the discrete empirical interpolation method
(DEIM) \cite{chaturantabut2010nonlinear}.

When applying DEIM or related hyper-reduction, numerical stability can be a
concern if the sampling points are not well chosen or if the number of sampling points $q$ is too close to the number of viscosity basis functions
$s$. We adopt the GappyPOD+R strategy of Peherstorfer
\emph{et al.}~\cite{peherstorfer2020stability} to improve stability.
The idea is to oversample the viscosity field by selecting more sampling
points than the number of basis functions (\ie, $q > s$), and then determine
the nonlinear coefficients via least-squares fitting instead of exact
interpolation.

Concretely, we proceed in two steps. First, we use a QR factorization with
column pivoting on $\mat{\Psi}^\top$ (where $\mat{\Psi}$ contains the leading
$s$ basis vectors for the viscosity) to pick $s$ important points, a
procedure called QDEIM \cite{drmac2016new}. Next, we randomly select
(hence ``+R'' for randomness) an additional $q-s$ points uniformly from the nodal index set $\{1, \ldots, N_{\eta}\}$, which, on an adaptively refined mesh, results in a spatially denser sampling near the refined regions. This oversampling improves the conditioning of the GappyPOD regression matrix
$\mat{P}^\top\mat{\Psi}$ relative to the square case $q = s$. The cost
increase is negligible since $q-s$ is small. The procedure is given in
Algorithm~\ref{DEIM} for completeness.

\begin{algorithm}[H]
\caption{GappyPOD+R sampling point selection. Initialization by
QDEIM~\cite{drmac2016new}; oversampling as in
\cite[Sec.~6.1]{peherstorfer2020stability}.}
\label{DEIM}
\begin{algorithmic}[1]
\State \textbf{Input:} Viscosity basis matrix
  $\mat{\Psi} = [\zeta_1, \ldots, \zeta_s] \in \mathbb{R}^{N_{\eta} \times s}$,
  oversampling count $q > s$.
\State \textbf{Output:} Selection matrix
  $\mat{P} \in \mathbb{R}^{N_{\eta} \times q}$.
\State Compute the QR factorization with column pivoting of $\mat{\Psi}^\top$
  to obtain $s$ pivot indices $\{\wp_1, \ldots, \wp_s\}$.
\State Initialize $\mat{P} = [\mat{I}_{\wp_1}, \ldots, \mat{I}_{\wp_s}]$
  (the $s$ corresponding columns of the identity matrix).
\State Select $q - s$ additional distinct indices
  $\{\wp_{s+1}, \ldots, \wp_q\}$ uniformly at random from
  $\{1, \ldots, N_{\eta}\}$.
\State Augment $\mat{P} \leftarrow
  [\mat{P} \;\; \mat{I}_{\wp_{s+1}} \;\; \cdots \;\; \mat{I}_{\wp_q}]$.
\end{algorithmic}
\end{algorithm}

The GappyPOD+R approximation of $\hat{\mat{\upeta}}$, required in the
solution of Eq.~\eqref{eq:rom_deim_solve}, is constructed as follows. Let
$\mat{\upeta}_\mathrm{ip} \in \mathbb{R}^q$ denote the viscosity values
evaluated at the $q$ sampling points encoded by $\mat{P}$. These values
are obtained as
\begin{equation}
    \mat{\upeta}_\mathrm{ip}
    = \mat{P}^\top \mat{\upeta}\bigl(\dot\gamma(\hat{\mat{u}};\newparams)\bigr)
    \in \mathbb{R}^q,
\end{equation}
which requires only $q$ point-wise evaluations of $\eta(\dot{\gamma})$
rather than a full-field assembly. The reduced viscosity coefficients
$\hat{\mat{\upeta}} \in \mathbb{R}^s$ are then obtained by solving the
overdetermined system in the least-squares sense:
\begin{equation}
    \hat{\mat{\upeta}} = (\mat{P}^\top \mat{\Psi})^{+}\,
      \mat{\upeta}_\mathrm{ip},
\end{equation}
where $(\cdot)^{+}$ denotes the Moore-Penrose pseudo-inverse, which can
be precomputed in the \offline phase.
 During the \online phase, the only cost
associated with the viscosity is the $q$ point-wise evaluations
$\eta(\dot{\gamma})$ at the selected interpolation points. The \online ROM-DEIM algorithm is
outlined in Algorithm~\ref{alg:rom_deim}.

To evaluate $\eta(\dot{\gamma})$ at the DEIM points, we need the shear rate at
those points. We approximate the velocity gradient field using the reduced basis:
\begin{align}
    \ten{D}(\vec{u}) \; & \approx\; \ten{D}(\hat{u}_i\,\vec{\xi}_i) + \ten{D}(\vec{u}_\mathrm{lift})\,, \\
    & = \hat{u}_i\,\ten{D}(\vec{\xi}_i) + \ten{D}(\vec{u}_\mathrm{lift})\,,
\end{align}
and then compute $\dot{\gamma}(\vec{x})$ from Eq.~\eqref{eq:shear_rate}. The terms
$\ten{D}(\vec{\xi}_i)$ and $\ten{D}(\vec{u}_\mathrm{lift})$ can be precomputed in
the \offline phase using a projection, similar to Eq.~\eqref{eq:l2proj}.

By using DEIM with randomized oversampling, ROM-DEIM significantly reduces the \online cost: instead of evaluating the nonlinear viscosity field over all
$N_{\eta}$ degrees of freedom, only $q$ point-wise evaluations are required (with
$q \ll N_{\eta}$).

\begin{algorithm}[H]
\caption{ROM-DEIM: \online phase (Picard iteration)}
\label{alg:rom_deim}
\begin{algorithmic}[1]
\State \textbf{Input:} Reduced basis $\mat{\Phi} \in \mathbb{R}^{N_u \times r}$,
viscosity basis $\mat{\Psi} \in \mathbb{R}^{N_\eta \times s}$,
precomputed tensor $T_{ijk}$,
pseudo-inverse $\mat{M} = (\mat{P}^\top\mat{\Psi})^{+} \in \mathbb{R}^{s \times q}$,
selection matrix $\mat{P} \in \mathbb{R}^{N_\eta \times q}$,
precomputed gradients $\ten{D}(\vec{\xi}_i)\big|_\mathrm{ip}$ and
$\ten{D}(\mat{u}_\mathrm{lift})\big|_\mathrm{ip}$, lifting function $\mat{u}_\mathrm{lift} \in \mathbb{R}^{N_u}$,
new parameter values $\newparams$,
tolerance $\varepsilon_\mathrm{break} = 10^{-6}$.
\State \textbf{Output:} $\mat{u}_\mathrm{rom} \in \mathbb{R}^{N_u}$.
\State Initialize $\hat{\mat{u}}^{(0)} = \mat{0} \in \mathbb{R}^r$ and $\mat{u}^{(0)} = \mat{0} \in \mathbb{R}^{N_u}$.
\For{$m = 0, 1, 2, \ldots$}
    \State Evaluate $\dot{\gamma}^{(m)}$ at the $q$ DEIM points using
    $\hat{u}_i^{(m)}\,\ten{D}(\vec{\xi}_i)\big|_\mathrm{ip} +
    \ten{D}(\mat{u}_\mathrm{lift})\big|_\mathrm{ip}$
    and Eq.~\eqref{eq:shear_rate}.
    \State Sample the viscosity: $\mat{\upeta}_\mathrm{ip}^{(m)} =
    \eta\!\left(\dot{\gamma}^{(m)}\right) \in \mathbb{R}^q$.
    \State Recover the reduced viscosity coefficients:
    $\hat{\mat{\upeta}}^{(m)} = \mat{M}\,\mat{\upeta}_\mathrm{ip}^{(m)}$.
    \State Assemble and solve the reduced system:
    $\bigl[\hat{\mat{A}}\bigr]_{ij} = 2\,\hat{\upeta}_k^{(m)}\,T_{ijk}$,
    \quad $\hat{\mat{A}}\,\hat{\mat{u}}^{(m+1)} = \hat{\mat{b}}$.
    \State Reconstruct the full nodal values of the velocity field:
    \[
        \mat{u}^{(m+1)} = \mat{\Phi}\,\hat{\mat{u}}^{(m+1)}
        + \mat{u}_\mathrm{lift}.
    \]
    \State Check convergence:
    \[
        \varepsilon^{(m)} =
       \|\mat{u}^{(m+1)} - \mat{u}^{(m)}\|_2.
    \]
    \If{$\varepsilon^{(m)} < \varepsilon_\mathrm{break}$}
        \State \textbf{break}
    \EndIf
\EndFor
\State Reconstruct: $\mat{u}_\mathrm{rom} =
\mat{\Phi}\,\hat{\mat{u}}^{(m+1)} + \mat{u}_\mathrm{lift}$.
\end{algorithmic}
\end{algorithm}

\subsection{ROM-RBF}
\label{sec:rom-rbf}
In many applications, the governing equations or the internals of the full-order solvers are not accessible, for example, when using commercial CFD software or when
constructing reduced models from experimental data. In such settings, a
non-intrusive reduced-order model is the natural choice. The ROM-RBF approach is
fully non-intrusive: rather than projecting and solving a reduced system of
governing equations, it constructs a surrogate that maps the parameters
$\params$ directly to the reduced coefficients $\hat{\mat{u}}(\params)$ through
radial basis function interpolation.

For each snapshot $\mat{u}_i \in \mathbb{R}^{N_u}$ computed at the parameter
sample $\params_i$ during the \offline phase, the reduced coefficients are
obtained by projecting the centered snapshot onto the POD basis,
\begin{equation}
    \hat{\mat{u}}_i
    =
    \mat{\Phi}^{\top}
    \left(\mat{u}_i - \mat{u}_\mathrm{lift}\right),
    \qquad i = 1, \ldots, M,
\end{equation}
which yields the training dataset $\{(\params_i,\, \hat{\mat{u}}_i)\}_{i=1}^{M}$.
Prior to interpolation, each parameter component $\mu^{(j)}$ is mapped to the
unit interval by min-max scaling,
\begin{equation}\label{eq:minmax}
    \tilde{\mu}^{(j)} =
    \frac{\mu^{(j)} - \mu^{(j)}_{\min}}{\mu^{(j)}_{\max} - \mu^{(j)}_{\min}}\,,
\end{equation}
where $\mu^{(j)}_{\min}$ and $\mu^{(j)}_{\max}$ are the minimum and maximum
values of the $j$-th component over the training set. This scaling ensures that
all parameter components contribute equally to the Euclidean distance,
independent of their physical units and magnitudes.

Each reduced component $\hat{u}_k$, $k = 1, \ldots, r$, is approximated by a
radial basis function surrogate
\begin{equation}\label{eq:rbf_surrogate}
    \hat{u}_k(\params) = \sum_{i=1}^{M} w_{k,i}\,
    \phi\!\left(\lVert \tilde{\params} - \tilde{\params}_i \rVert_2\right),
\end{equation}
where $\phi\colon \mathbb{R}_{\geq 0} \to \mathbb{R}$ is a radial kernel and the
$w_{k,i}$ are scalar interpolation weights. The kernels considered in this work
are listed in Table~\ref{RBF_kernels}. Collecting the weights into a matrix
$\mat{W} \in \mathbb{R}^{r \times M}$ and the training coefficients into
$\hat{\mat{U}} = [\hat{\mat{u}}_1, \ldots, \hat{\mat{u}}_M]
\in \mathbb{R}^{r \times M}$, the weights are determined by enforcing exact
interpolation at the training samples,
\begin{equation}\label{eq:rbf_system}
    \mat{W}\,\mat{K} = \hat{\mat{U}}\,,
    \qquad
    K_{ij} = \phi\!\left(\lVert \tilde{\params}_i - \tilde{\params}_j \rVert_2\right),
    \quad i, j = 1, \ldots, M,
\end{equation}
where $\mat{K} \in \mathbb{R}^{M \times M}$ is the symmetric kernel matrix. The
solution $\mat{W} = \hat{\mat{U}}\,\mat{K}^{-1}$ requires a single $M \times M$
linear solve, performed once during the \offline phase.

\begin{table}[!ht]
\centering
\renewcommand{\arraystretch}{1.2}
\begin{tabular}{lc}
\toprule
\textbf{Kernel} & \textbf{Radial function $\phi(\rho)$} \\
\midrule
\multicolumn{2}{l}{\textit{Parameter-independent kernels}} \\
\midrule
Linear              & $\phi(\rho) = \rho$ \\
Cubic               & $\phi(\rho) = \rho^{3}$ \\
Quintic             & $\phi(\rho) = \rho^{5}$ \\
Thin plate spline   & $\phi(\rho) = \rho^{2}\,\log(\rho)$ \\
\midrule
\multicolumn{2}{l}{\textit{Shape-dependent kernels} ($\theta > 0$)} \\
\midrule
Gaussian             & $\phi(\rho) = \exp\!\big(-(\theta \rho)^{2}\big)$ \\
Multiquadric         & $\phi(\rho) = \sqrt{1 + (\theta \rho)^{2}}$ \\
Inverse multiquadric & $\phi(\rho) = 1/\sqrt{1 + (\theta \rho)^{2}}$ \\
\bottomrule
\end{tabular}
\caption{Radial basis function kernels considered in this work.
$\rho = \lVert \tilde{\params} - \tilde{\params}_{i} \rVert_{2}$ is the Euclidean
distance in the normalized parameter space and $\theta > 0$ is the shape
parameter.}
\label{RBF_kernels}
\end{table}

The kernels in Table~\ref{RBF_kernels} are first compared by their mean
prediction error to identify suitable candidates for ROM-RBF interpolation. The
parameter-independent kernels (linear, cubic, quintic, thin plate spline) are
evaluated directly, whereas the shape-dependent kernels (Gaussian, multiquadric,
inverse multiquadric) are assessed over a range of shape parameter values. The
results of this sensitivity study are reported in Section~\ref{sec:results} (see
Figures~\ref{fig:rbf_kernel_comp_lid} and~\ref{fig:rbf_kernel_comp_axi}). The
best-performing kernel from each family is then retained for the subsequent
ROM-RBF analysis.

Prediction at a new parameter value $\newparams$ proceeds as follows. The
parameter is first scaled via Eq.~\eqref{eq:minmax} to obtain
$\tilde{\params}^*$. The kernel vector $\mat{k}(\tilde{\params}^*) \in
\mathbb{R}^M$, with entries $k_i = \phi(\lVert \tilde{\params}^* -
\tilde{\params}_i \rVert_2)$, is evaluated, and the reduced coefficients follow
from $\hat{\mat{u}}(\newparams) = \mat{W}\,\mat{k}(\tilde{\params}^*)$. The nodal
velocity field is reconstructed as $\mat{u}_\mathrm{rom} =
\mat{\Phi}\,\hat{\mat{u}}(\newparams) + \mat{u}_\mathrm{lift}$. The complete
procedure is summarized in Algorithm~\ref{alg:rom_rbf}.

\begin{algorithm}[H]
\caption{ROM-RBF: \online phase}
\label{alg:rom_rbf}
\begin{algorithmic}[1]
\State \textbf{Input:} New parameter values $\newparams$, scaling bounds
       $\{\mu^{(j)}_{\min}, \mu^{(j)}_{\max}\}$, precomputed weight matrix
       $\mat{W} \in \mathbb{R}^{r \times M}$, scaled training samples
       $\{\tilde{\params}_i\}_{i=1}^{M}$, POD basis
       $\mat{\Phi} \in \mathbb{R}^{N_u \times r}$, lifting function
       $\mat{u}_\mathrm{lift} \in \mathbb{R}^{N_u}$.
\State \textbf{Output:} Reconstructed velocity
       $\mat{u}_\mathrm{rom} \in \mathbb{R}^{N_u}$.
\State Scale the query parameter via Eq.~\eqref{eq:minmax} to obtain
       $\tilde{\params}^*$.
\State Evaluate the kernel vector
       $k_i = \phi\!\left(\lVert \tilde{\params}^* - \tilde{\params}_i
       \rVert_2 \right)$, $\quad i = 1, \ldots, M$.
\State Compute the reduced coefficients
       $\hat{\mat{u}}(\newparams) = \mat{W}\,\mat{k}(\tilde{\params}^*)
       \in \mathbb{R}^r$.
\State Reconstruct $\mat{u}_\mathrm{rom} =
       \mat{\Phi}\,\hat{\mat{u}}(\newparams) + \mat{u}_\mathrm{lift}$.
\end{algorithmic}
\end{algorithm}

\FloatBarrier

\section{Numerical results}
\label{sec:results}
This section assesses the accuracy, convergence, and robustness of ROM-FULL,
ROM-DEIM and ROM-RBF on two benchmark problems: a lid-driven cavity and a
sphere settling in a closed container. These cases represent boundary-driven and
force-driven flows, respectively.

Throughout, ROM errors are measured in the relative $\ell^2$ norm. For a
parameter value $\params$, the relative velocity and viscosity errors are
\begin{equation}
\varepsilon_{u}(\params) =
\frac{\|\,\mat{u}_\mathrm{rom}(\params) - \mat{u}_\mathrm{fom}(\params)\,\|_2}
     {\|\,\mat{u}_\mathrm{fom}(\params)\,\|_2}\,,
     \qquad
\varepsilon_{\eta}(\params) =
\frac{\|\,\mat{\upeta}_\mathrm{rom}(\params) - \mat{\upeta}_\mathrm{fom}(\params)\,\|_2}
     {\|\,\mat{\upeta}_\mathrm{fom}(\params)\,\|_2}\,,
\label{eq:relative_error_velocity}
\end{equation}
where $\|\cdot\|_2$ denotes the Euclidean norm over all degrees of freedom. For a
test set $\mathcal{P} = \{\params_1, \ldots, \params_{M_\mathrm{test}}\}$ of size
$M_\mathrm{test}$, the mean relative errors are
\begin{equation}
\bar{\varepsilon}_{u}(\mathcal{P}) =
\frac{1}{M_\mathrm{test}} \sum_{i=1}^{M_\mathrm{test}}
\varepsilon_{u}(\params_i)\,, \qquad
\bar{\varepsilon}_{\eta}(\mathcal{P}) =
\frac{1}{M_\mathrm{test}} \sum_{i=1}^{M_\mathrm{test}}
\varepsilon_{\eta}(\params_i)\,.
\end{equation}
The generation of the training and test sets for each problem is described in the
respective sections.

\subsection{Lid-driven cavity}
\begin{figure}[!ht]
\centering
\includegraphics[width=0.33\linewidth]{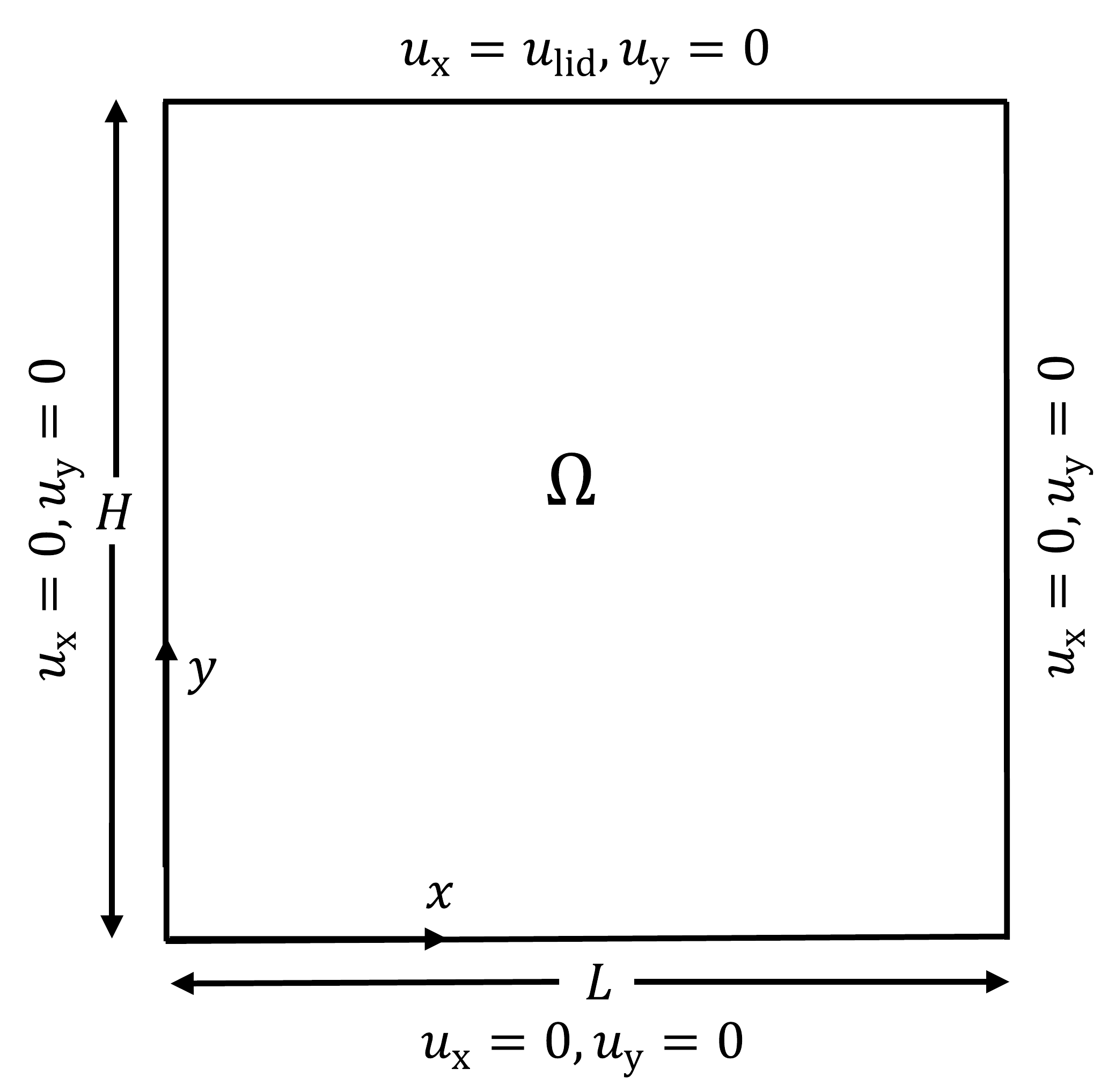}
\caption{Geometry and boundary conditions of the lid-driven cavity benchmark.}
\label{fig:lid_schematic}
\end{figure}

\begin{figure}[!ht]
    \centering
    \begin{subfigure}[t]{0.32\linewidth}
        \centering
        \includegraphics[width=\linewidth]{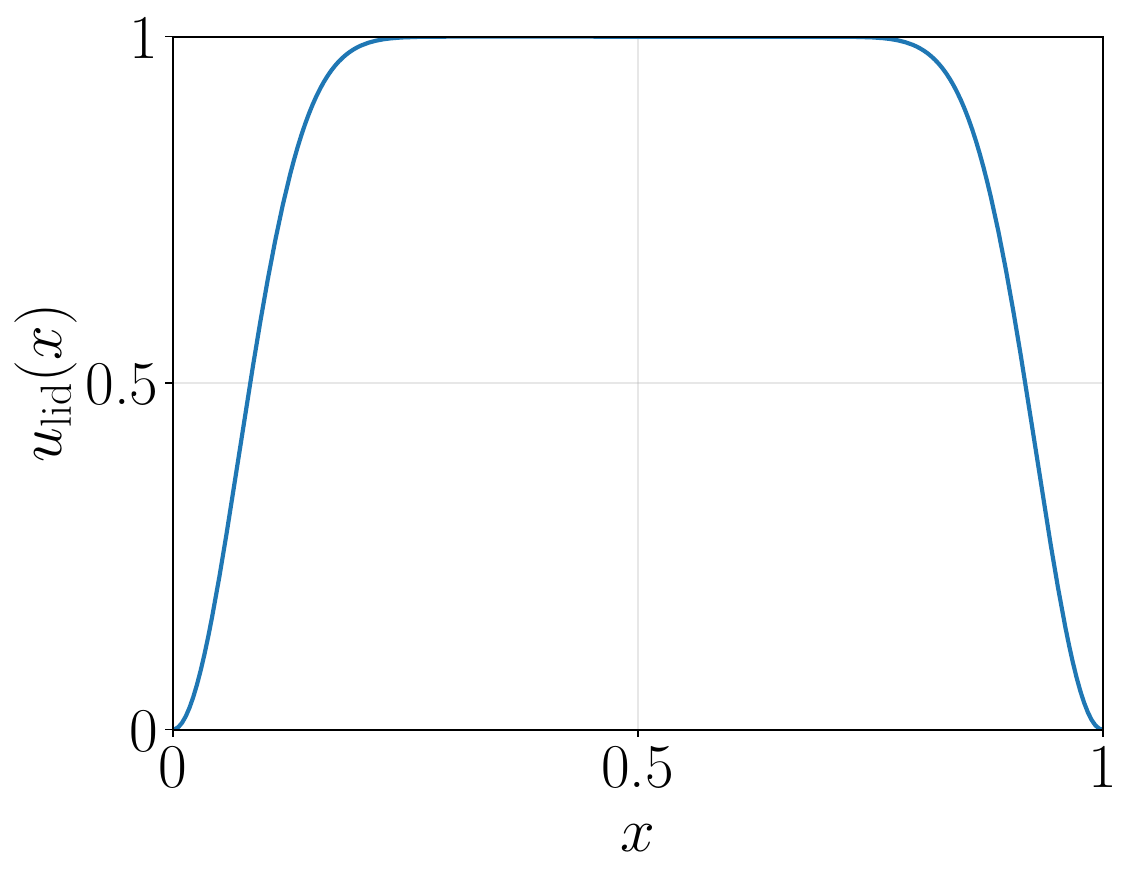}
        \caption{velocity profile}
        \label{lid_velocity_profile}
    \end{subfigure}
    \begin{subfigure}[t]{0.32\linewidth}
        \centering
        \includegraphics[width=\linewidth]{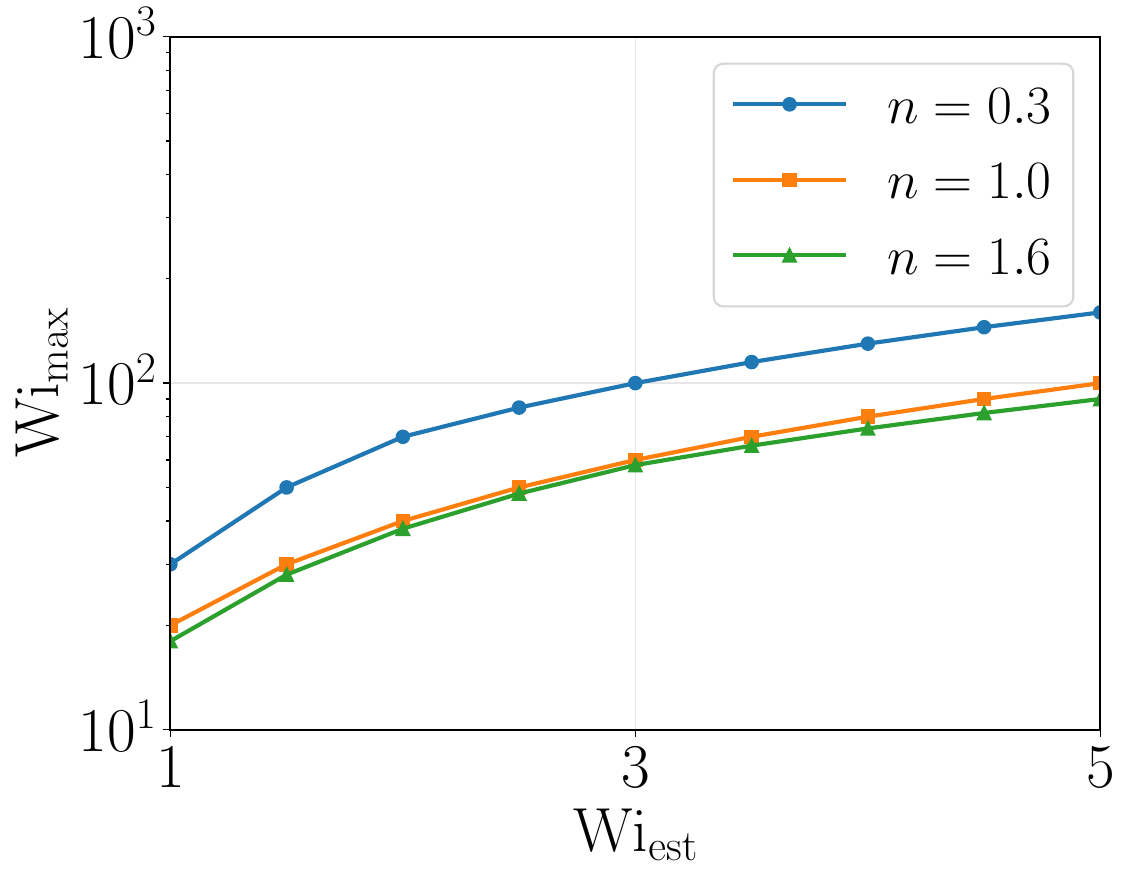}
        \caption{$\wi_{\mathrm{est}}$ vs.\ $\wi_{\mathrm{max}}$}
        \label{true_wi_lid}
    \end{subfigure}
    \begin{subfigure}[t]{0.33\linewidth}
        \centering
        \includegraphics[width=\linewidth]{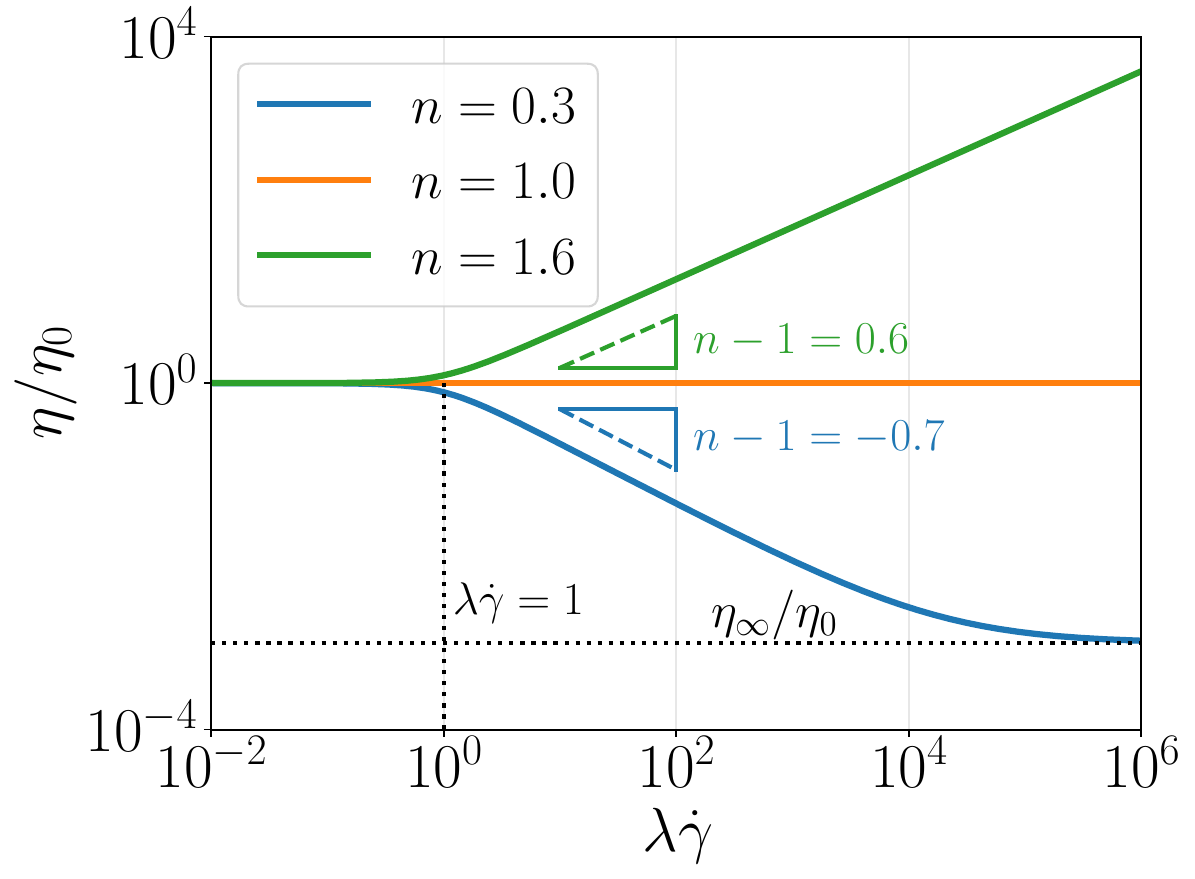}
        \caption{normalized Carreau viscosity}
        \label{carreau_model}
    \end{subfigure}
\caption{(a) Prescribed tangential velocity profile on the moving lid.
(b) True maximum Weissenberg number $\wi_\mathrm{max}$, evaluated from the actual
maximum shear rate, compared with the estimate $\wi_\mathrm{est} = \lambda U/L$
for $n \in \{0.3,\,1.0,\,1.6\}$, covering the shear-thinning, Newtonian, and
shear-thickening regimes. (c) Normalized Carreau viscosity $\eta/\eta_0$ versus
$\lambda\dot{\gamma}$ for the same values of $n$.}
\end{figure}

The lid-driven cavity is a classical flow benchmark \cite{bruneau2006cavity},
characterized by recirculation induced by a moving wall. The problem is posed in
Cartesian coordinates $(x,y)$ on a unit-square domain ($L = H = 1$), as
illustrated in Figure~\ref{fig:lid_schematic}. Let $\Omega = [0,1]\times[0,1]$
with boundary $\partial\Omega$, and let
$\Gamma_\mathrm{lid} = \{(x,1) \mid x \in [0,1]\}$ denote the moving lid. The
boundary conditions are
\begin{align}
\vec{u}(x,y) &= u_\mathrm{lid}(x)\,\vec{e}_x && \text{on } \Gamma_\mathrm{lid}, \\
\vec{u}(x,y) &= \vec{0} && \text{on } \partial\Omega \setminus \Gamma_\mathrm{lid},
\end{align}
where $\vec{e}_x$ is the unit vector in the $x$-direction. The lid velocity
profile is prescribed as
\begin{equation}
u_\mathrm{lid}(x) =
1 - \left[ \tfrac{1}{2} - \tfrac{1}{2}\cos\!\big(\pi(2x - 1)\big) \right]^{10},
\quad x \in [0,1],
\end{equation}
which, as plotted in Figure~\ref{lid_velocity_profile}, has a maximum of one at
the center and decays to zero at the edges ($x=0$ and $x=1$). The pressure level
is fixed by imposing $p(0,0) = 0$.
The meshes are generated using Gmsh \cite{geuzaine2009gmsh}. Based on a
mesh-convergence study reported in the Supplementary Material, which compares five
successively refined meshes (M1 to M5), we use mesh M2 for all high-fidelity
lid-driven cavity simulations; this mesh is refined along the moving lid and near
the top corners, where the velocity gradients are largest.

We use the Carreau model of Eq.~\eqref{eq:carreau}, fixing the zero-shear and infinite-shear viscosities to $\eta_0=1$ and $\eta_\infty=10^{-3}$, respectively. The
parametric study varies the power-law index $n$ and the characteristic time scale
$\lambda$, \ie $\params = (n,\lambda)$, while the remaining parameters are kept
fixed. The ranges $n\in[0.3,1.6]$ and $\lambda\in[1,5]$ span shear-thinning
($n<1$), near-Newtonian ($n\approx 1$), and shear-thickening ($n>1$) behavior.
To quantify the degree of shear-rate dependence across the chosen $(n,\lambda)$
range, we use the Weissenberg number $\wi$, defined as the product of the
characteristic time $\lambda$ and a characteristic shear rate. From the
simulations, we define the true maximum value
\begin{equation}
\wi_\mathrm{max} = \lambda\,\max_{\Omega}\dot{\gamma},
\label{eq:wi_max}
\end{equation}
and, for reference, the standard estimate based on the maximum lid speed $U$ and
the cavity length $L$,
\begin{equation}
\wi_\mathrm{est} = \frac{\lambda U}{L},
\end{equation}
which reduces to $\wi_\mathrm{est}=\lambda$ in the present non-dimensional setting,
since $U=L=1$. Figure~\ref{true_wi_lid} compares the estimated Weissenberg number
$\wi_\mathrm{est}$ with the maximum value obtained from the FOM,
$\wi_\mathrm{max}$. For all three rheological
regimes, $\wi_\mathrm{max}$ is substantially larger than $\wi_\mathrm{est}$,
indicating that the simple global estimate $U/L$ underestimates the largest
local shear rates in the cavity. The discrepancy is largest for the
shear-thinning case ($n=0.3$), where the viscosity reduction near the moving lid localizes the deformation and produces higher maximum shear rates. The
Newtonian and shear-thickening cases show lower and more comparable values of
$\wi_\mathrm{max}$ over the same range of $\wi_\mathrm{est}$.

The role of each Carreau parameter is illustrated in
Figure~\ref{carreau_model}: $\lambda$ sets the onset of non-Newtonian behavior at
$\lambda\dot{\gamma} = 1$, separating the zero-shear plateau $\eta_0$ from the
infinite-shear plateau $\eta_\infty$, while the power-law index $n$ fixes the
asymptotic log-log slope $n-1$, with $n<1$, $n=1$, and $n>1$ corresponding to
shear-thinning, Newtonian, and shear-thickening regimes, respectively.

\begin{figure}[!ht]
  \centering
  \begin{subfigure}[t]{0.22\linewidth}
    \centering
    \includegraphics[width=\linewidth]{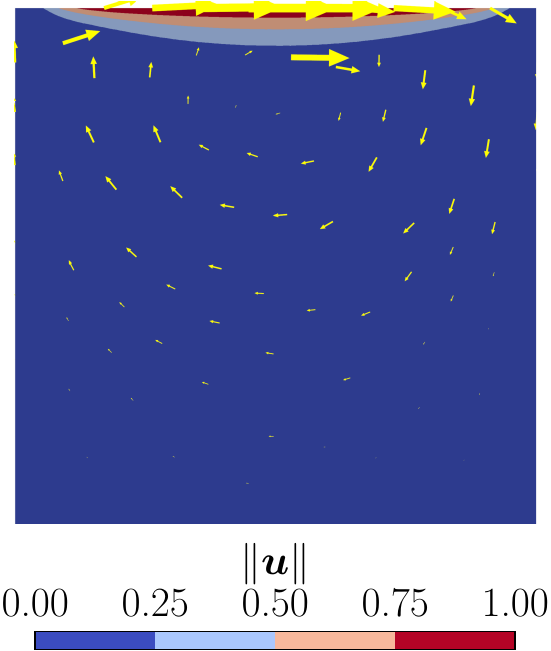}
    \caption{Velocity magnitude, $n = 0.3$, $\lambda = 5.0$}
  \end{subfigure}
  \hspace{0.4cm}
  \begin{subfigure}[t]{0.22\linewidth}
    \centering
    \includegraphics[width=\linewidth]{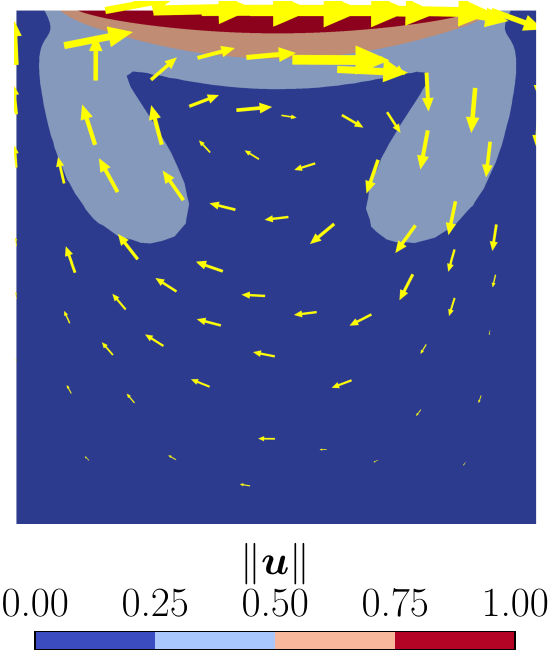}
    \caption{Velocity magnitude, $n = 1.0$, $\lambda = 5.0$}
  \end{subfigure}
  \hspace{0.4cm}
  \begin{subfigure}[t]{0.22\linewidth}
    \centering
    \includegraphics[width=\linewidth]{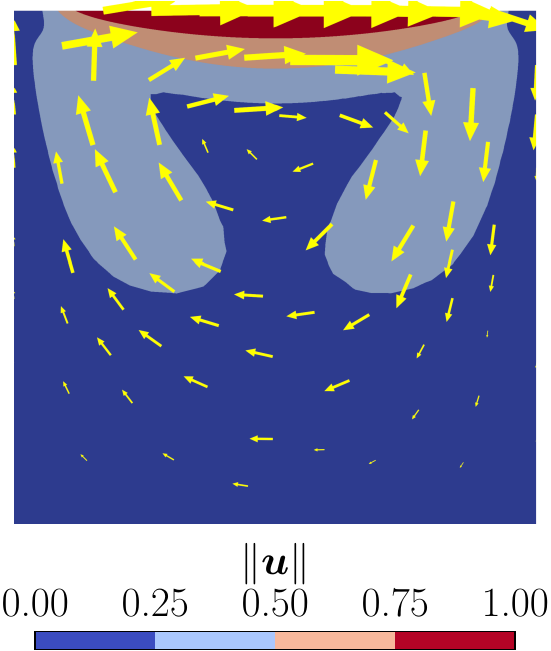}
    \caption{Velocity magnitude, $n = 1.6$, $\lambda = 5.0$}
  \end{subfigure}
  
  \vspace{0.5em}
  
  \begin{subfigure}[t]{0.22\linewidth}
    \centering
    \includegraphics[width=\linewidth]{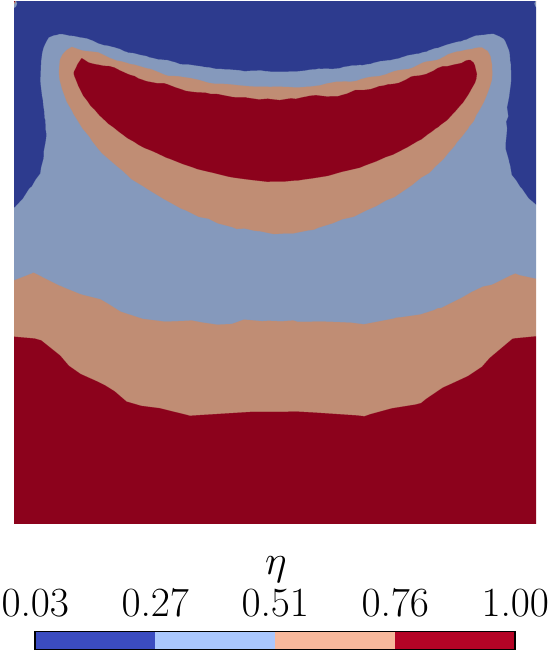}
    \caption{Viscosity field, $n = 0.3$, $\lambda = 5.0$}
  \end{subfigure}
  \hspace{0.4cm}
  \begin{subfigure}[t]{0.22\linewidth}
    \centering
    \includegraphics[width=\linewidth]{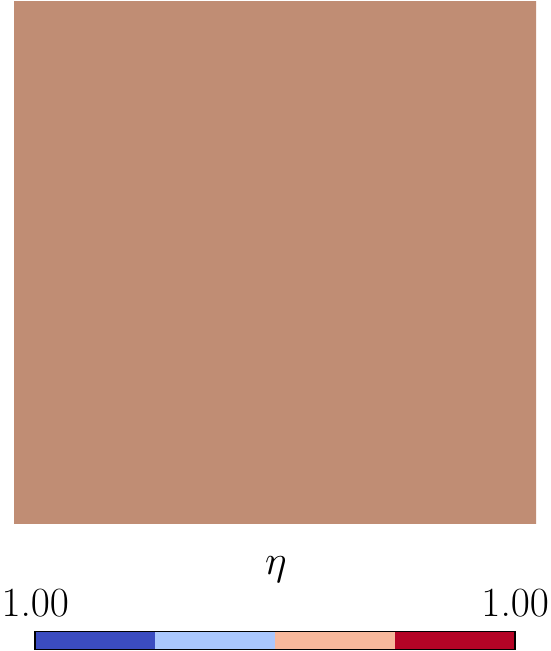}
    \caption{Viscosity field, $n = 1.0$, $\lambda = 5.0$}
  \end{subfigure}
  \hspace{0.4cm}
  \begin{subfigure}[t]{0.22\linewidth}
    \centering
    \includegraphics[width=\linewidth]{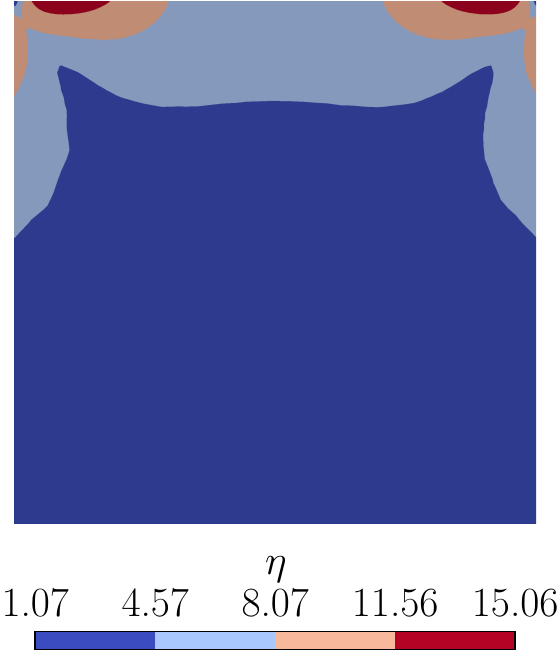}
    \caption{Viscosity field, $n = 1.6$, $\lambda = 5.0$}
  \end{subfigure}
\caption{FOM solutions for the lid-driven cavity at $\lambda = 5.0$. Top row,
(a) to (c): velocity magnitude, with yellow arrows indicating the velocity
magnitude and direction. Bottom row, (d) to (f): viscosity field. Columns
correspond to a shear-thinning fluid ($n=0.3$, left), the Newtonian case
($n=1$, middle), and a shear-thickening fluid ($n=1.6$, right).}
  \label{fig:fom_lid_velocity_viscosity}
\end{figure}

The FOM velocity and viscosity fields are presented in
Figure~\ref{fig:fom_lid_velocity_viscosity} for $\lambda = 5.0$ and the
shear-thinning ($n=0.3$), Newtonian ($n=1$), and shear-thickening ($n=1.6$)
fluids. In the Newtonian case, a region of high shear is visible near the moving lid. For the shear-thinning fluid, this region becomes significantly thinner, owing to the local reduction in viscosity. In the shear-thickening case, the velocity spreads over a larger region, and the gradients are
reduced because the viscosity increases near the lid, especially at the corners
where the tangential velocity gradient is steepest (Figure~\ref{lid_velocity_profile}).

\subsubsection{POD basis generation}
To generate the training data, the parameter space $\params=(n,\lambda)$ is
sampled using uniform grids of increasing resolution over the domain
$n \in [0.3,\,1.6]$ and $\lambda \in [1,\,5]$. Three nested training sets are
considered: $S_1$ ($3\times3$, 9 samples), $S_2$ ($5\times5$, 25 samples), and
$S_3$ ($9\times9$, 81 samples), with $S_1 \subset S_2 \subset S_3$.

\begin{figure}[!ht]
    \centering
    \begin{subfigure}[t]{0.33\linewidth}
        \centering
        \includegraphics[width=\linewidth]{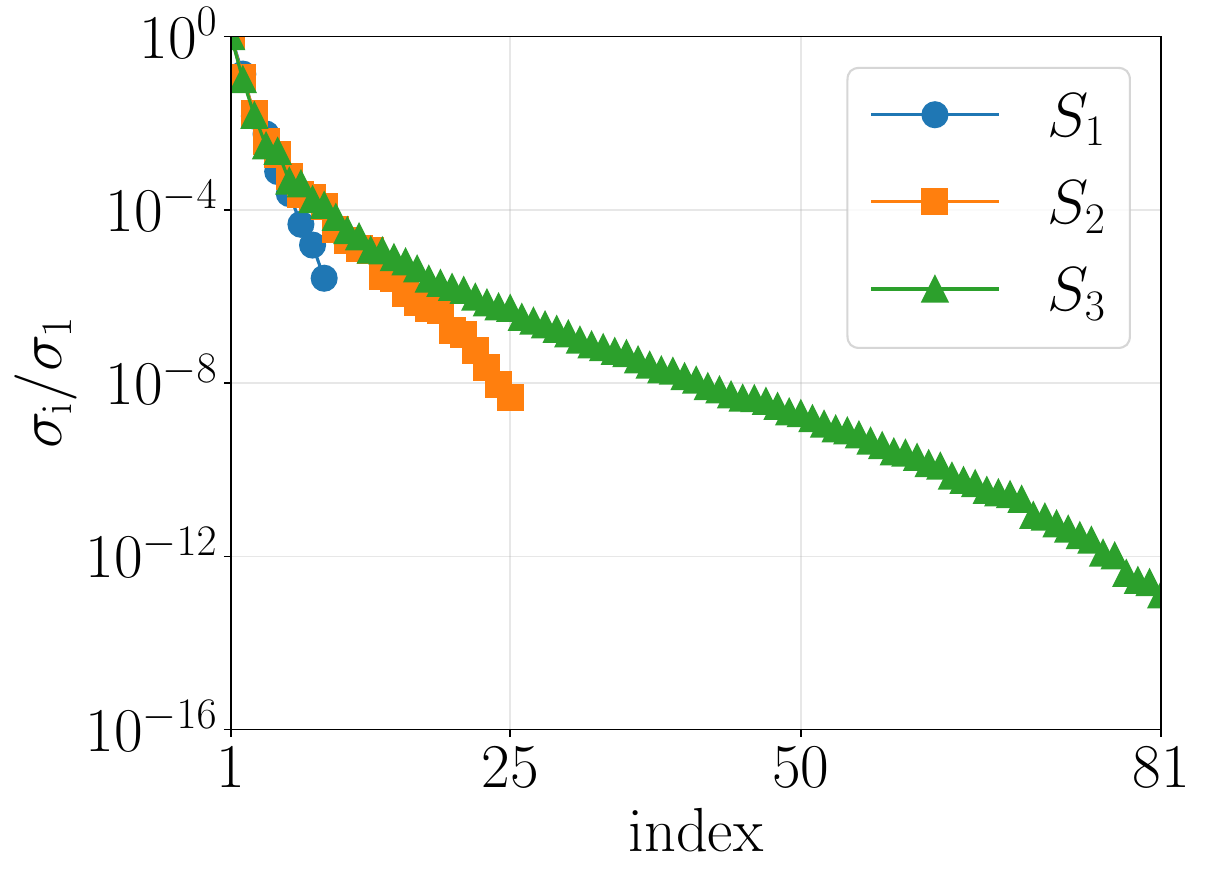}
        \caption{velocity}
    \end{subfigure}
    \begin{subfigure}[t]{0.33\linewidth}
        \centering
        \includegraphics[width=\linewidth]{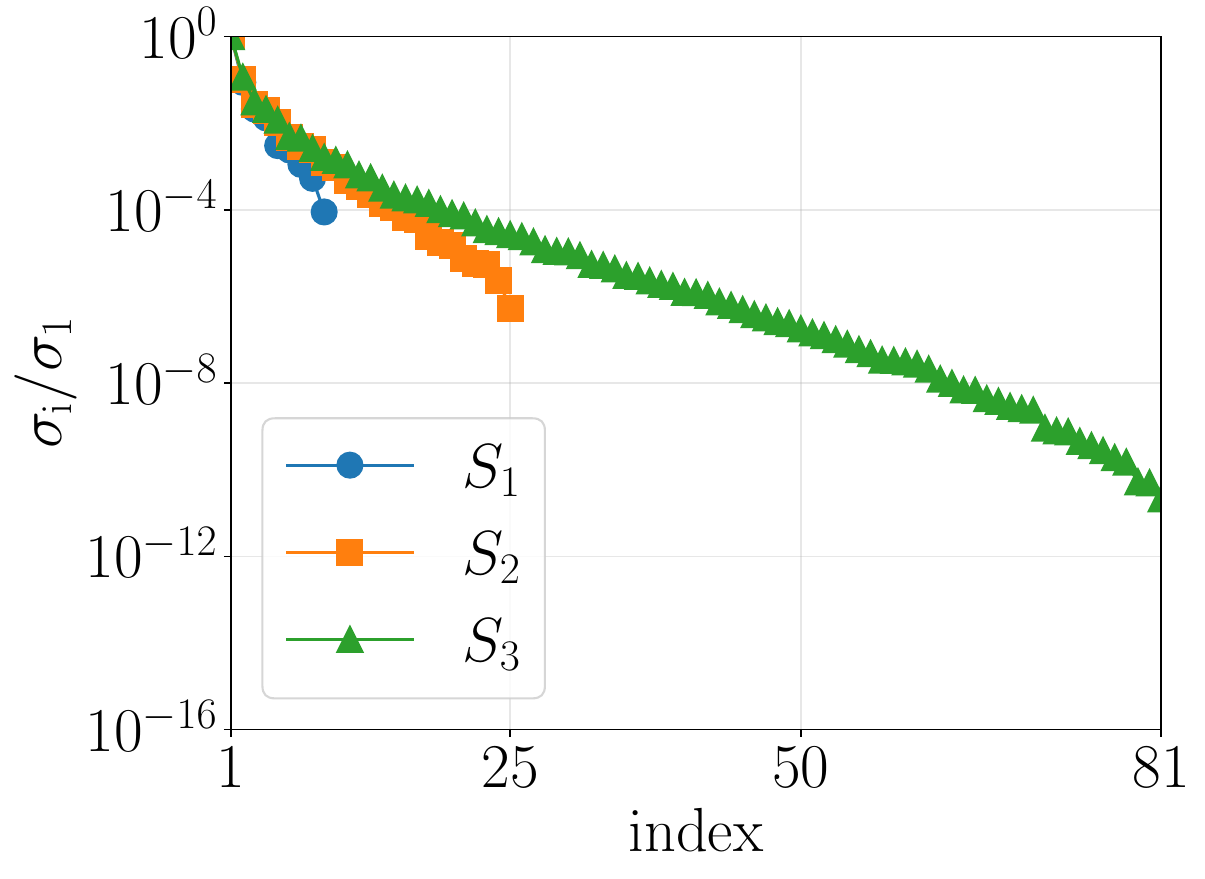}
        \caption{viscosity}
    \end{subfigure}
\caption{Normalized singular value decay for the lid-driven cavity snapshot
sets, scaled by the leading singular value. (a) Singular values of the velocity
snapshots. (b) Singular values of the viscosity snapshots.}
    \label{svd_velocity_lid}
\end{figure}

\begin{figure}[!ht]
    \centering
    \begin{subfigure}[t]{0.22\linewidth}
        \centering
\includegraphics[width=\textwidth]{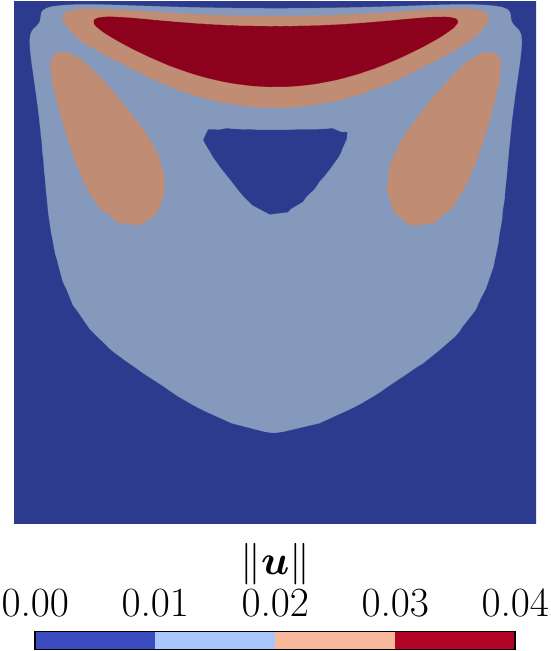}
        \caption{Velocity mode 1}
    \end{subfigure}
    \begin{subfigure}[t]{0.22\linewidth}
        \centering
        \includegraphics[width=\textwidth]{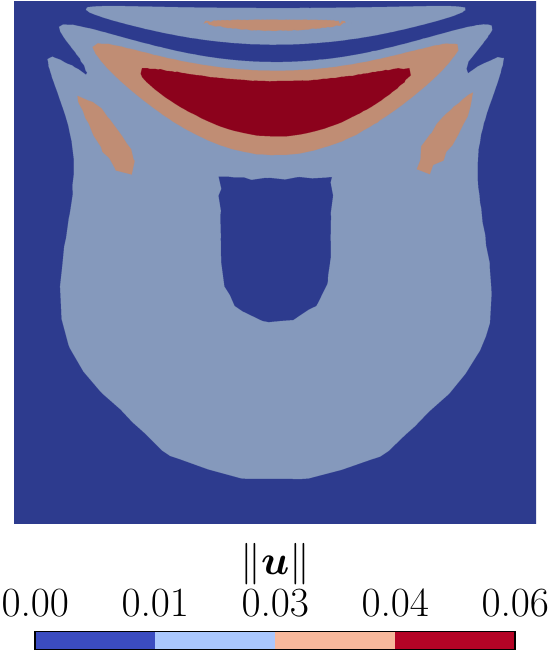}
        \caption{Velocity mode 2}
    \end{subfigure}
    \begin{subfigure}[t]{0.22\linewidth}
        \centering
        \includegraphics[width=\textwidth]{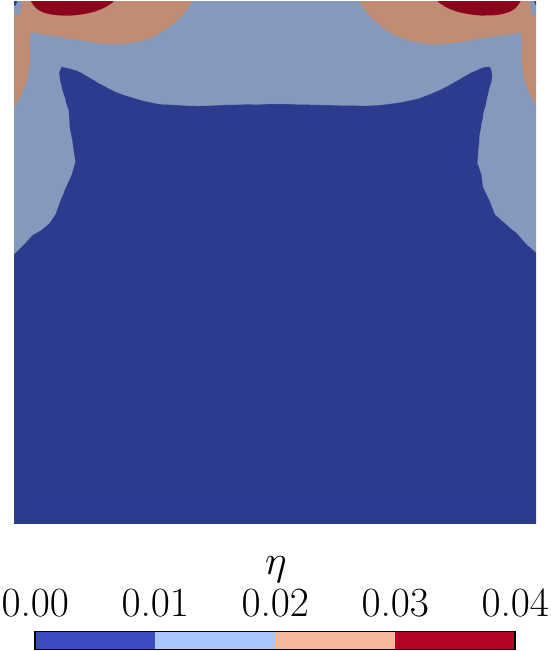}
        \caption{Viscosity mode 1}
    \end{subfigure}
    \begin{subfigure}[t]{0.22\linewidth}
        \centering
        \includegraphics[width=\textwidth]{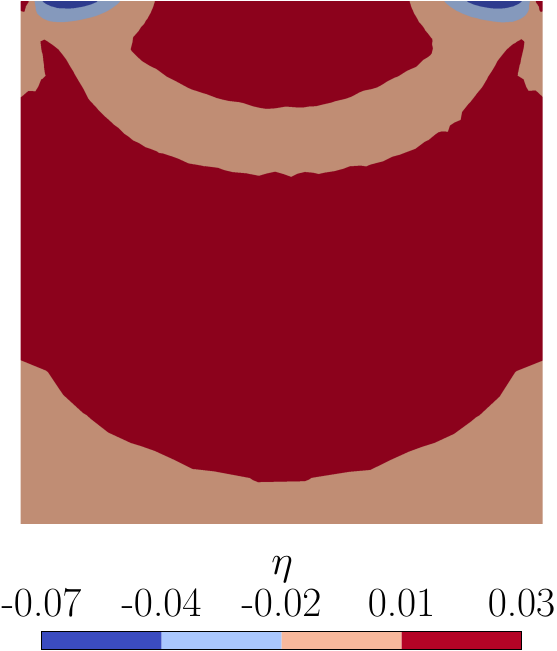}
        \caption{Viscosity mode 2}
    \end{subfigure}
\caption{Dominant POD modes for the lid-driven cavity, extracted from the
finest training ensemble $S_3$. Panels (a) and (b) show the first two centered
velocity modes, while panels (c) and (d) display the first two viscosity modes.}
    \label{svd_basis_lid}
\end{figure}

Separate POD bases are constructed for the velocity and viscosity fields by
singular value decomposition of the respective snapshot matrices
(Section~\ref{sec:svd}). The normalized singular values, shown in
Figure~\ref{svd_velocity_lid}, decay slowly for both fields, indicating that a
large number of modes is required to represent the snapshots accurately. The
viscosity spectrum decays more slowly than the velocity spectrum, reflecting the
more localized, nonlinear structure of the viscosity field under shear-thinning
conditions. The leading POD modes (Figure~\ref{svd_basis_lid}) are consistent
with this observation: the centered velocity modes are smooth and globally
distributed, whereas the viscosity modes are concentrated near the moving lid,
where the shear rates are highest.

Using these POD bases, we construct the ROM-FULL, ROM-DEIM, and ROM-RBF models.
Their performance is evaluated on an independent test set of
$M_\mathrm{test}=100$ parameter values, generated in the $(n,\lambda)$ space by
Latin hypercube sampling (LHS).

We consider two
regimes: an interpolation regime, in which the test points lie within the training
domain, and an extrapolation regime, in which $\lambda$ is extended beyond the
training range. 

\subsubsection{Hyperparameter tuning}
While ROM-FULL has no method-specific hyperparameters beyond the POD truncation tolerance, both ROM-DEIM and ROM-RBF have additional
hyperparameters that must be chosen carefully. Here, we investigate their effect on the ROM accuracy evaluated on the test set.

A hyperparameter of ROM-DEIM is the number of viscosity basis functions $s$ used
to approximate the nonlinear viscosity field. To isolate the effect of $s$, the
POD tolerance for the velocity field is fixed at $\epsilon_\mathrm{pod} = 10^{-8}$,
\ie the number of velocity basis vectors $r$ is the smallest value for which the
corresponding normalized singular value falls below this threshold (see
Figure~\ref{svd_velocity_lid}); for cases in which all normalized singular values
exceed the threshold, all basis vectors are retained ($r=M$). The number of (oversampled) DEIM interpolation points is set to $q = 2s$. This is a pragmatic choice, supported by numerical experiments: it provides sufficient oversampling to ensure the stability of the GappyPOD+R reconstruction without significantly increasing the \online computational cost.

\begin{figure}[!ht]
    \centering
    \begin{subfigure}[t]{0.33\linewidth}
        \centering
        \includegraphics[width=\linewidth]{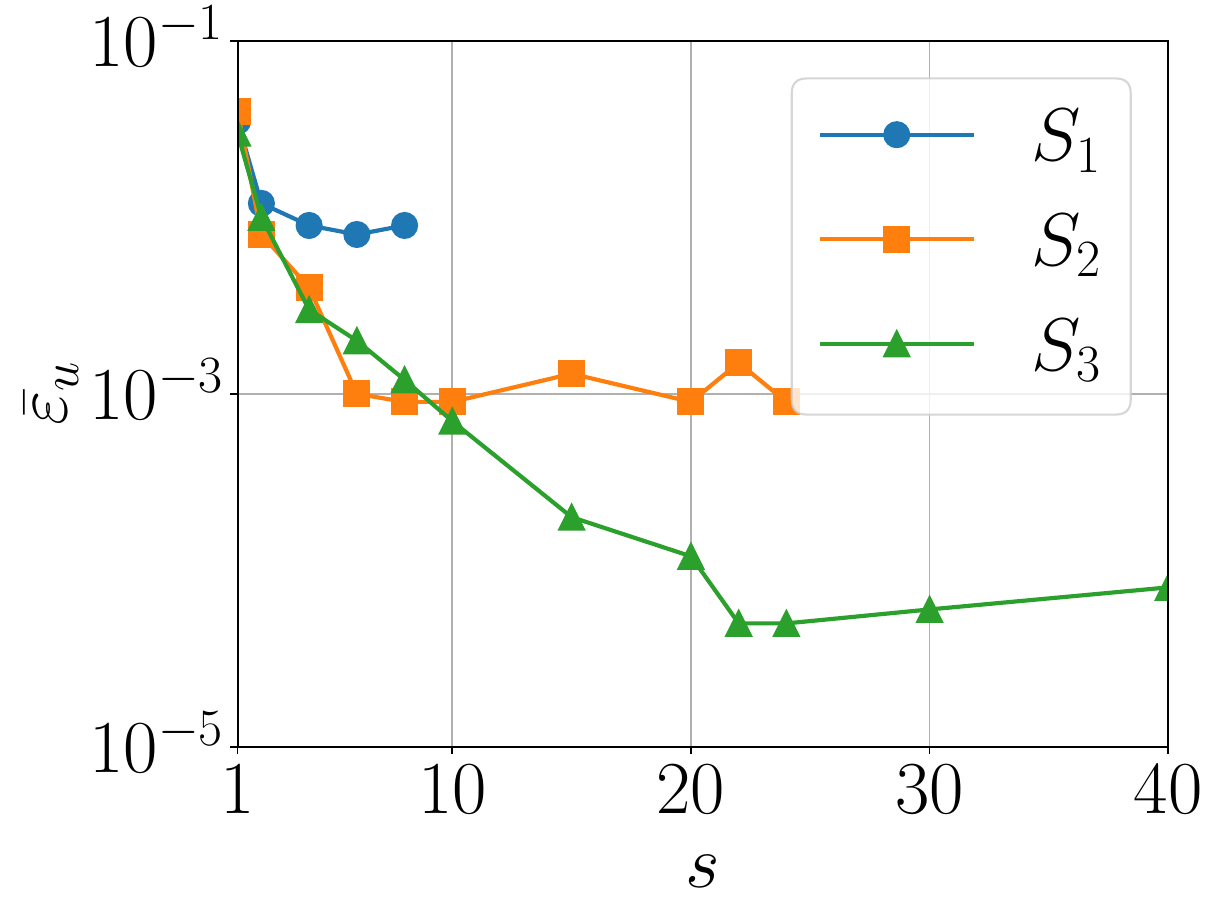}
        \caption{velocity}
    \end{subfigure}
    \begin{subfigure}[t]{0.33\linewidth}
        \centering
        \includegraphics[width=\linewidth]{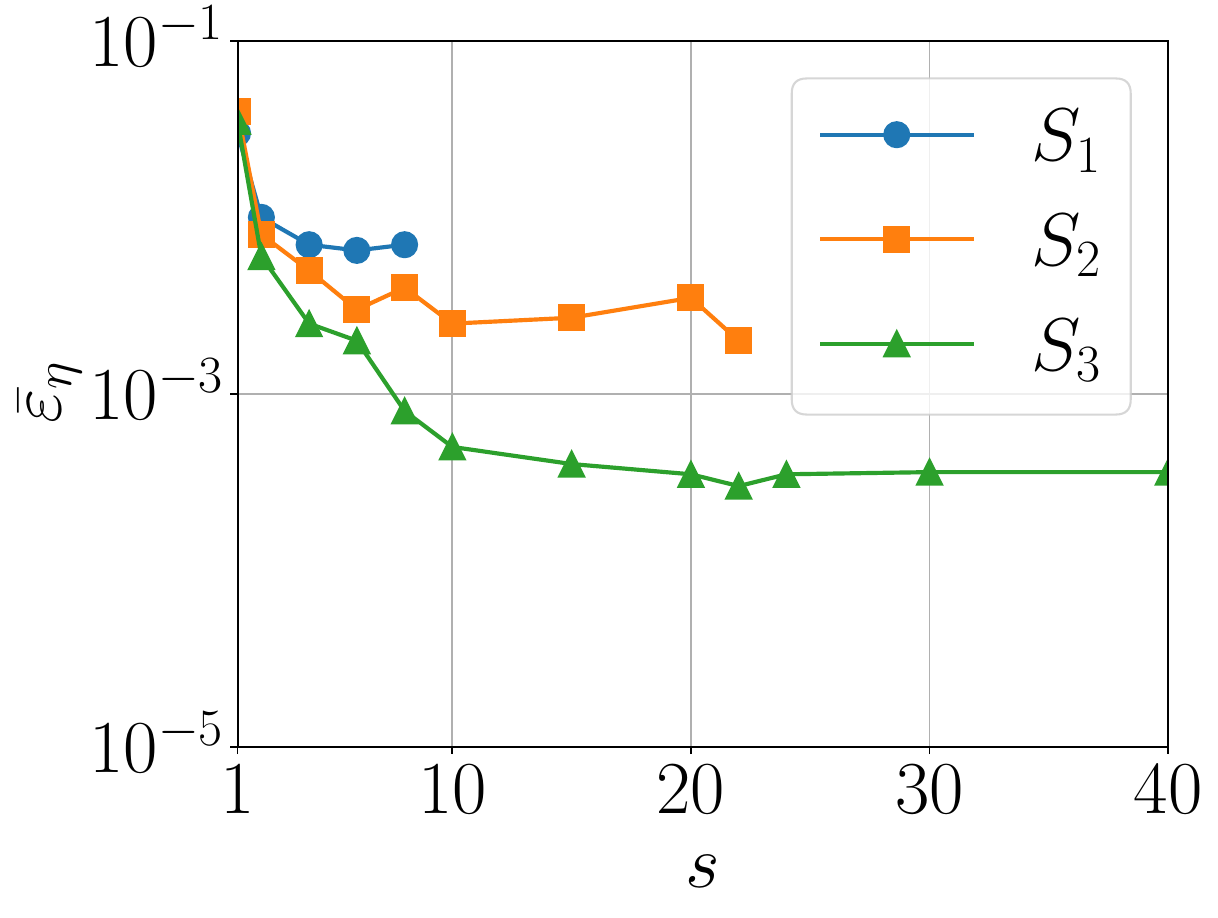}
        \caption{viscosity}
    \end{subfigure}
\caption{Convergence of the ROM-DEIM error for the lid-driven cavity with
respect to the number of viscosity basis functions $s$, where the number of
DEIM interpolation points is set to $q = 2s$. The POD truncation tolerance is
fixed at $\epsilon_\mathrm{pod} = 10^{-8}$. Panel (a) shows the mean velocity
error $\bar{\varepsilon}_{u}$, while panel (b) shows the mean viscosity
error $\bar{\varepsilon}_{\eta}$, each evaluated over the three training
samples $S_1$, $S_2$, and $S_3$.}
    \label{deim-conv-lid}
\end{figure}

Figure~\ref{deim-conv-lid} shows the mean relative error as a function of $s$ for
each training set. The error decreases with $s$ and then saturates once $s$ is
large enough that the DEIM approximation is no longer the dominant error source.
The saturation threshold grows with the training-set size, approximately
$s \approx 8$ for $S_1$, $s \approx 20$ for $S_2$, and $s \approx 24$ for $S_3$,
reflecting the increasing complexity of the viscosity field captured by larger
snapshot sets. These values are therefore adopted as the default ROM-DEIM
settings throughout this section (with $q = 2s$).

\begin{figure}[!ht]
    \centering
    \includegraphics[width=0.70\linewidth]{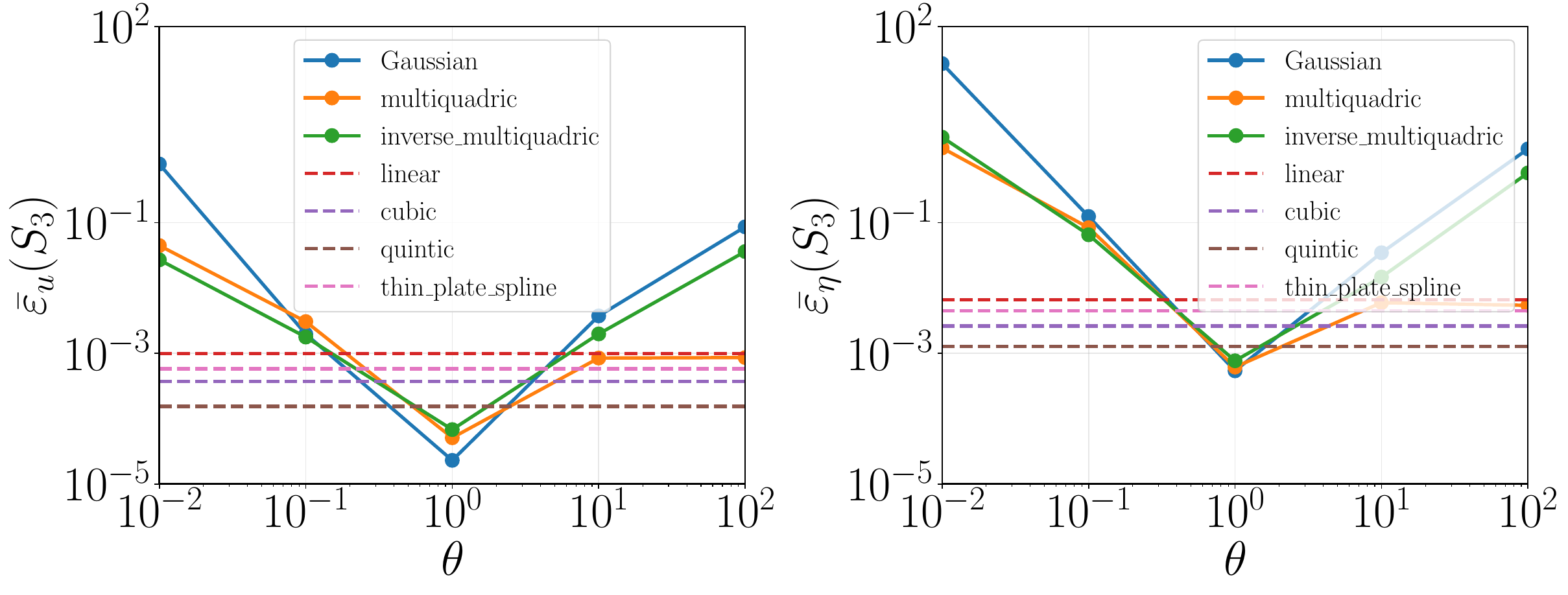}
\caption{Kernel comparison for the lid-driven cavity ROM-RBF, trained on the 
 training sample $S_{3}$. The mean velocity error 
$\bar{\varepsilon}_{u}(S_{3})$ (left) and the mean viscosity error 
$\bar{\varepsilon}_{\eta}(S_{3})$ (right) are reported as functions of the 
shape parameter $\theta$ for the Gaussian, multiquadric, and inverse 
multiquadric kernels. The linear, cubic, quintic, and thin plate spline 
kernels are reported as parameter-independent references.}
    \label{fig:rbf_kernel_comp_lid}
\end{figure}

For ROM-RBF, the shape-dependent kernels (Gaussian, multiquadric, inverse
multiquadric) are compared with the parameter-independent kernels (linear, cubic,
quintic, thin plate spline) listed in Table~\ref{RBF_kernels}, using the training
sample $S_{3}$ as the reference dataset. Figure~\ref{fig:rbf_kernel_comp_lid}
shows that the shape-dependent kernels are strongly sensitive to the shape
parameter $\theta$, with the minimum mean velocity and viscosity errors attained
near $\theta = 1$. This optimum is consistent with the min-max normalization of
the parameter space (Section~\ref{sec:rom-rbf}), which fixes the characteristic
length scale of the inputs to unity. At $\theta = 1$, the Gaussian kernel gives
the lowest velocity error of all candidates and a viscosity error at least as low
as the other kernels, so the Gaussian kernel with $\theta = 1$ is adopted as the
default ROM-RBF kernel for the lid-driven cavity benchmark.

\subsubsection{Comparison of ROMs}
\begin{figure}[!ht]
    \centering
    \includegraphics[width=\linewidth]{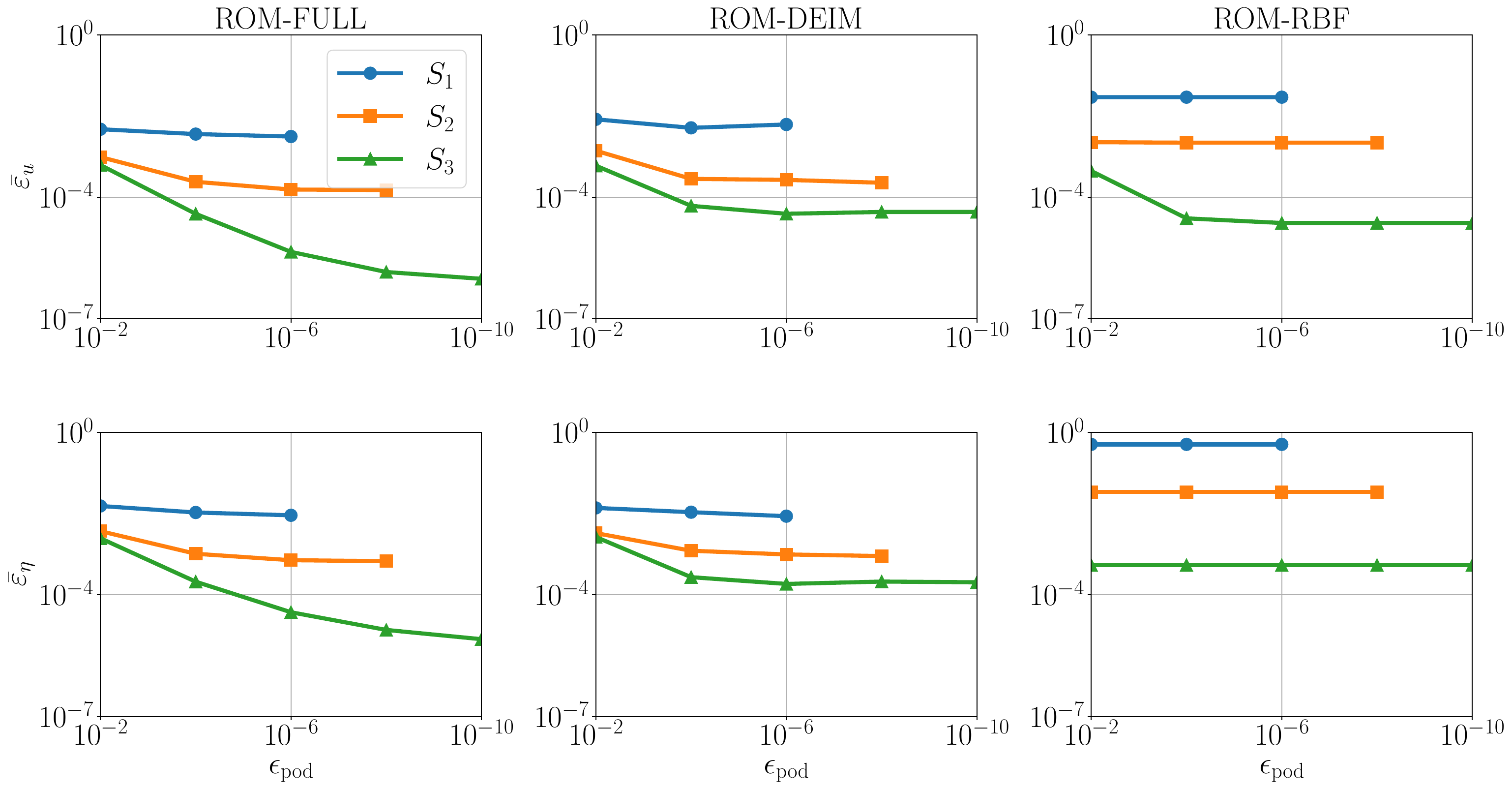}
\caption{Convergence of the mean relative error with respect to the POD
truncation tolerance $\epsilon_\mathrm{pod}$ for the lid-driven cavity ROMs.
Columns correspond to the three ROM strategies (ROM-FULL, ROM-DEIM, ROM-RBF);
the top row reports the mean velocity error $\bar{\varepsilon}_{u}$ and the
bottom row the mean viscosity error $\bar{\varepsilon}_{\eta}$. Within each panel, the curves correspond to
training sets of increasing size, $S_1 \subset S_2 \subset S_3$.}
    \label{fig:eps-convergence-lid}
\end{figure}

\begin{figure}[!ht]
    \centering
    \includegraphics[width=\linewidth]{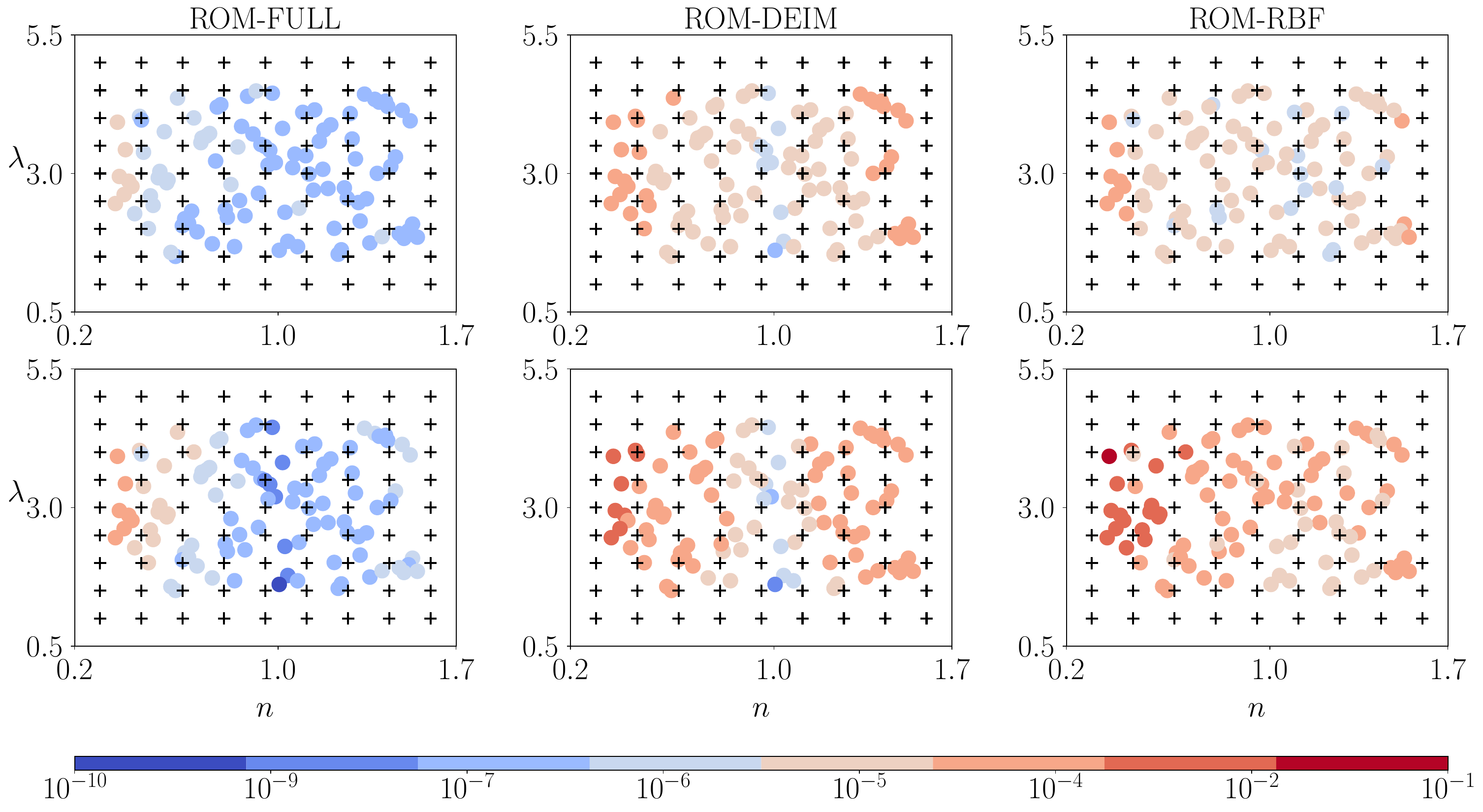}
\caption{Point-wise relative errors over the parameter space
$\params = (n, \lambda)$ for the lid-driven cavity ROMs, obtained with the
training sample $S_3$ and $\epsilon_\mathrm{pod} = 10^{-8}$ for both velocity and
viscosity. Columns correspond to the three ROM strategies (ROM-FULL, ROM-DEIM,
ROM-RBF); the top row shows the velocity error, and the bottom row shows the viscosity
error. Colored dots mark test parameter locations, while plus signs ($+$)
indicate the training sample points.}
    \label{pointwise-error-lid-cavity}
\end{figure}

Figure~\ref{fig:eps-convergence-lid} reports the mean relative velocity error
$\bar{\varepsilon}_{u}$ (top row) and viscosity error
$\bar{\varepsilon}_{\eta}$ (bottom row) versus the POD truncation tolerance
$\epsilon_\mathrm{pod}$ for the three ROM strategies and the training sets
$S_1$, $S_2$, and $S_3$. Denser training sets consistently improve accuracy, with
$S_3$ outperforming $S_2$ and $S_1$ in every panel.
For ROM-FULL, the convergence depends strongly on the training set: with $S_1$ and
$S_2$, the error plateaus at $\epsilon_\mathrm{pod}\approx 10^{-6}$, whereas with $S_3$ it continues to decrease as more POD modes are retained. ROM-DEIM follows
the same trend until $\epsilon_\mathrm{pod} \approx 10^{-6}$, beyond which the
DEIM approximation error dominates, and no further improvement is obtained,
regardless of the number of POD modes retained. For ROM-RBF, the error is almost
independent of $\epsilon_\mathrm{pod}$: the curves are nearly flat across the
entire range, showing that its accuracy is limited by the interpolation of the
reduced coefficients rather than by POD truncation. The training-set size is the
controlling factor, and with the densest set $S_3$, the ROM-RBF errors become
comparable to those of ROM-DEIM, indicating that sufficiently dense sampling of
the parameter space can substantially improve the accuracy of the data-driven
method.
Figure~\ref{pointwise-error-lid-cavity} shows the point-wise relative errors over
the $(n, \lambda)$ parameter space, evaluated with the training set $S_3$.
ROM-FULL achieves the lowest errors throughout the domain. Both its velocity and viscosity errors are smallest near $n = 1$, the Newtonian limit where the problem is linear, and increase toward the shear-thinning boundary $n \approx 0.3$, consistent with the observation of Reyes \emph{et al.}~\cite{reyes2023reduced}
that the POD basis becomes less efficient as the nonlinearity increases. ROM-DEIM
reproduces the overall error distribution of ROM-FULL at a uniformly higher level. 
ROM-RBF
shows its largest errors in the strongly shear-thinning regime, where the solution
manifold undergoes the most pronounced parametric variations, while in the Newtonian and shear-thickening regions, its errors remain comparatively uniform.
\begin{figure}
    \centering
    \includegraphics[width=\linewidth]{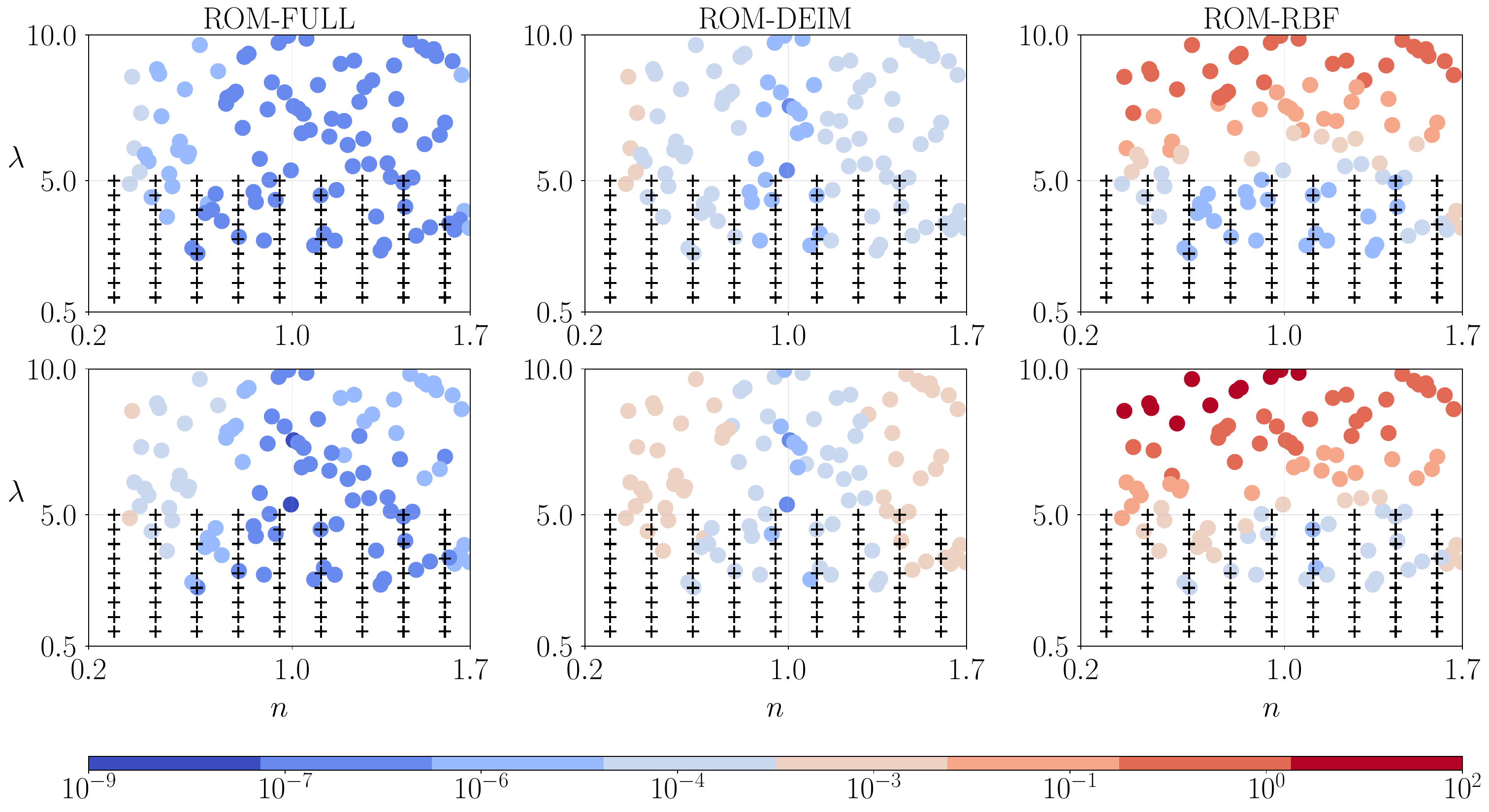}
\caption{Point-wise relative errors over the parameter space
$\params = (n, \lambda)$ for the lid-driven cavity ROMs in the extrapolation
regime, where the test parameters extend beyond the training range in $\lambda$.
Columns correspond to the three ROM strategies (ROM-FULL, ROM-DEIM, ROM-RBF); the top row shows the velocity error, and the bottom row shows the viscosity error.
Colored dots mark test parameter locations, while plus signs ($+$) indicate the
training sample points.}
    \label{pointwise-lid-cavity-extrapolaion}
\end{figure}
\begin{figure}
    \centering
    \begin{subfigure}{0.85\linewidth}
        \centering
        \includegraphics[width=\linewidth]{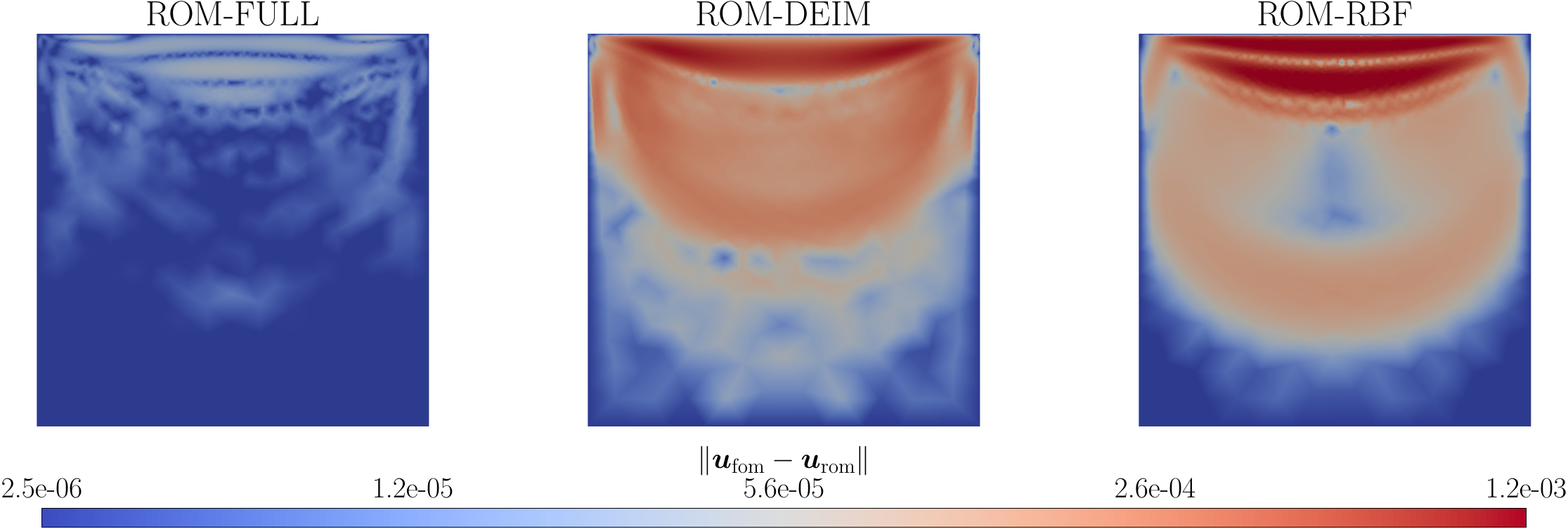}
    \end{subfigure}

    \vspace{0.2cm}

    \begin{subfigure}{0.85\linewidth}
        \centering
        \includegraphics[width=\linewidth]{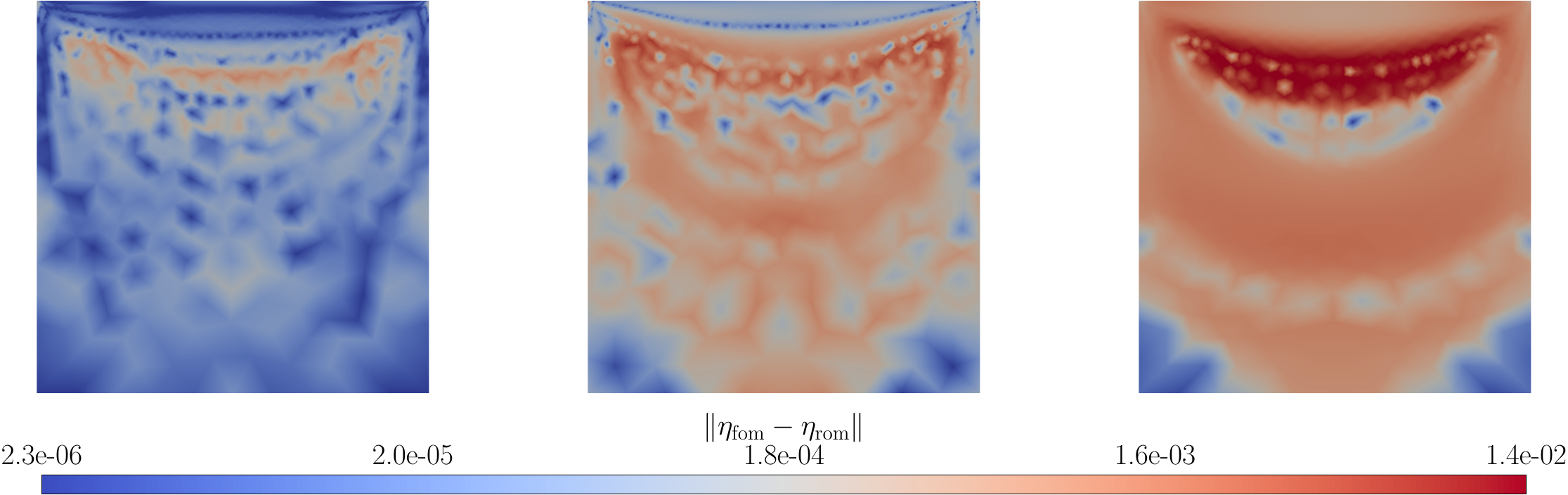}
    \end{subfigure}
\caption{Spatial distribution of ROM errors for the lid-driven cavity at the
parameter $\params = (n, \lambda) = (0.35,\, 4.5)$, plotted on a $\log_{10}$
scale. The top row shows the velocity magnitude error
$\|\vec{u}_\fom - \vec{u}_\rom\|$ and the bottom row the viscosity error
$\|\eta_\fom - \eta_\rom\|$. Columns correspond to the three ROM strategies
(ROM-FULL, ROM-DEIM, ROM-RBF). For the velocity field, errors on the Dirichlet
boundaries are zero up to numerical precision because the boundary values are
imposed exactly; the colorbar range has been adjusted to enhance the visibility
of the interior error distribution.}
    \label{fig:rom-comp-velocity-lid}
\end{figure}
We next examine the performance of each ROM in extrapolation, using a new LHS
test set in which $\lambda$ is extended up to $10$ while the training data cover
only $\lambda \in [1,5]$. The point-wise errors for this test are shown in
Figure~\ref{pointwise-lid-cavity-extrapolaion}. ROM-FULL and ROM-DEIM remain
robust, exhibiting error patterns similar to those in the interpolation regime.
ROM-RBF, by contrast, deteriorates rapidly outside the training range, with the
largest errors in the high-$\lambda$ band farthest from the sampled region.
Although ROM-RBF is accurate for interpolation, given a sufficiently dense training
set, this behavior reflects the limited extrapolation capability inherent to
purely data-driven reduced models, in agreement with the observations of Czech \emph{et al.}~\cite{czech2022data}.
Figure~\ref{fig:rom-comp-velocity-lid} shows the spatial distribution of the ROM
errors for the velocity magnitude and viscosity in a shear-thinning case,
$\params = (n,\lambda) = (0.35,\, 4.5)$. For all three ROMs, the errors are
concentrated near the moving lid and in the vortex core, where the flow varies
most rapidly.

Finally, we briefly address computational times. The online cost of the three ROMs is summarized in the Supplementary Material for a
representative shear-thinning ($\params = (0.35,\,4.5)$) and shear-thickening
($\params = (1.5,\,4.5)$) case. ROM-FULL, which was only marginally numerically optimized, requires on the order of $10$\,s per
query, since it reassembles the full-order operator at every Picard iteration, whereas ROM-DEIM and ROM-RBF reduce the online cost to a few milliseconds by
removing the dependence on the full mesh, ROM-RBF, which performs a single regression evaluation, is the cheapest of the three.
A direct wall-clock comparison with the FOM is not reported because the FOM
and ROMs use different implementations (e.g., programming languages) with 
different levels of optimization; such timings would, therefore, reflect implementation
details rather than intrinsic algorithmic costs. 

\FloatBarrier

\subsection{Settling sphere}
\begin{figure}[!ht]
  \centering
  \includegraphics[width=0.7\linewidth]{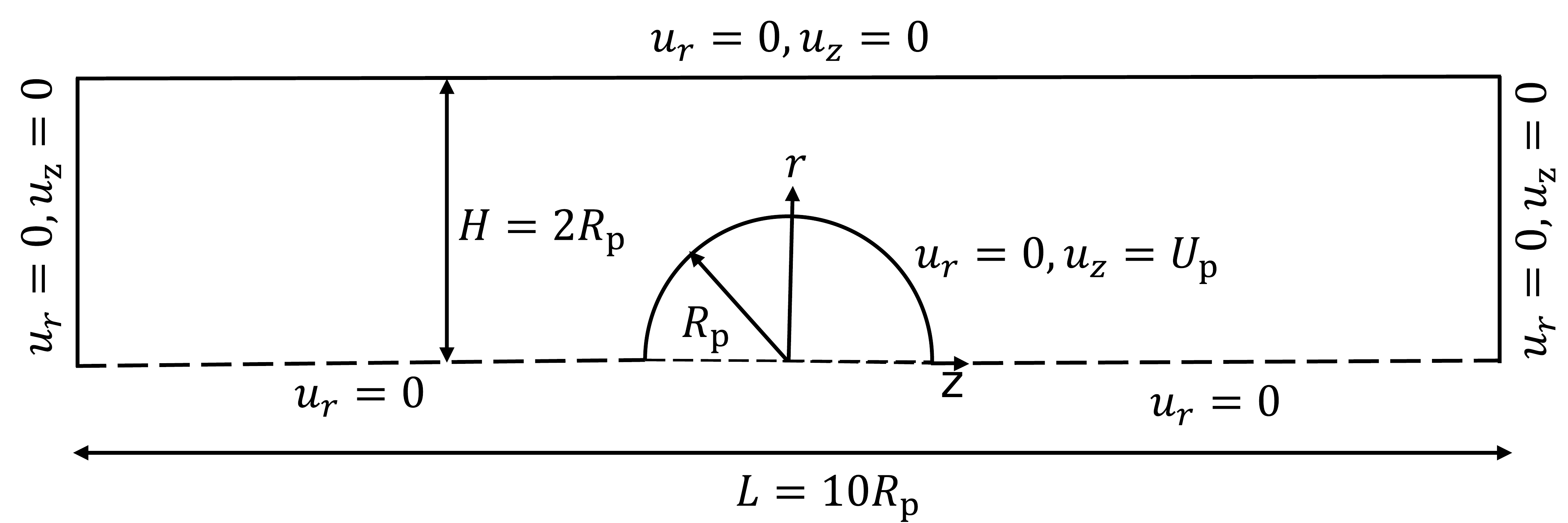}
\caption{Settling sphere benchmark: schematic of the axisymmetric computational
domain and boundary conditions. The sphere of radius $R_\mathrm{p}$ translates
with the unknown axial velocity $U_\mathrm{p}$ induced by the prescribed force
$F$ in a cylindrical channel of radius $H = 2R_\mathrm{p}$ and length
$L = 10R_\mathrm{p}$. The dashed line denotes the axis of symmetry.}
  \label{fig:settling_sphere_setup}
\end{figure}

The second benchmark considers the force-driven settling of a rigid spherical
particle in a closed cylindrical container filled with a generalized Newtonian
fluid. The problem is axisymmetric and is therefore solved in cylindrical
coordinates on the two-dimensional meridional $(z,r)$ domain. All quantities are
non-dimensionalized using the particle radius as the length scale and the
zero-shear viscosity as the viscosity scale, so that $R_\mathrm{p}=1$ and
$\eta_0=1$. The sphere is centered on the axis of symmetry, and the cylinder has
a length $L=10$ and radius $H=2$. In contrast to the lid-driven cavity, where the
motion is prescribed through a boundary velocity, here the total force on the
particle is prescribed, and the settling velocity is recovered as part of the
solution.

Let $\Omega \subset \mathbb{R}^2$ denote the meridional $(z,r)$ domain, with
boundary
\[
\partial\Omega = \Gamma_\mathrm{axis} \cup \Gamma_\mathrm{walls} \cup \Gamma_\mathrm{p},
\]
where $\Gamma_\mathrm{axis}$, $\Gamma_\mathrm{walls}$, and $\Gamma_\mathrm{p}$
denote the axis of symmetry, the outer cylinder walls, and the particle surface,
respectively. On the axis of symmetry ($r=0$) and on the particle boundary, the
radial velocity component $u_r$ vanishes,
\begin{equation}
u_r = 0 \quad \text{on } \Gamma_\mathrm{axis} \cup \Gamma_\mathrm{p}.
\end{equation}
On the outer walls (top, bottom, and lateral boundaries), no-slip conditions are
prescribed:
\begin{equation}
\vec{u} = \vec{0} \quad \text{on } \Gamma_\mathrm{walls},
\end{equation}
and on the particle surface, the velocity is constrained to rigid-body
translation,
\begin{equation}
\vec{u} = U_\mathrm{p}\,\vec{e}_z \quad \text{on } \Gamma_\mathrm{p},
\end{equation}
where $U_\mathrm{p}$ is the unknown settling velocity and $\vec{e}_z$ is the unit
vector in the $z$-direction. The particle motion is determined by a force
balance: the total hydrodynamic force on the particle satisfies
\begin{equation}
\int_{\Gamma_\mathrm{p}} \ten{\sigma}\cdot\vec{n}\,\mathrm{d}A = F\,\vec{e}_z,
\end{equation}
where $\ten{\sigma}$ is the Cauchy stress tensor, $\vec{n}$ is the outward unit
normal on $\Gamma_\mathrm{p}$, and $F$ is the prescribed force magnitude. The
pressure level is fixed by imposing $p(-L/2,0) = 0$. A schematic of the
configuration is shown in Figure~\ref{fig:settling_sphere_setup}.

The spatial discretization was selected based on a mesh-convergence study,
reported in the Supplementary Material, which compares five successively refined
meshes (M1 to M5) \cite{geuzaine2009gmsh}. Based on this study, we use the intermediate mesh M2 for all
high-fidelity settling-sphere simulations, as it provides sufficient accuracy at a
manageable computational cost; this mesh is locally refined near the particle
surface, where the largest velocity gradients and viscosity variations occur.

With the discretization fixed, we now specify the rheology and the parameter
space explored in this benchmark. We use the Carreau model of
Eq.~\eqref{eq:carreau} with $\eta_0=1$ and $\eta_\infty=10^{-3}$. The
characteristic time scale is fixed at $\lambda=1$, and the parametric study
varies the power-law index $n$ and the non-dimensional force magnitude $F$, \ie
$\params=(n,F)$. The range $n\in[0.3,1.6]$ and $F\in[1,100]$ spans shear-thinning
($n<1$), near-Newtonian ($n\approx 1$), and shear-thickening ($n>1$) behavior.
Because the problem is force-driven, increasing $F$ raises the settling velocity
and hence the shear-rate level around the particle.

\begin{figure}[!ht]
    \centering
    \includegraphics[width=0.33\linewidth]{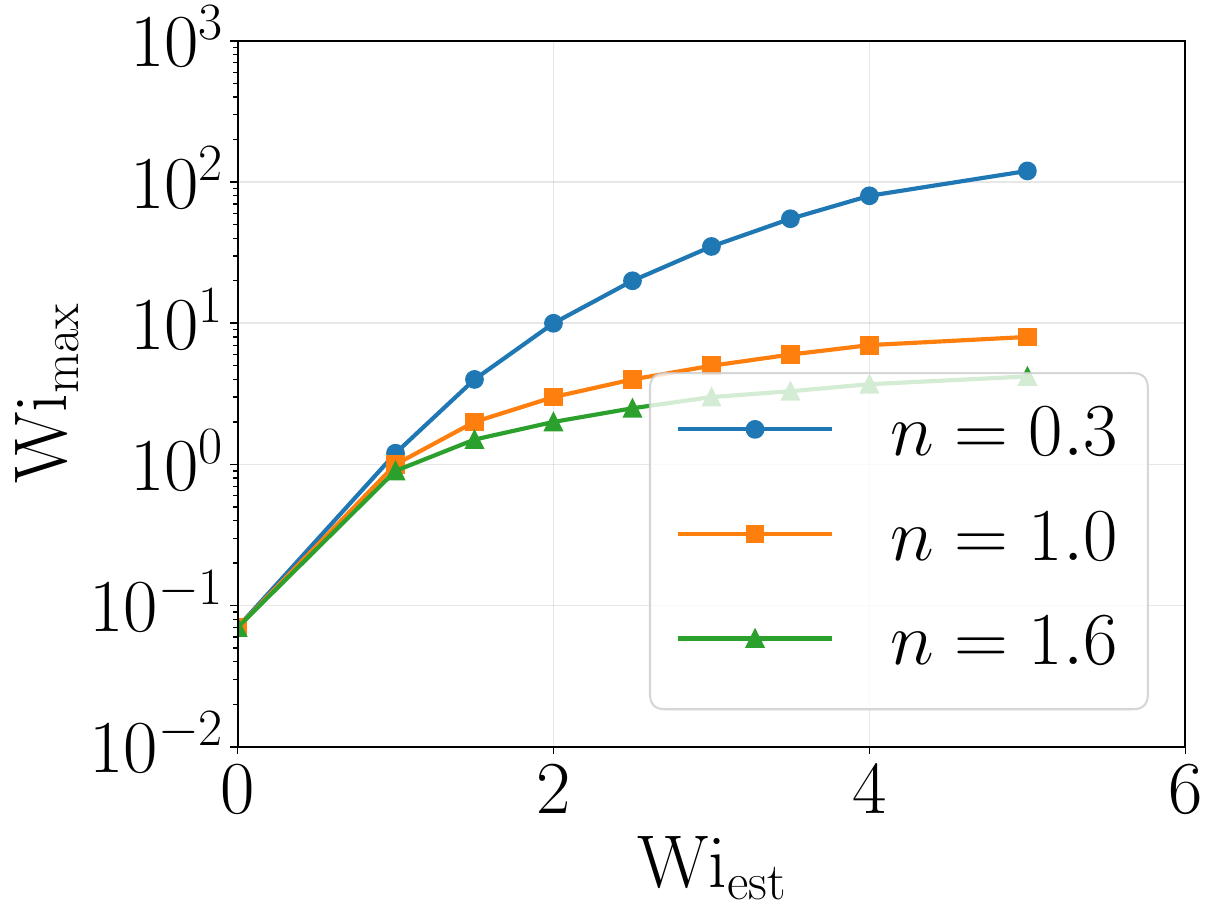}
\caption{True maximum Weissenberg number $\wi_\mathrm{max}$ versus the estimated
Weissenberg number $\wi_\mathrm{est}$ for the settling sphere, for three
rheological regimes: shear-thinning ($n = 0.3$), Newtonian ($n = 1.0$), and
shear-thickening ($n = 1.6$).}
    \label{wi-axi}
\end{figure}

\begin{figure}[!ht]
  \centering
  \begin{subfigure}[t]{0.4\linewidth}
    \centering
    \includegraphics[width=\linewidth]{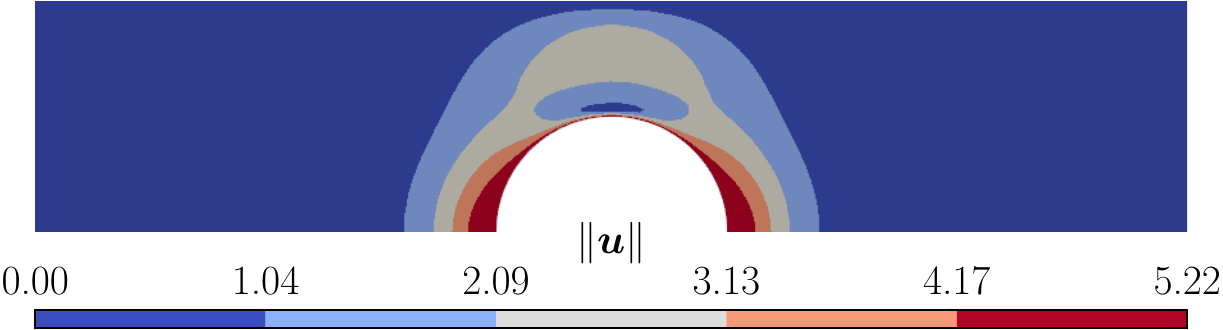}
    \caption{Velocity magnitude, $n = 0.3$, $F = 100$}
  \end{subfigure}
  \begin{subfigure}[t]{0.4\linewidth}
    \centering
    \includegraphics[width=\linewidth]{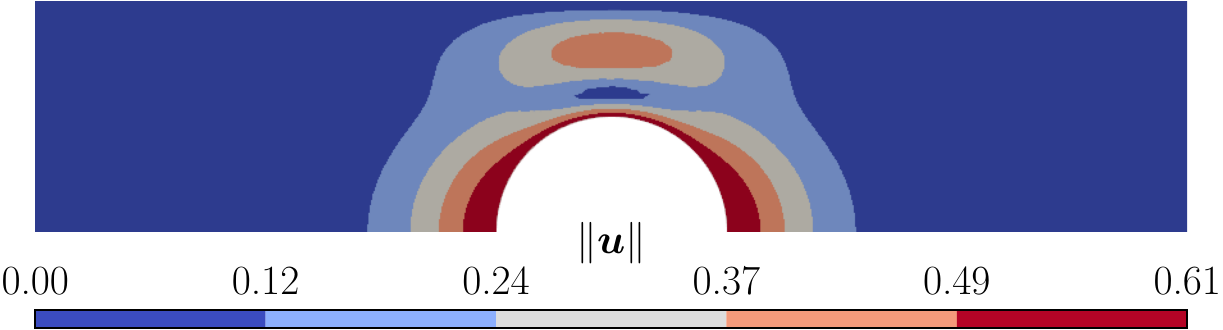}
    \caption{Velocity magnitude, $n = 1.6$, $F = 100$}
  \end{subfigure}
  \vspace{0.8em}
  \begin{subfigure}[t]{0.4\linewidth}
    \centering
    \includegraphics[width=\linewidth]{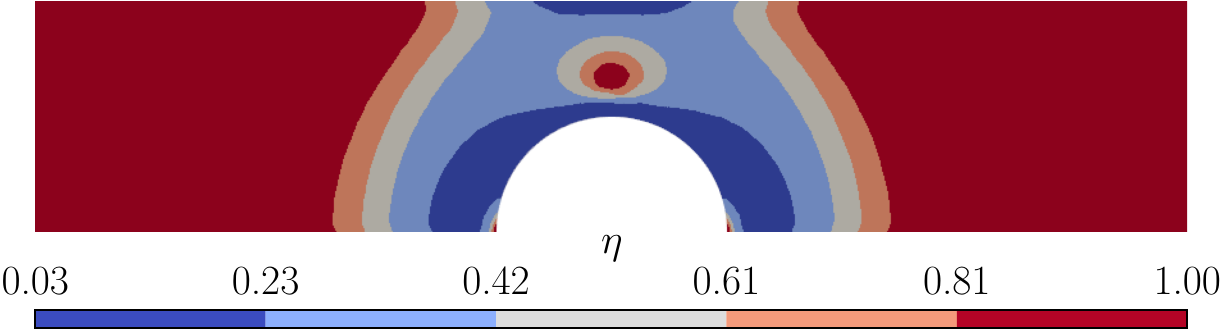}
    \caption{Viscosity field, $n = 0.3$, $F = 100$}
  \end{subfigure}
  \begin{subfigure}[t]{0.4\linewidth}
    \centering
    \includegraphics[width=\linewidth]{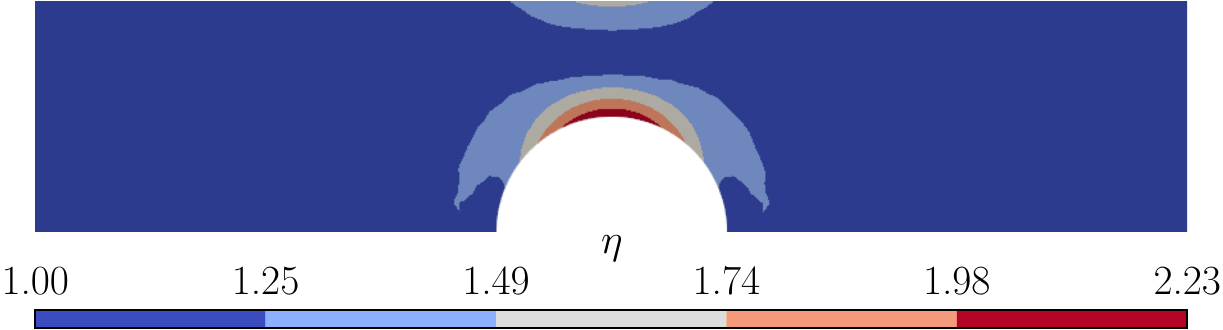}
    \caption{Viscosity field, $n = 1.6$, $F = 100$}
  \end{subfigure}
\caption{Full-order solutions for the settling sphere at $F = 100$. The left
column corresponds to a strongly shear-thinning fluid ($n = 0.3$) and the right
column to a shear-thickening fluid ($n = 1.6$). Panels (a) and (b) show the
velocity magnitude, and panels (c) and (d) the viscosity field.}
  \label{fig:fom_axi_fields}
\end{figure}

To characterize the degree of shear-rate dependence over the selected $(n, F)$
range, we again use the maximum Weissenberg number computed from the FOM
solution (see Eq.~\eqref{eq:wi_max}).
Because the problem is force-driven, estimating the Weissenberg number a priori is
more difficult than for a velocity-driven problem. We therefore use a pragmatic
estimate based on the Stokes prediction for the particle velocity with the
zero-shear viscosity, $U_\mathrm{p} \sim F/(6\pi\eta_0 R_\mathrm{p})$, which gives
\begin{equation}
\wi_\mathrm{est} = \frac{\lambda F}{6\pi},
\end{equation}
reducing to $\wi_\mathrm{est}=F/(6\pi)$ for $\lambda=\eta_0=R_\mathrm{p}=1$.
Figure~\ref{wi-axi} compares the two quantities as $F$ varies for representative
power-law indices $n$, and shows that $\wi_\mathrm{est}$ substantially
underestimates $\wi_\mathrm{max}$, particularly in the shear-thinning regime
($n = 0.3$).

Representative FOM velocity and viscosity fields are shown in
Figure~\ref{fig:fom_axi_fields} for $F=100$ in two rheological regimes: strongly
shear-thinning ($n=0.3$) and shear-thickening ($n=1.6$). In the shear-thinning
case, the particle motion creates a narrow high-shear layer around the sphere, in
which the viscosity drops toward the infinite-shear plateau; this reduces the
resistance to motion and yields a large settling velocity, with a maximum velocity magnitude exceeding $5$. In the shear-thickening case, the viscosity increases
near the particle and remains close to the zero-shear value in the far field. The
resulting flow is weaker and smoother, with a maximum velocity magnitude below $1$
and gradients spread over a larger region around the sphere.

\subsubsection{POD basis generation}
To generate the training data, the parameter space $\params=(n, F)$ is sampled
using uniform grids of increasing resolution. Three nested training sets are
considered: $T_1$ ($3\times3$, 9 samples), $T_2$ ($5\times5$, 25 samples), and
$T_3$ ($9\times9$, 81 samples), with $T_1 \subset T_2 \subset T_3$, constructed
over $n\in[0.3,1.6]$ and $F\in[1,100]$. As for the lid-driven cavity, separate
POD bases are constructed for the velocity and viscosity fields.

\begin{figure}[!ht]
    \centering
    \begin{subfigure}[t]{0.33\linewidth}
        \centering
        \includegraphics[width=\linewidth]{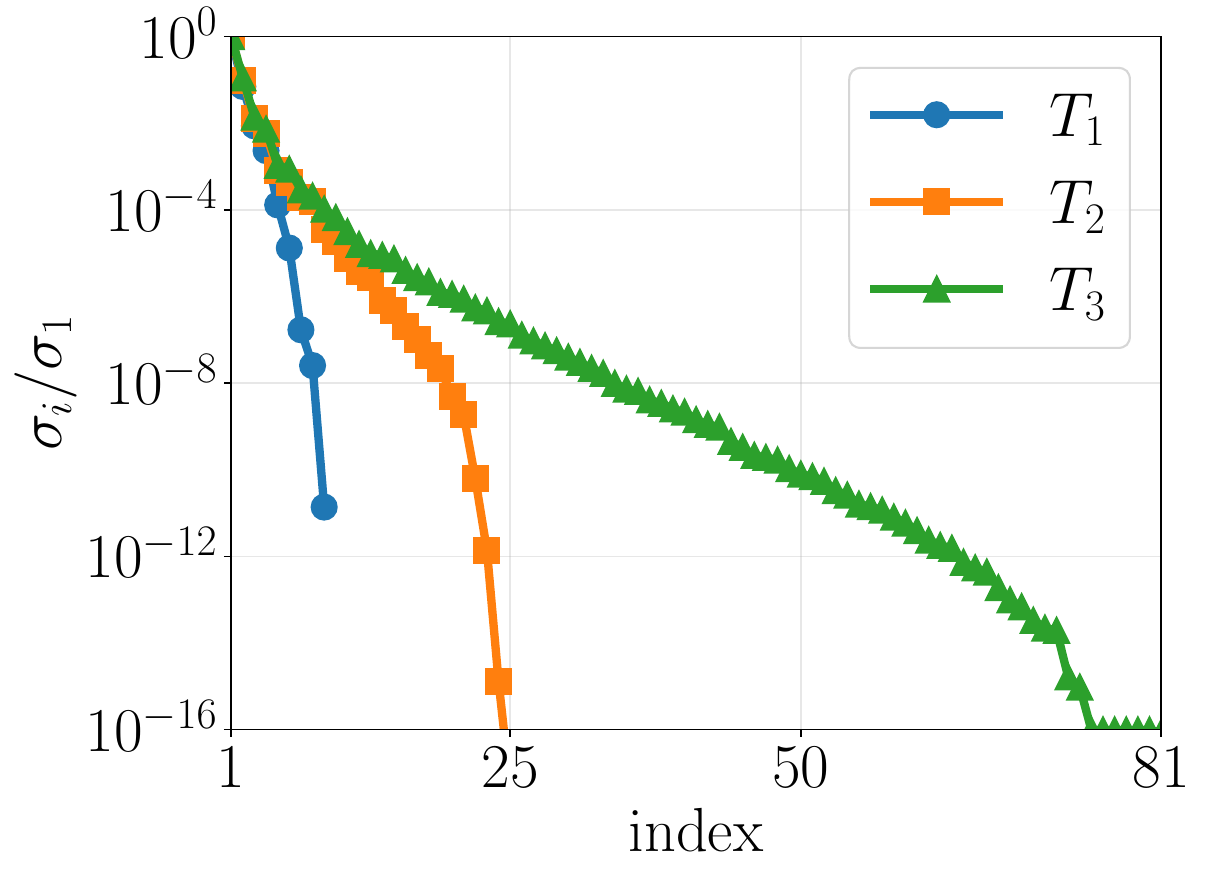}
        \caption{velocity}
    \end{subfigure}
    \begin{subfigure}[t]{0.33\linewidth}
        \centering
        \includegraphics[width=\linewidth]{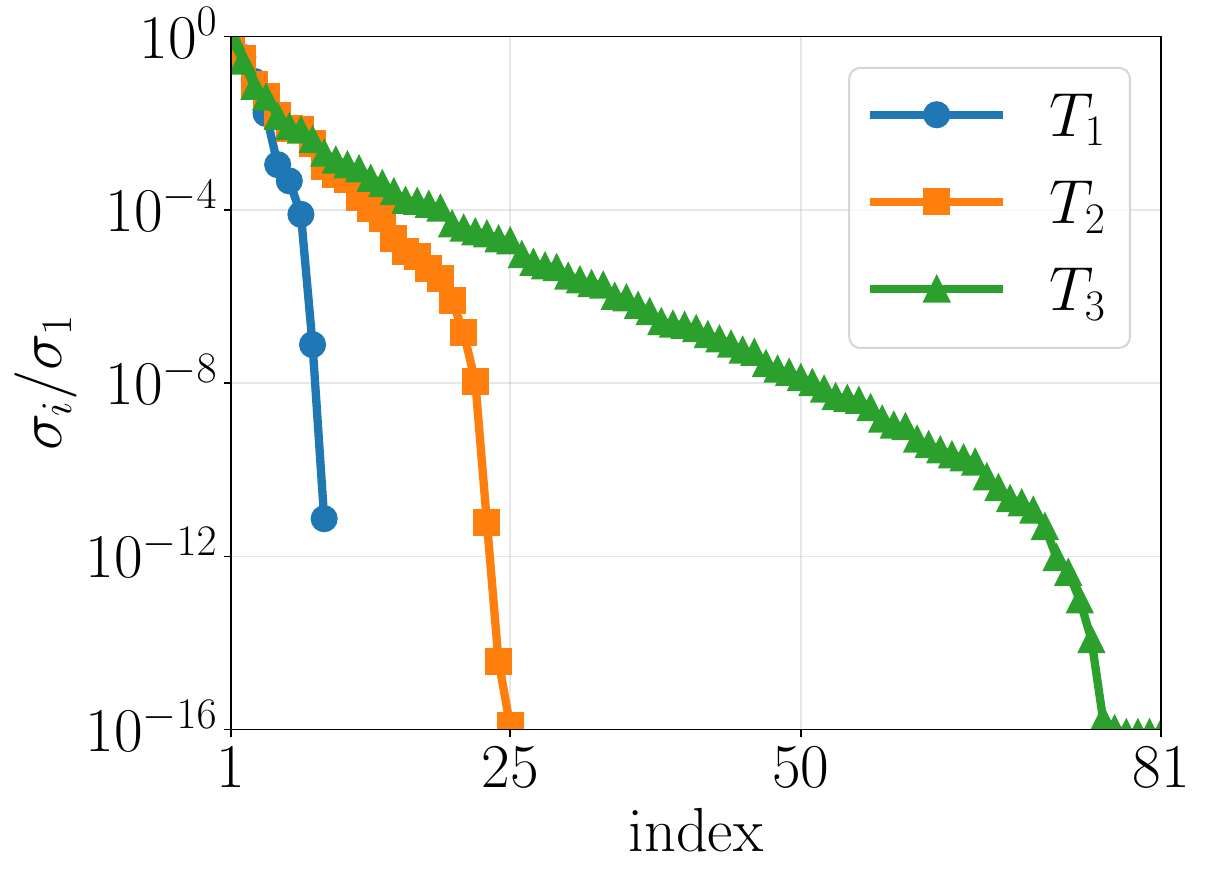}
        \caption{viscosity}
    \end{subfigure}
\caption{Normalized singular value decay (scaled by the leading singular value
$\sigma_1$) for the settling sphere, evaluated on the three training samples
$T_1$, $T_2$, and $T_3$. (a) Singular values of the velocity snapshots.
(b) Singular values of the viscosity snapshots.}
    \label{fig:singular-axi}
\end{figure}

The normalized singular-value spectra in Figure~\ref{fig:singular-axi} decay
rapidly for both fields, in contrast to the slower decay observed for the
lid-driven cavity benchmark, indicating that the settling-sphere snapshots can be
represented accurately with fewer modes. This faster decay is consistent
with the smaller maximum Weissenberg numbers (Figure~\ref{wi-axi}) and indicates
that the selected force range samples a less nonlinear region of the solution
manifold than the lid-driven cavity case.

\begin{figure}[!ht]
    \centering
    \begin{subfigure}[b]{0.45\linewidth}
        \centering
        \includegraphics[width=\textwidth]{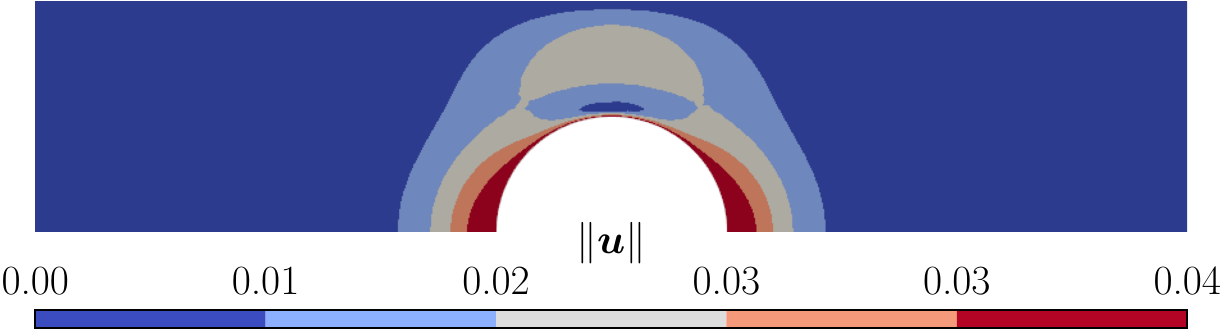}
        \caption{Velocity mode 1}
    \end{subfigure}
    \begin{subfigure}[b]{0.45\linewidth}
        \centering
        \includegraphics[width=\textwidth]{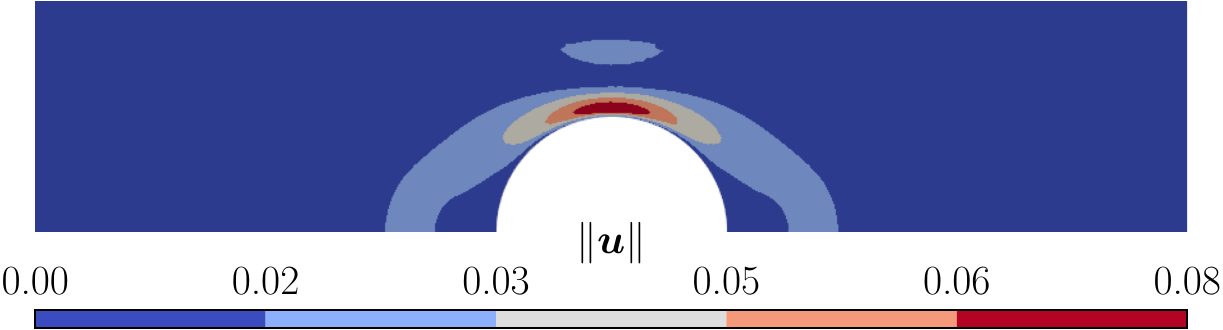}
        \caption{Velocity mode 2}
    \end{subfigure}

    \vspace{1em}

    \begin{subfigure}[b]{0.45\linewidth}
        \centering
        \includegraphics[width=\textwidth]{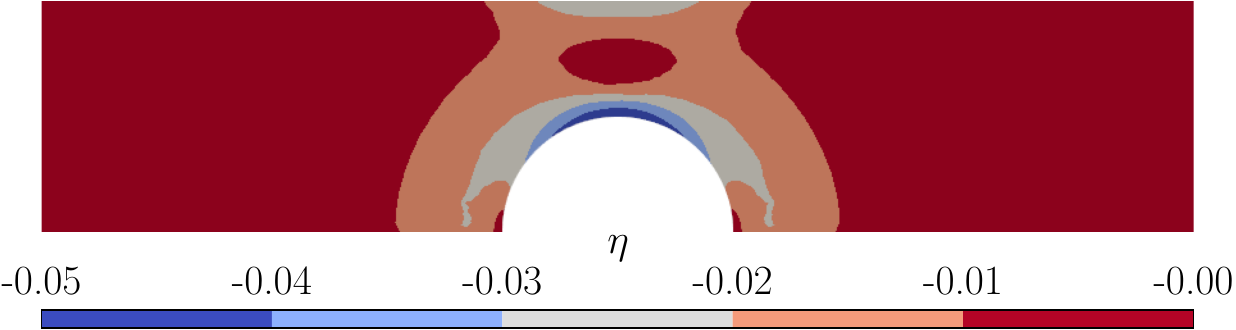}
        \caption{Viscosity mode 1}
    \end{subfigure}
    \begin{subfigure}[b]{0.45\linewidth}
        \centering
        \includegraphics[width=\textwidth]{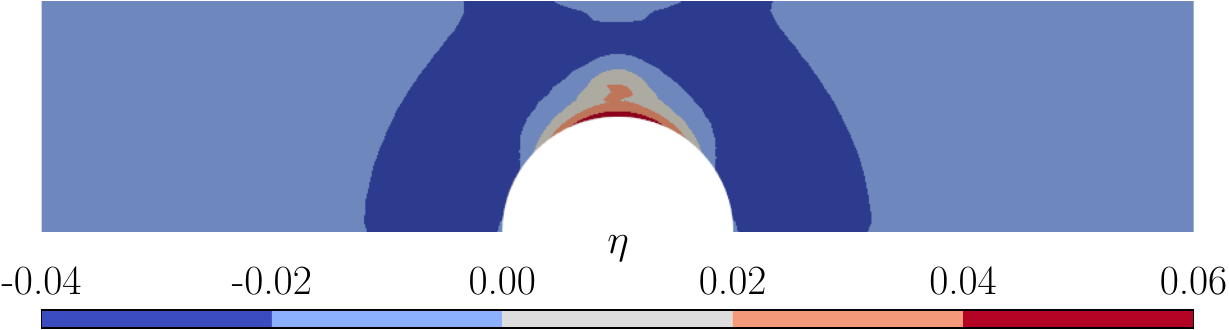}
        \caption{Viscosity mode 2}
    \end{subfigure}
\caption{Dominant POD modes for the settling sphere problem, computed from the
training sample $T_3$. Panels (a) and (b) show the first two velocity modes,
while panels (c) and (d) show the first two viscosity modes.}
    \label{svd_basis_axi}
\end{figure}

The leading POD modes in Figure~\ref{svd_basis_axi} reflect the localized
character of the settling flow. The velocity modes are concentrated near the
particle and decay quickly away from its surface. The viscosity modes are
likewise localized around the high-shear region: the first mode captures the
dominant viscosity change around the sphere, while the second mode adds structure
in the wake and near-surface regions where the first mode alone is insufficient.

Using these POD bases, we construct the ROM-FULL, ROM-DEIM, and ROM-RBF models.
Their performance is evaluated on an independent test set of
$M_\mathrm{test}=100$ parameter values, generated by Latin hypercube sampling
within the same training domain $n \in [0.3,1.6]$ and $F \in [1,100]$.

\subsubsection{Hyperparameter tuning}
As in the lid-driven cavity benchmark, ROM-DEIM and ROM-RBF require
hyperparameter tuning before the three ROM strategies can be compared. For
ROM-DEIM, the key parameter is the number of viscosity basis functions $s$ used
to approximate the nonlinear viscosity field. To isolate the effect of this
choice, the POD tolerance for the velocity field is fixed at
$\epsilon_\mathrm{pod} = 10^{-8}$, and the number of (oversampled) DEIM interpolation points is set to $q = 2s$.

\begin{figure}[!ht]
    \centering
    \begin{subfigure}[t]{0.33\linewidth}
        \centering
        \includegraphics[width=\linewidth]{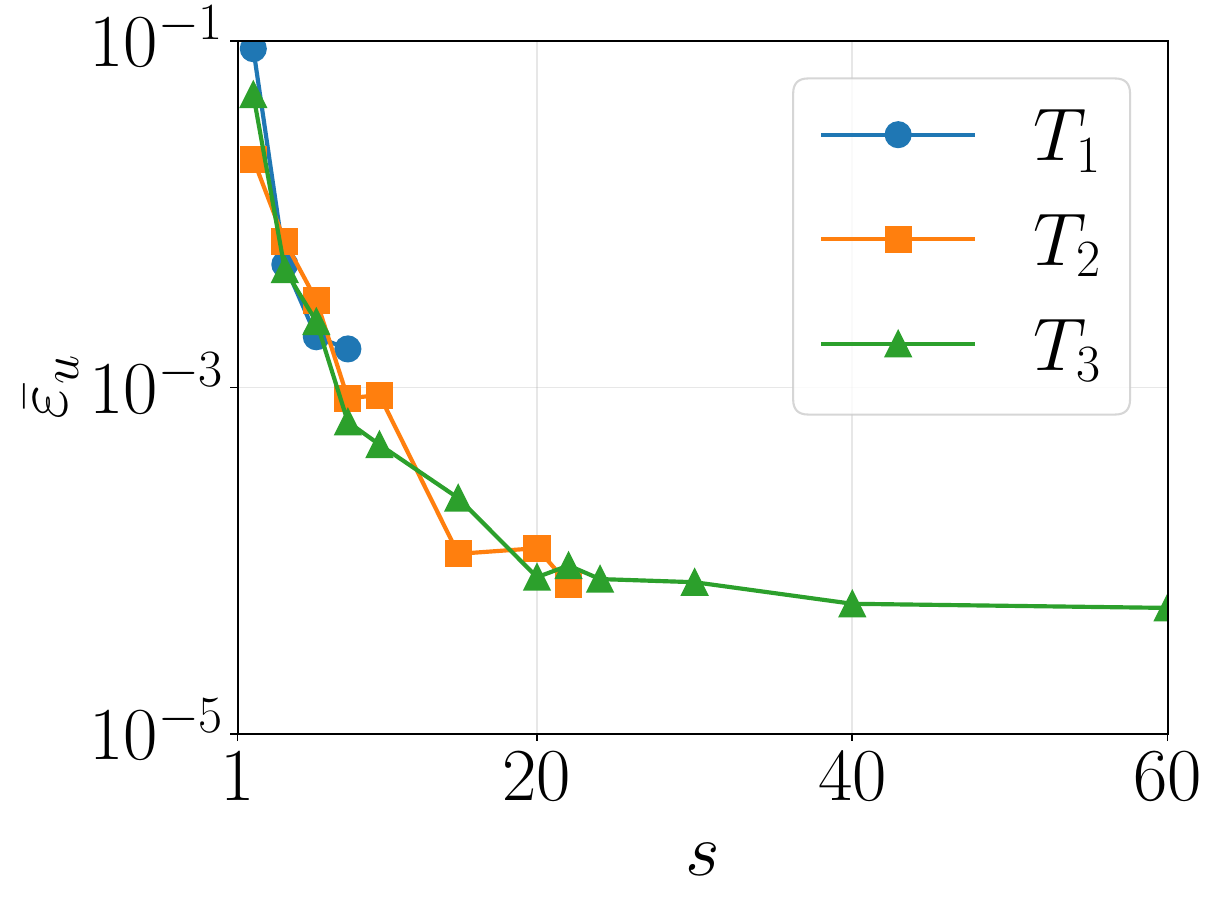}
        \caption{velocity}
    \end{subfigure}
    \begin{subfigure}[t]{0.33\linewidth}
        \centering
        \includegraphics[width=\linewidth]{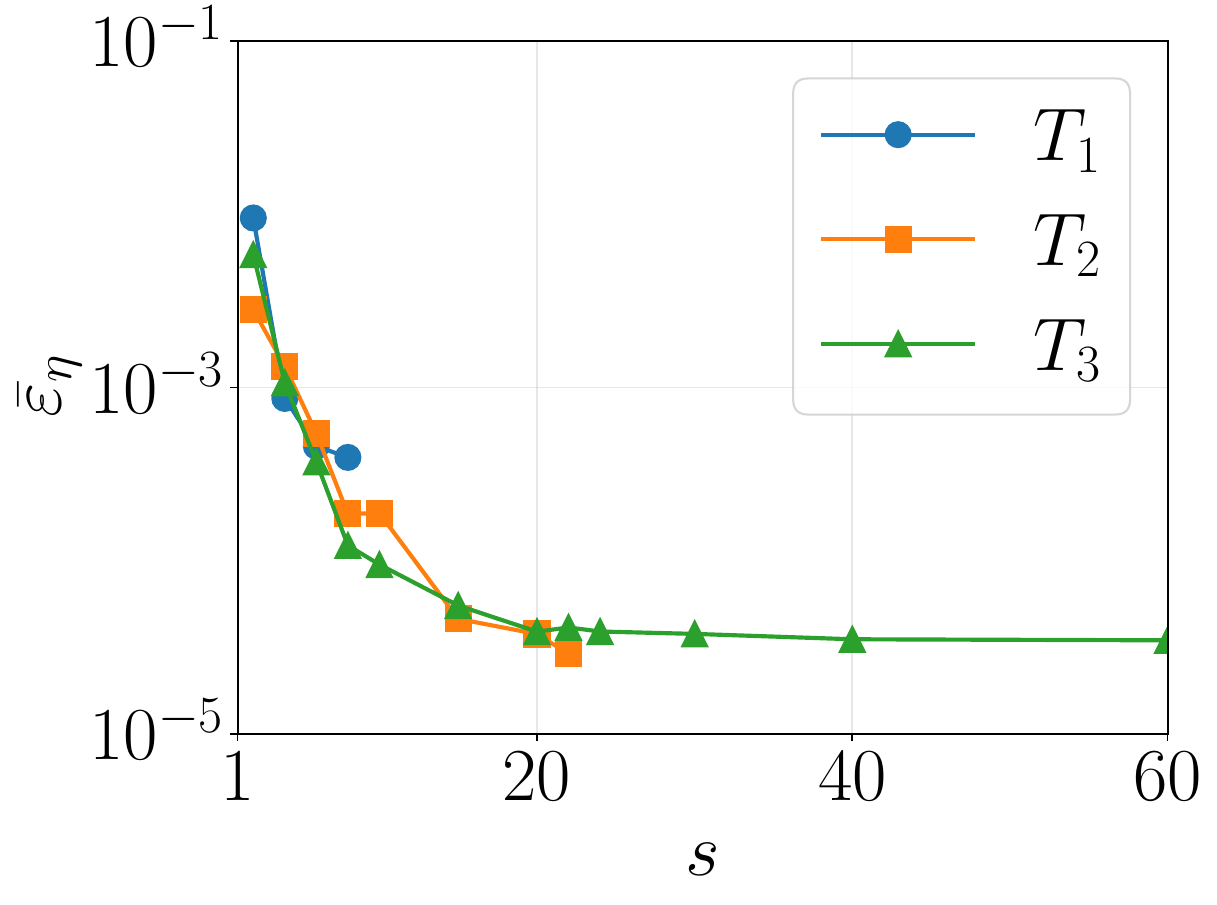}
        \caption{viscosity}
    \end{subfigure}
\caption{Convergence of the ROM-DEIM accuracy for the settling sphere with
respect to the number of viscosity basis functions $s$, where the number of
DEIM interpolation points is set to $q = 2s$. Panel (a) shows the mean velocity
error $\bar{\varepsilon}_{u}$ and panel (b) the mean viscosity error
$\bar{\varepsilon}_{\eta}$, each evaluated over the three training samples
$T_1$, $T_2$, and $T_3$.}
    \label{fig:deim-convergence-axi}
\end{figure}

The mean relative errors in Figure~\ref{fig:deim-convergence-axi} decrease as more viscosity basis functions are retained and then reach a plateau at approximately $s \approx 8$ for $T_1$ and $s \approx 20$ for $T_2$ and $T_3$. This indicates that beyond these $s$ values, the dominant error is no longer due to the DEIM approximation. These values are therefore adopted as the default ROM-DEIM
settings for the settling-sphere benchmark.

\begin{figure}[!ht]
    \centering
    \includegraphics[width=0.66\linewidth]{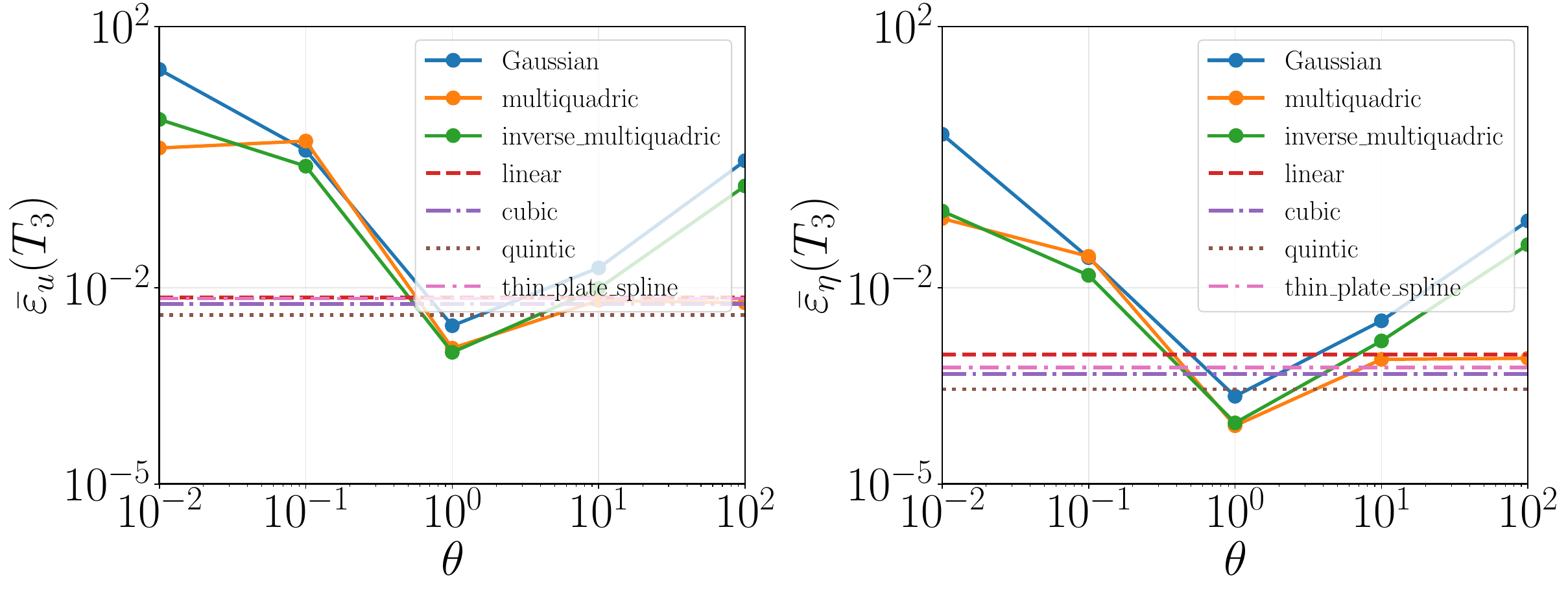}
\caption{Kernel comparison for the settling sphere ROM-RBF, trained on the 
training sample $T_3$. The mean velocity error $\bar{\varepsilon}_{u}(T_3)$ 
(left) and the mean viscosity error $\bar{\varepsilon}_{\eta}(T_3)$ (right) are 
reported as functions of the shape parameter $\theta$ for the Gaussian, 
multiquadric, and inverse multiquadric kernels. The linear, cubic, quintic, and 
thin plate spline kernels are reported as parameter-independent references.}
    \label{fig:rbf_kernel_comp_axi}
\end{figure}

For ROM-RBF, the shape-dependent kernels (Gaussian, multiquadric, inverse
multiquadric) are compared with the parameter-independent kernels (linear,
cubic, quintic, thin plate spline) listed in Table~\ref{RBF_kernels}. For the
training sample $T_3$, Figure~\ref{fig:rbf_kernel_comp_axi} shows that the
shape-dependent kernels are sensitive to the shape parameter $\theta$, with the
minimum mean velocity and viscosity errors attained near $\theta = 1$. At this
value, the multiquadric kernel matches or exceeds the accuracy of the
parameter-independent alternatives for both $\bar{\varepsilon}_{u}$ and
$\bar{\varepsilon}_{\eta}$. Accordingly, the multiquadric kernel with shape
parameter $\theta = 1$ is adopted as the default ROM-RBF kernel for the settling
sphere benchmark.

\begin{figure}[!ht]
    \centering
    \includegraphics[width=\linewidth]{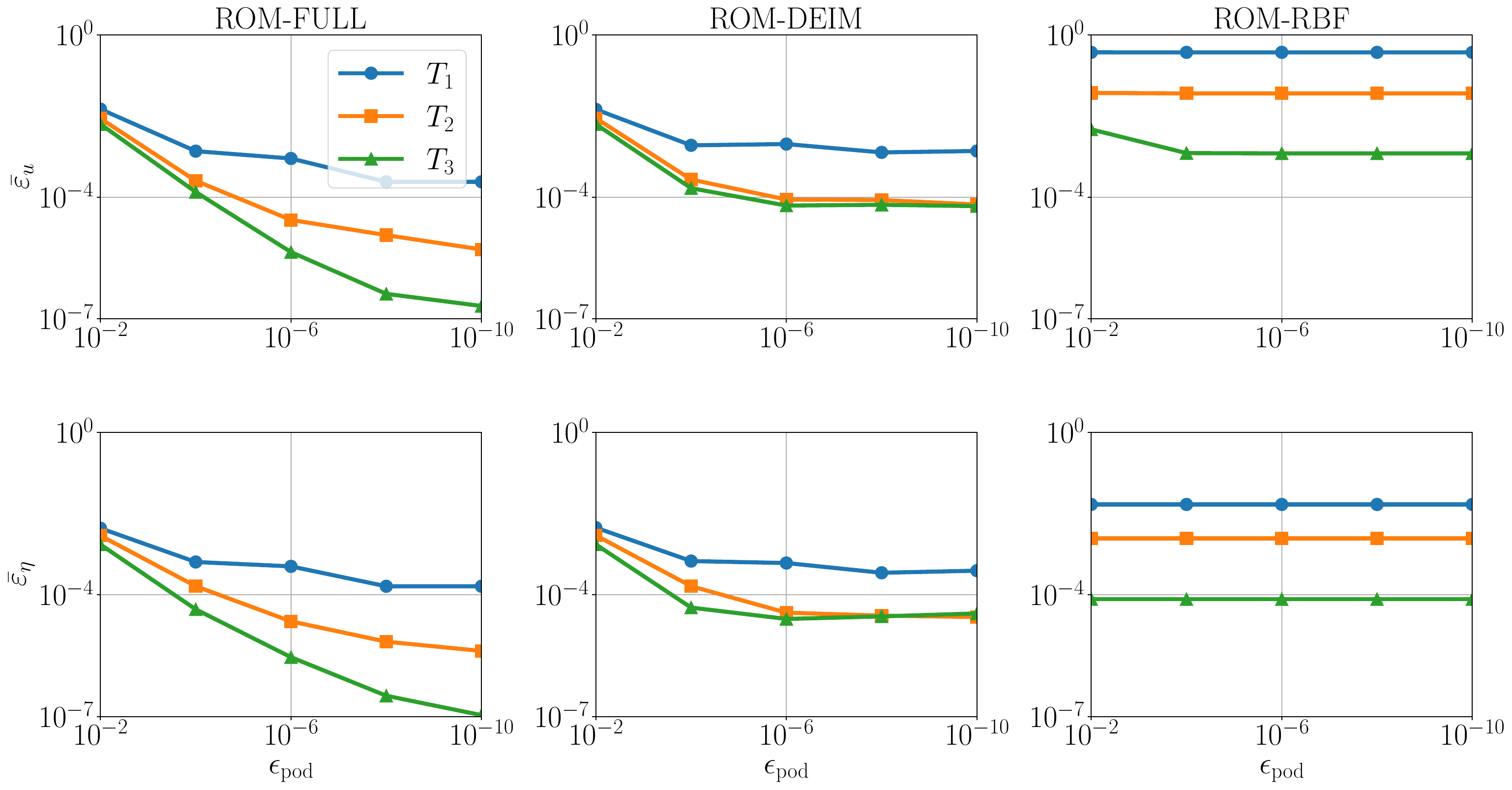}
\caption{Convergence of the mean relative error with respect to the POD
truncation tolerance $\epsilon_\mathrm{pod}$ for the settling sphere ROMs.
Columns correspond to the three ROM strategies (ROM-FULL, ROM-DEIM, ROM-RBF);
the top row reports the mean velocity error $\bar{\varepsilon}_{u}$ and
the bottom row the mean viscosity error $\bar{\varepsilon}_{\eta}$. Within each panel, the curves correspond to
training sets of increasing size, $T_1 \subset T_2 \subset T_3$.}
    \label{fig:mean-error-axi}
\end{figure}

\subsubsection{Comparison of ROMs}
The mean relative velocity error $\bar{\varepsilon}_u$ (top row) and viscosity
error $\bar{\varepsilon}_{\eta}$ (bottom row) are plotted in
Figure~\ref{fig:mean-error-axi} as functions of the POD truncation tolerance
$\epsilon_\mathrm{pod}$ for the three ROM strategies and the training sets $T_1$,
$T_2$, and $T_3$. Increasing the training set size improves the accuracy of all
three ROMs. ROM-FULL converges steadily as more POD modes are retained, while
ROM-DEIM follows the same trend until the error saturates at approximately
$\epsilon_\mathrm{pod}=10^{-6}$; beyond this point, retaining additional velocity
modes does not reduce the error, because the DEIM approximation becomes the
limiting contribution. ROM-RBF plateaus much earlier, at around
$\epsilon_\mathrm{pod}=10^{-4}$, indicating that its accuracy is limited by the interpolation of the reduced coefficients rather than by the POD truncation.

\begin{figure}[!ht]
    \centering
    \includegraphics[width=\linewidth]{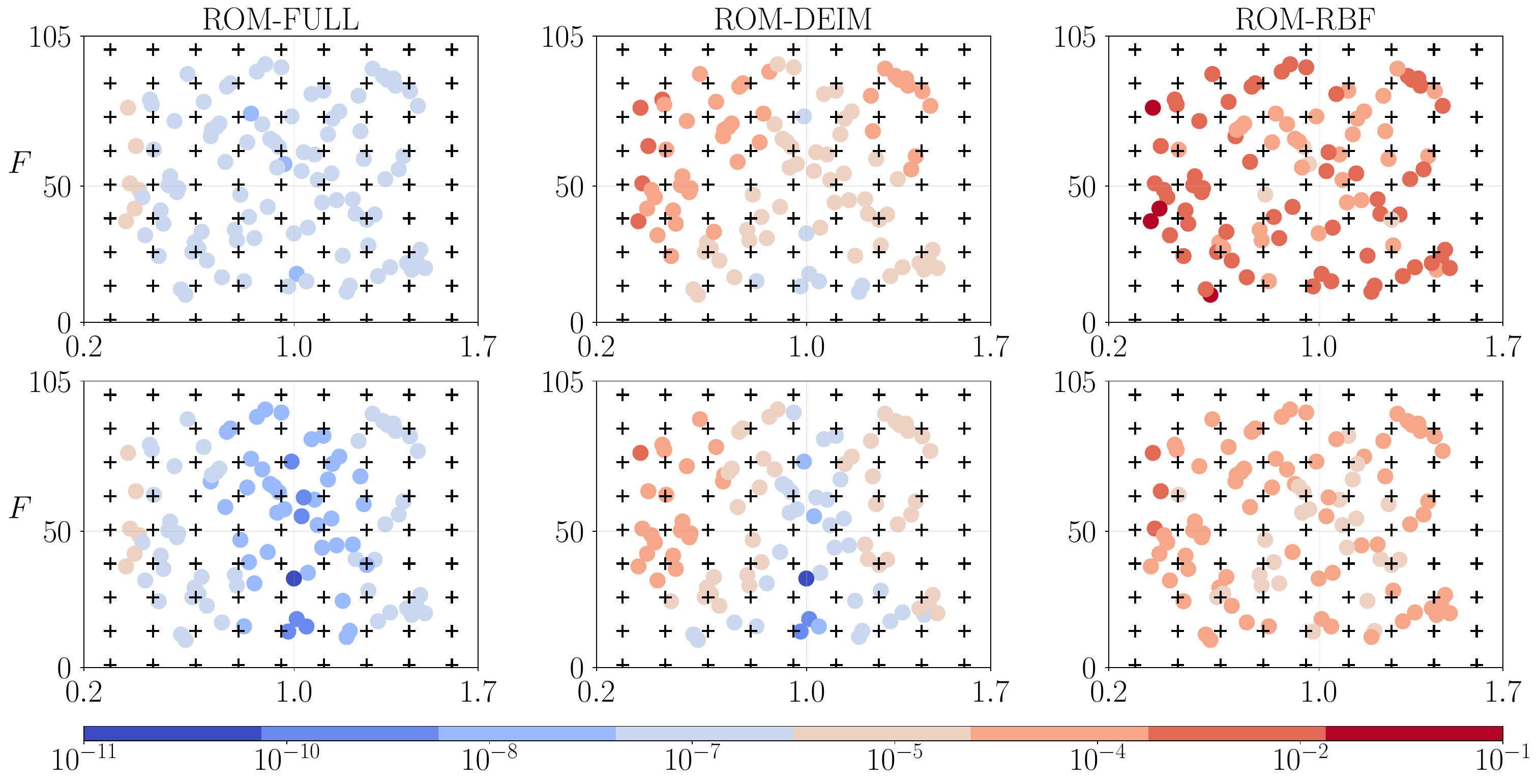}
\caption{Point-wise relative errors over the parameter space $\params = (n, F)$
for the settling sphere ROMs in the interpolation regime, where the test
parameters lie within the training range. Columns correspond to the three ROM
strategies (ROM-FULL, ROM-DEIM, ROM-RBF); the top row shows the relative
velocity error, and the bottom row shows the relative viscosity error. Colored dots
mark test parameter locations, while plus signs ($+$) indicate the training
sample points.}
    \label{relative-error-axi}
\end{figure}

The point-wise error maps in Figure~\ref{relative-error-axi}, obtained with the
training set $T_3$, show the corresponding distribution over the $(n,F)$
parameter space. ROM-FULL gives the lowest errors throughout the test parameter space, and ROM-DEIM reproduces the same overall pattern at slightly higher error levels.
Both projection-based ROMs are most accurate near the Newtonian limit ($n=1$),
where the viscosity is independent of the shear rate, and their errors increase
toward the strongly shear-thinning regime. ROM-RBF again exhibits higher velocity
errors in the shear-thinning regime, whereas its viscosity errors remain largely
uniform across the $(n, F)$ space. 

\begin{figure}[!ht]
    \centering
    \begin{subfigure}{0.9\linewidth}
        \centering
        \includegraphics[width=\linewidth]{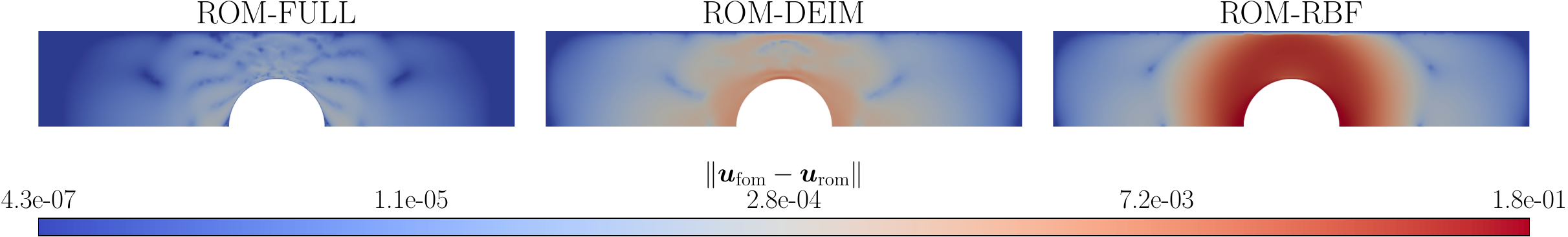}
    \end{subfigure}

\vspace{0.2cm}
    \begin{subfigure}{0.9\linewidth}
        \centering
        \includegraphics[width=\linewidth]{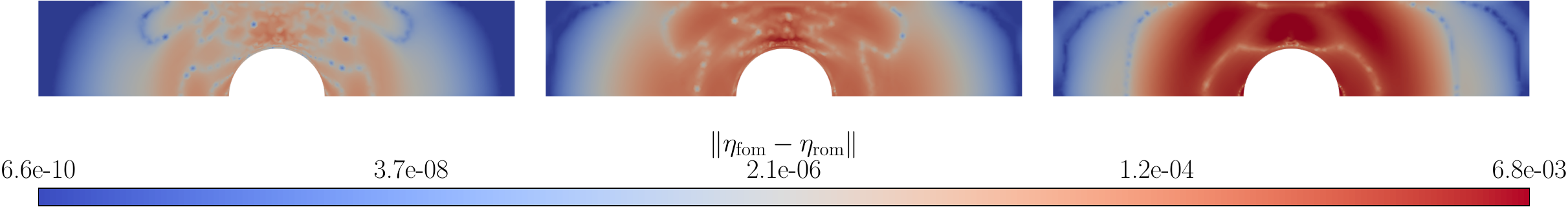}
    \end{subfigure}
\caption{Spatial distribution of ROM errors for the settling sphere at the
parameter $\params = (n, F) = (0.35,\, 95)$, plotted on a $\log_{10}$ scale.
The top row shows the velocity magnitude error
$\|\vec{u}_\mathrm{fom} - \vec{u}_\mathrm{rom}\|$ and the bottom row shows the
viscosity error $\|\eta_\mathrm{fom} - \eta_\mathrm{rom}\|$. Columns correspond
to the three ROM strategies (ROM-FULL, ROM-DEIM, ROM-RBF).}
    \label{fig:rom-comp-axi}
\end{figure}

The spatial error distributions for a strongly shear-thinning test case
$(n,F)=(0.35,95)$ are presented in Figure~\ref{fig:rom-comp-axi}, analogous to
the lid-driven cavity comparison in
Figure~\ref{fig:rom-comp-velocity-lid}. The largest errors occur near the
particle surface and in the wake, where the velocity gradients and viscosity
variations are strongest. ROM-FULL produces the smallest errors in both fields.
ROM-DEIM remains close to ROM-FULL, but its hyper-reduction error is most visible
near the high-shear layer around the sphere. ROM-RBF gives the largest and most spatially diffuse errors, reflecting the fact that it does not enforce the governing equations during prediction.

\begin{figure}[!ht]
    \centering
    \includegraphics[width=\linewidth]{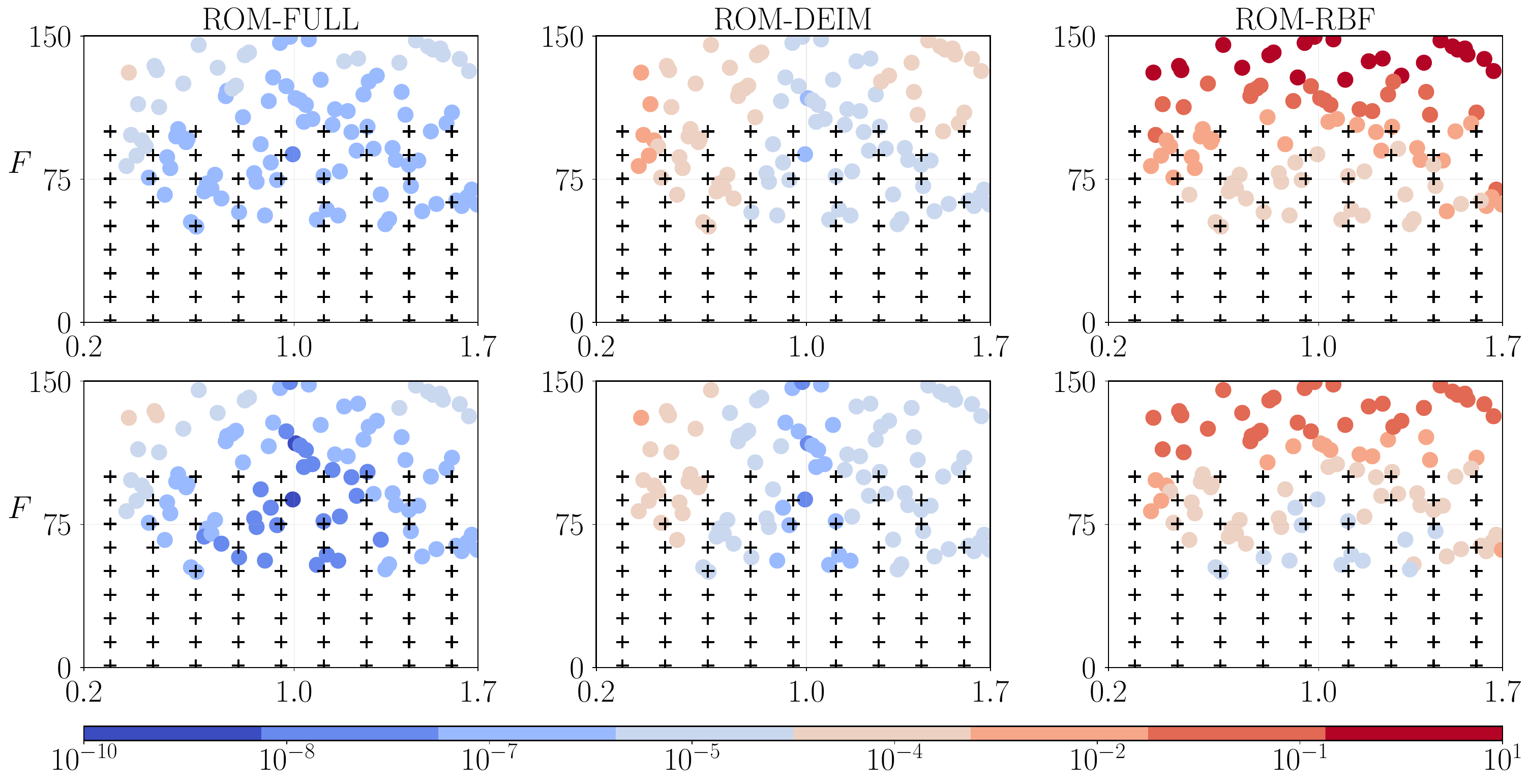}
\caption{Point-wise relative errors over the parameter space $\params = (n, F)$
for the settling sphere ROMs in the extrapolation regime, where the test range in $F$ is extended beyond the training range. Columns correspond to the
three ROM strategies (ROM-FULL, ROM-DEIM, ROM-RBF); the top row shows the
velocity error, and the bottom row shows the viscosity error. Colored dots mark the test parameter locations, while plus signs ($+$) indicate the training sample
points.}
    \label{fig:relative-error-axi-extrapolation}
\end{figure}

Finally, we examine robustness outside the force range used for training. The
ROMs are trained with $T_3$, which spans $F\in[1,100]$, and are evaluated on an
independent LHS test set with the force range extended to $F\in[50,150]$ while
keeping $n\in[0.3,1.6]$. The extrapolation error maps in
Figure~\ref{fig:relative-error-axi-extrapolation} show that ROM-FULL and ROM-DEIM
remain accurate as the force increases beyond the training range, although their
errors grow in the more nonlinear shear-thinning regime. ROM-RBF deteriorates
more strongly once the query points move outside the sampled region, confirming
that the non-intrusive method is reliable primarily for interpolation.

The corresponding \online costs for the settling sphere are reported in the
Supplementary Material for a representative shear-thinning ($\params = (0.35,\,95)$)
and shear-thickening ($\params = (1.5,\,95)$) case, and follow the same trend as
the lid-driven cavity.
\FloatBarrier

\section{Conclusions}\label{sec:conclusion}

This work compared three POD-based reduced-order modeling strategies for steady
generalized Newtonian flows governed by the bounded Carreau viscosity model:
an intrusive Galerkin projection with full operator reassembly (ROM-FULL), an
intrusive hyper-reduced Galerkin model using DEIM (ROM-DEIM), and a fully
non-intrusive RBF interpolation model (ROM-RBF). The comparison was performed on
two representative benchmarks, a lid-driven cavity and a settling sphere,
covering both boundary-driven and force-driven flows under interpolative and
extrapolative parameter queries.

The results show that ROM-FULL
provides the most accurate predictions in all cases and remains robust under
extrapolation since it retains the projected governing equations. The resulting
error is a combination of the number of POD modes as well as the amount of training data
available. When compared to ROM-FULL, the accuracy of ROM-DEIM is comparable for the cases where
little training data is used, but the DEIM hyper-reduction introduces additional
errors that cannot be reduced by either more POD modes or more training data.
Finally, ROM-RBF can achieve competitive accuracy when the training set
is sufficiently dense, especially in interpolation, but its accuracy deteriorates
more strongly outside the sampled parameter range.

The main distinction between the three approaches is the trade-off
between solver access, computational efficiency, and predictive robustness.
ROM-FULL is the most reliable option when accuracy is the priority, but its
\online cost remains affected by full-order operator assembly. ROM-DEIM offers
the best overall compromise by evaluating the nonlinear viscosity only at a small
set of DEIM points while preserving a physics-based projection structure, especially
when little data are available and/or when predicting outside the training range.
ROM-RBF is the fastest, reducing the \online phase to a regression evaluation, and is therefore attractive when data are abundant and/or the governing equations
or solver internals are unavailable, as in commercial, legacy, or experimental
data-driven settings.
The \offline cost of all three approaches is dominated by the generation of the
high-fidelity FOM snapshots and the construction of the POD bases, although
ROM-DEIM and ROM-RBF require additional method-specific preprocessing. 

Future work should extend the comparison to more complex rheologies, including
viscoelastic, viscoplastic, and elastoviscoplastic flows, where reduced-order models may face
additional stability challenges. For non-intrusive modeling, Gaussian process
regression, neural networks, and adaptive sampling strategies could improve
accuracy while reducing the number of required training simulations. Finally,
applying these ROM frameworks to practical problems, such as optimization, 
uncertainty quantification, parameter inference, and model selection, would 
further test their usefulness in computationally demanding settings. This will be 
especially relevant for problems that require three-dimensional simulations
coupled with complex physics, where 
the cost of snapshot generation is prohibitively high.

\section*{Declaration of generative AI and AI-assisted technologies in the manuscript preparation process}

During the preparation of this work the author(s) used Anthropic's Claude in order to improve the language and grammar of the manuscript. After using this tool/service, the author(s) reviewed and edited the content as needed and take(s) full responsibility for the content of the published article.

\section*{Data availability}

The datasets generated and analyzed in this study, including the full-order
simulation snapshots, parameter samples, and mesh files for the lid-driven
cavity and settling-sphere benchmarks are openly available on Zenodo at
\url{https://doi.org/10.5281/zenodo.20272964}.

\section*{Acknowledgments}
The authors thank M.A. Hulsen at the Eindhoven University of Technology (TU/e) for access to the TFEM software libraries.
\appendix
\FloatBarrier

\section{Imposing a constant force in ROM-FULL and ROM-DEIM}
\label{sec:force_constraint}

In the settling-sphere benchmark, the particle motion is driven by a prescribed
external force $F$, while the particle velocity $U_\mathrm{p}$ is unknown and
must be determined as part of the solution. Physically, this enforces rigid-body
motion of the particle, together with a global balance between the applied force
and the hydrodynamic reaction.

The weak formulation, including the rigid-body constraint, reads: find
$\vec{u} \in H^1(\Omega)^2$, $p \in L^2(\Omega)$, $U_\mathrm{p} \in \mathbb{R}$,
and $\uplambda \in L^2(\Gamma_\mathrm{p})$ such that
\begin{align}
    \int_\Omega 2\,\eta(\dot{\gamma})\,\ten{D}(\vec{u}):\ten{D}(\vec{v})\,\mathrm{d}V
    - \int_\Omega p\,(\nabla\cdot\vec{v})\,\mathrm{d}V
    + \int_{\Gamma_\mathrm{p}} (v_z - V_\mathrm{p})\,\uplambda\,\mathrm{d}A
    &= V_\mathrm{p}\,F, \\[4pt]
    \int_\Omega q\,(\nabla\cdot\vec{u})\,\mathrm{d}V &= 0, \\[4pt]
    \int_{\Gamma_\mathrm{p}} \chi\,(u_z - U_\mathrm{p})\,\mathrm{d}A &= 0,
\end{align}
for all test functions $\vec{v} \in H^1(\Omega)^2$, $q \in L^2(\Omega)$,
$V_\mathrm{p} \in \mathbb{R}$, and $\chi \in L^2(\Gamma_\mathrm{p})$.

At the full-order level, rigid-body motion is imposed by collocation
constraints at the $N_\mathrm{c}$ nodes on the particle boundary $\Gamma_\mathrm{p}$,
requiring the axial fluid velocity at those nodes to equal $U_\mathrm{p}$.
Introducing Lagrange multipliers $\uplambda \in \mathbb{R}^{N_\mathrm{c}}$ and treating
$U_\mathrm{p}$ as an additional unknown leads to the saddle-point system
\cite{glowinski1999distributed, patankar2000new}:
\begin{equation} \label{eq:fom_constrained}
  \begin{bmatrix}
    \mat{A} & -\mat{L}^\top & \mat{C}^\top & \mat{0} \\
   -\mat{L} &  \mat{0}      & \mat{0}      & \mat{0} \\
    \mat{C} &  \mat{0}      & \mat{0}      & -\mat{1}_{N_\mathrm{c}} \\
    \mat{0} &  \mat{0}      & -\mat{1}^\top_{N_\mathrm{c}} & 0
  \end{bmatrix}
  \begin{bmatrix} \mat{u} \\ \mat{p} \\ \uplambda \\ U_\mathrm{p}
  \end{bmatrix}
  =
  \begin{bmatrix} \mat{0} \\ \mat{0} \\ \mat{0} \\ F \end{bmatrix},
\end{equation}
where $\mat{C} \in \mathbb{R}^{N_{\mathrm{c}} \times N_u}$ is the collocation matrix
selecting the axial-velocity degrees of freedom on $\Gamma_\mathrm{p}$, and
$\mat{1}_{N_\mathrm{c}}$ is a vector of ones of length $N_\mathrm{c}$. The third block row enforces the rigid-body constraint
$\mat{C}\,\mat{u} = U_\mathrm{p}\,\mat{1}_{N_\mathrm{c}}$, and the fourth row reduces
to the scalar force balance $-\sum_{i=1}^{N_\mathrm{c}} \uplambda_i = F$.

For the reduced-order model, the velocity field is approximated as
$\mat{u} \approx \mat{\Phi}\,\hat{\mat{u}}$. Because the POD basis is
constructed from full-order snapshots that already satisfy the rigid-body
constraint, both the constraint and the particle velocity are implicitly
embedded in the reduced space and need not be enforced explicitly. The pressure
block is dropped using the same approximate-divergence-free argument as in
Section~\ref{sec:rom-full}.

To include the forcing by $F$ in the reduced model, we write the Lagrange
multipliers as
\begin{equation} \label{eq:lagrange_multiplier}
  \uplambda = -\frac{F}{N_\mathrm{c}}\,\mat{1}_{N_\mathrm{c}},
\end{equation}
which satisfies the force balance exactly. Substituting
Eq.~\eqref{eq:lagrange_multiplier} into the projected momentum equation yields
the reduced system:
\begin{equation}\label{eq:rom_online}
  \hat{\mat{A}}\,\hat{\mat{u}} = F\,\hat{\mat{b}},
\end{equation}
where
\begin{equation}\label{eq:rom_online2}
  \hat{\mat{A}} = \mat{\Phi}^\top \mat{A}(\mat{\upeta})\,\mat{\Phi},
  \qquad \text{and} \qquad
  \hat{\mat{b}} = \frac{1}{N_\mathrm{c}}\,\mat{\Phi}^\top \mat{C}^\top
                  \mat{1}_{N_\mathrm{c}} \;\in\; \mathbb{R}^{r}.
\end{equation}
The vector $\hat{\mat{b}}$ is parameter-independent and is precomputed
\offline. The reduced system is still nonlinear in $\hat{\mat{u}}$ through
$\mat{\upeta} = \mat{\upeta}(\dot{\gamma}(\mat{\Phi}\,\hat{\mat{u}}))$ and is
solved using Picard iteration (ROM-FULL and ROM-DEIM). In the \online phase,
varying $F$ at fixed rheology amounts to scaling the right-hand side and
requires no re-assembly. After solving the reduced system, the particle
velocity can be obtained from the reconstructed velocity field at the nodes on
the particle boundary. Note that in the ROM-RBF approach, the above is not
required since the force $F$ is treated as an additional parameter, and its effect is accounted for in the learned mapping
$\params \mapsto \hat{\mat{u}}(\params)$.

\FloatBarrier
\bibliographystyle{elsarticle-num}
\bibliography{references}

@article{geuzaine2009gmsh,
  title={Gmsh: A 3-D finite element mesh generator with built-in pre-and post-processing facilities},
  author={Geuzaine, Christophe and Remacle, Jean-Fran{\c{c}}ois},
  journal={International journal for numerical methods in engineering},
  volume={79},
  number={11},
  pages={1309--1331},
  year={2009},
  publisher={Wiley Online Library}
}

@article{stabile2018finite,
  title={Finite volume POD-Galerkin stabilised reduced order methods for the parametrised incompressible Navier--Stokes equations},
  author={Stabile, Giovanni and Rozza, Gianluigi},
  journal={Computers \& Fluids},
  volume={173},
  pages={273--284},
  year={2018},
  publisher={Elsevier}
}

@article{patankar2000new,
  title={A new formulation of the distributed Lagrange multiplier/fictitious domain method for particulate flows},
  author={Patankar, Neelesh A and Singh, Pushpendra and Joseph, Daniel D and Glowinski, Roland and Pan, T-W},
  journal={International Journal of Multiphase Flow},
  volume={26},
  number={9},
  pages={1509--1524},
  year={2000},
  publisher={Elsevier}
}

@article{glowinski1999distributed,
  title={A distributed Lagrange multiplier/fictitious domain method for flows around moving rigid bodies: application to particulate flow},
  author={Glowinski, Roland and Pan, Tsorng-Whay and Hesla, Todd I and Joseph, Daniel D and Periaux, Jacques},
  journal={International journal for numerical methods in fluids},
  volume={30},
  number={8},
  pages={1043--1066},
  year={1999},
  publisher={Wiley Online Library}
}

@article{chetry2023comparing,
  title={Comparing different stabilization strategies for reduced order modeling of viscoelastic fluid flow problems},
  author={Chetry, Manisha and Borzacchiello, Domenico and D’avino, Gaetano and Da Silva, Luisa Rocha},
  journal={Computers \& Fluids},
  volume={265},
  pages={106013},
  year={2023},
  publisher={Elsevier}
}

@article{rinkens2026bayesian,
  title={Bayesian Model Selection for Complex Flows of Yield Stress Fluids},
  author={Rinkens, Aricia and Verhoosel, Clemens V and Alicke, Alexandra and Anderson, Patrick D and Jaensson, Nick O},
  journal={arXiv preprint arXiv:2601.10115},
  year={2026}
}

@book{rozza2022advanced,
  title={Advanced reduced order methods and applications in computational fluid dynamics},
  author={Rozza, Gianluigi and Stabile, Giovanni and Ballarin, Francesco},
  year={2022},
  publisher={SIAM}
}

@book{brenner2008mathematical,
  title={The mathematical theory of finite element methods},
  author={Brenner, Susanne C and Scott, L Ridgway},
  year={2008},
  publisher={Springer}
}

@book{larson2013finite,
  title={The finite element method: theory, implementation, and applications},
  author={Larson, Mats G and Bengzon, Fredrik},
  volume={10},
  year={2013},
  publisher={Springer Science \& Business Media}
}

@book{john2016finite,
  title={Finite element methods for incompressible flow problems},
  author={John, Volker},
  volume={51},
  year={2016},
  publisher={Springer}
}

@article{peherstorfer2018survey,
  title={Survey of multifidelity methods in uncertainty propagation, inference, and optimization},
  author={Peherstorfer, Benjamin and Willcox, Karen and Gunzburger, Max},
  journal={Siam Review},
  volume={60},
  number={3},
  pages={550--591},
  year={2018},
  publisher={SIAM}
}

@book{ern2004theory,
  title={Theory and practice of finite elements},
  author={Ern, Alexandre and Guermond, Jean-Luc},
  volume={159},
  year={2004},
  publisher={Springer}
}

@book{boffi2013mixed,
  title={Mixed finite element methods and applications},
  author={Boffi, Daniele and Brezzi, Franco and Fortin, Michel and others},
  volume={44},
  year={2013},
  publisher={Springer}
}

@article{xiao2019domain,
  title={A domain decomposition non-intrusive reduced order model for turbulent flows},
  author={Xiao, D and Heaney, CE and Fang, F and Mottet, L and Hu, R and Bistrian, DA and Aristodemou, E and Navon, IM and Pain, CC},
  journal={Computers \& Fluids},
  volume={182},
  pages={15--27},
  year={2019},
  publisher={Elsevier}
}

@article{hajisharifi2023non,
  title={A non-intrusive data-driven reduced order model for parametrized CFD-DEM numerical simulations},
  author={Hajisharifi, Arash and Romano, Francesco and Girfoglio, Michele and Beccari, Andrea and Bonanni, Domenico and Rozza, Gianluigi},
  journal={Journal of Computational Physics},
  volume={491},
  pages={112355},
  year={2023},
  publisher={Elsevier}
}

@article{balajewicz2012stabilization,
  title={Stabilization of projection-based reduced order models of the Navier--Stokes},
  author={Balajewicz, Maciej and Dowell, Earl H},
  journal={Nonlinear Dynamics},
  volume={70},
  number={2},
  pages={1619--1632},
  year={2012},
  publisher={Springer}
}

@book{quarteroni2014reduced,
  title={Reduced order methods for modeling and computational reduction},
  author={Quarteroni, Alfio and Rozza, Gianluigi and others},
  volume={9},
  year={2014},
  publisher={Springer}
}

@article{rinkens2023uncertainty,
  title={Uncertainty quantification for the squeeze flow of generalized Newtonian fluids},
  author={Rinkens, Aricia and Verhoosel, Clemens V and Jaensson, Nick O},
  journal={Journal of Non-Newtonian Fluid Mechanics},
  volume={322},
  pages={105154},
  year={2023},
  publisher={Elsevier}
}

@article{benner2015survey,
  title={A survey of projection-based model reduction methods for parametric dynamical systems},
  author={Benner, Peter and Gugercin, Serkan and Willcox, Karen},
  journal={SIAM review},
  volume={57},
  number={4},
  pages={483--531},
  year={2015},
  publisher={SIAM}
}

@article{abbasian2020effects,
  title={Effects of different non-Newtonian models on unsteady blood flow hemodynamics in patient-specific arterial models with in-vivo validation},
  author={Abbasian, Majid and Shams, Mehrzad and Valizadeh, Ziba and Moshfegh, Abouzar and Javadzadegan, Ashkan and Cheng, Shaokoon},
  journal={Computer methods and programs in biomedicine},
  volume={186},
  pages={105185},
  year={2020},
  publisher={Elsevier}
}

@article{hernandez2017dimensional,
  title={Dimensional hyper-reduction of nonlinear finite element models via empirical cubature},
  author={Hernandez, Joaquin Alberto and Caicedo, Manuel Alejandro and Ferrer, Alex},
  journal={Computer methods in applied mechanics and engineering},
  volume={313},
  pages={687--722},
  year={2017},
  publisher={Elsevier}
}

@article{barrault2004empirical,
  title={An ‘empirical interpolation’method: application to efficient reduced-basis discretization of partial differential equations},
  author={Barrault, Maxime and Maday, Yvon and Nguyen, Ngoc Cuong and Patera, Anthony T},
  journal={Comptes Rendus Mathematique},
  volume={339},
  number={9},
  pages={667--672},
  year={2004},
  publisher={Elsevier}
}

@article{xiao2015non,
  title={Non-intrusive reduced-order modelling of the Navier--Stokes equations based on RBF interpolation},
  author={Xiao, Dunhui and Fang, Fangxin and Pain, Christopher and Hu, Guangwei},
  journal={International Journal for Numerical Methods in Fluids},
  volume={79},
  number={11},
  pages={580--595},
  year={2015},
  publisher={Wiley Online Library}
}

@article{reyes2023reduced,
  title={Reduced order modeling for parametrized generalized Newtonian fluid flows},
  author={Reyes, Ricardo and Ruz, Oscar and Bayona-Roa, Camilo and Castillo, Ernesto and Tello, Alexis},
  journal={Journal of Computational Physics},
  volume={484},
  pages={112086},
  year={2023},
  publisher={Elsevier}
}

@article{gunzburger2007reduced,
  title={Reduced-order modeling of time-dependent PDEs with multiple parameters in the boundary data},
  author={Gunzburger, Max D and Peterson, Janet S and Shadid, John N},
  journal={Computer methods in applied mechanics and engineering},
  volume={196},
  number={4-6},
  pages={1030--1047},
  year={2007},
  publisher={Elsevier}
}

@article{hesthaven2018non,
  title={Non-intrusive reduced order modeling of nonlinear problems using neural networks},
  author={Hesthaven, Jan S and Ubbiali, Stefano},
  journal={Journal of Computational Physics},
  volume={363},
  pages={55--78},
  year={2018},
  publisher={Elsevier}
}

@article{guo2018reduced,
  title={Reduced order modeling for nonlinear structural analysis using Gaussian process regression},
  author={Guo, Mengwu and Hesthaven, Jan S},
  journal={Computer methods in applied mechanics and engineering},
  volume={341},
  pages={807--826},
  year={2018},
  publisher={Elsevier}
}

@article{krishnan2013simulation,
  title={Simulation of non-Newtonian fluid-food particle heat transfer in the holding tube used in aseptic processing operations},
  author={Krishnan, Suresh and Aravamudan, Kannan},
  journal={food and bioproducts processing},
  volume={91},
  number={2},
  pages={129--148},
  year={2013},
  publisher={Elsevier}
}

@article{liu2006laminar,
  title={Laminar mixing of shear thinning fluids in a SMX static mixer},
  author={Liu, Shiping and Hrymak, Andrew N and Wood, Philip E},
  journal={Chemical Engineering Science},
  volume={61},
  number={6},
  pages={1753--1759},
  year={2006},
  publisher={Elsevier}
}

@article{quarteroni2016geometric,
  title={Geometric multiscale modeling of the cardiovascular system, between theory and practice},
  author={Quarteroni, Alfio and Veneziani, Alessandro and Vergara, Christian},
  journal={Computer Methods in Applied Mechanics and Engineering},
  volume={302},
  pages={193--252},
  year={2016},
  publisher={Elsevier}
}

@article{chaturantabut2010nonlinear,
  title={Nonlinear model reduction via discrete empirical interpolation},
  author={Chaturantabut, Saifon and Sorensen, Danny C},
  journal={SIAM Journal on Scientific Computing},
  volume={32},
  number={5},
  pages={2737--2764},
  year={2010},
  publisher={SIAM}
}

@book{bird1987dynamics,
  title     = {Dynamics of Polymeric Liquids, Volume 1: Fluid Mechanics},
  author    = {Bird, R. Byron and Armstrong, Robert C. and Hassager, Ole},
  edition   = {2},
  year      = {1987},
  publisher = {John Wiley \& Sons},
  address   = {New York}
}

@book{Holmes_Lumley_Berkooz_Rowley_2012, place={Cambridge}, edition={2}, series={Cambridge Monographs on Mechanics}, title={Turbulence, Coherent Structures, Dynamical Systems and Symmetry}, publisher={Cambridge University Press}, author={Holmes, Philip and Lumley, John L. and Berkooz, Gahl and Rowley, Clarence W.}, year={2012}, collection={Cambridge Monographs on Mechanics}}

@article{kunisch2002galerkin,
  title={Galerkin proper orthogonal decomposition methods for a general equation in fluid dynamics},
  author={Kunisch, Karl and Volkwein, Stefan},
  journal={SIAM Journal on Numerical analysis},
  volume={40},
  number={2},
  pages={492--515},
  year={2002},
  publisher={SIAM}
}

@article{peherstorfer2020stability,
  title={Stability of discrete empirical interpolation and gappy proper orthogonal decomposition with randomized and deterministic sampling points},
  author={Peherstorfer, Benjamin and Drmac, Zlatko and Gugercin, Serkan},
  journal={SIAM Journal on Scientific Computing},
  volume={42},
  number={5},
  pages={A2837--A2864},
  year={2020},
  publisher={SIAM}
}

@article{drmac2016new,
  title={A new selection operator for the discrete empirical interpolation method---improved a priori error bound and extensions},
  author={Drmac, Zlatko and Gugercin, Serkan},
  journal={SIAM Journal on Scientific Computing},
  volume={38},
  number={2},
  pages={A631--A648},
  year={2016},
  publisher={SIAM}
}

@book{hesthaven2016certified,
  title     = {Certified Reduced Basis Methods for Parametrized Partial Differential Equations},
  author    = {Hesthaven, Jan S. and Rozza, Gianluigi and Stamm, Benjamin},
  series    = {SpringerBriefs in Mathematics},
  year      = {2016},
  publisher = {Springer Cham},
  edition   = {1},
  pages     = {XIII, 131},
  isbn      = {978-3-319-22470-1},
  issn      = {2191-8198}
}

@article{bruneau2006cavity,
  title        = {The 2D lid-driven cavity problem revisited},
  author       = {Bruneau, Charles-Henri and Saad, Mazen},
  journal      = {Computers \& Fluids},
  volume       = {35},
  number       = {3},
  pages        = {326--348},
  year         = {2006},
}

@article{ballarin2015supremizer,
  title        = {Supremizer stabilization of POD--Galerkin approximation of parametrized steady incompressible Navier--Stokes equations},
  author       = {Ballarin, Francesco and Manzoni, Andrea and Quarteroni, Alfio and Rozza, Gianluigi},
  journal      = {International Journal for Numerical Methods in Engineering},
  volume       = {102},
  number       = {5},
  pages        = {1136--1161},
  year         = {2015}
}

@article{oishi2024nonlinear,
  title={Nonlinear parametric models of viscoelastic fluid flows},
  author={Oishi, Cassio M. and Kaptanoglu, Alan A. and Kutz, J. Nathan and Brunton, Steven L.},
  journal={Royal Society Open Science},
  volume={11},
  number={10},
  pages={240995},
  year={2024},
  publisher={The Royal Society}
}

@article{farhat2015structure,
  title={Structure-preserving, stability, and accuracy properties of the energy-conserving sampling and weighting method for the hyper reduction of nonlinear finite element dynamic models},
  author={Farhat, Charbel and Chapman, Todd and Avery, Philip},
  journal={International journal for numerical methods in engineering},
  volume={102},
  number={5},
  pages={1077--1110},
  year={2015},
  publisher={Wiley Online Library}
}

@article{wang2019non,
  title={Non-intrusive reduced order modeling of unsteady flows using artificial neural networks with application to a combustion problem},
  author={Wang, Qian and Hesthaven, Jan S and Ray, Deep},
  journal={Journal of computational physics},
  volume={384},
  pages={289--307},
  year={2019},
  publisher={Elsevier}
}

@article{cicci2023uncertainty,
  title={Uncertainty quantification for nonlinear solid mechanics using reduced order models with Gaussian process regression},
  author={Cicci, Ludovica and Fresca, Stefania and Guo, Mengwu and Manzoni, Andrea and Zunino, Paolo},
  journal={Computers \& Mathematics with Applications},
  volume={149},
  pages={1--23},
  year={2023},
  publisher={Elsevier}
}

@article{czech2022data,
  title={Data-driven models for crashworthiness optimisation: intrusive and non-intrusive model order reduction techniques: C. Czech et al.},
  author={Czech, Catharina and Lesjak, Mathias and Bach, Christopher and Duddeck, Fabian},
  journal={Structural and Multidisciplinary Optimization},
  volume={65},
  number={7},
  pages={190},
  year={2022},
  publisher={Springer}
}

@article{padula2024brief,
  title={A brief review of reduced order models using intrusive and non-intrusive techniques},
  author={Padula, Guglielmo and Girfoglio, Michele and Rozza, Gianluigi},
  journal={PAMM},
  volume={24},
  number={4},
  pages={e202400210},
  year={2024},
  publisher={Wiley Online Library}
}

@article{lee2019simultaneous,
  author  = {Lee, Yong Hoon and Schuh, Jonathon K. and Ewoldt, Randy H. and Allison, James T.},
  title   = {Simultaneous design of non-{N}ewtonian lubricant and surface texture using surrogate-based multiobjective optimization},
  journal = {Structural and Multidisciplinary Optimization},
  volume  = {60},
  number  = {1},
  pages   = {99--116},
  year    = {2019},
}

@article{kim2019uncertainty,
  author  = {Kim, Jaekwang and Singh, Piyush K. and Freund, Jonathan B. and Ewoldt, Randy H.},
  title   = {Uncertainty propagation in simulation predictions of generalized {N}ewtonian fluid flows},
  journal = {Journal of Non-Newtonian Fluid Mechanics},
  volume  = {271},
  pages   = {104138},
  year    = {2019},
}

@article{kontogiannis2025learning,
  title={Learning rheological parameters of non-Newtonian fluids from velocimetry data},
  author={Kontogiannis, Alexandros and Hodgkinson, Richard and Reynolds, Steven and Manchester, Emily L},
  journal={Journal of Fluid Mechanics},
  volume={1011},
  pages={R3},
  year={2025},
  publisher={Cambridge University Press}
}

@misc{amor2026reduced,
  author        = {Amor, Christian and Corrochano, Adrián and Rosti, Marco Edoardo and Le Clainche, Soledad},
  title         = {Reduced-order modeling of a viscoelastic turbulent jet with hybrid machine learning models},
  year          = {2026},
  eprint        = {2604.26240},
  archivePrefix = {arXiv},
  primaryClass  = {physics.flu-dyn}
}

@article{Fonn2019,
  author  = {Fonn, Eivind and van Brummelen, Harald and Kvamsdal, Trond and Rasheed, Adil},
  title   = {Fast divergence-conforming reduced basis methods for steady {Navier}--{Stokes} flow},
  journal = {Computer Methods in Applied Mechanics and Engineering},
  volume  = {346},
  pages   = {486--512},
  year    = {2019},
  month   = apr,
  publisher = {North-Holland},
}

\end{document}


\maketitle
 
\tableofcontents

\newpage
 
\section{Mesh Convergence}
\label{app:mesh_convergence}

We performed a spatial mesh convergence study for the two benchmark problems
considered in this work: the lid-driven cavity and the settling sphere. The meshes are generated using Gmsh \cite{geuzaine2009gmsh}.
The full-order model employs Taylor-Hood ($\mathcal{P}_2/\mathcal{P}_1$)
finite elements, which yield quadratic velocity and linear pressure
approximations. Classical \emph{a priori} error analysis for this element pair
predicts convergence rates of order $\mathcal{O}(h^3)$ in the $L^2$-norm for
the velocity and $\mathcal{O}(h^2)$ for derived quantities such as the apparent
viscosity, where $h$ denotes the characteristic element size.

Spatial errors are quantified by computing relative $L^2$ errors along a fixed
sampling line $\Gamma_\mathrm{c}$ drawn through the domain interior (see
Figures~\ref{fig:sampling_lid} and~\ref{fig:sampling_axi}). Let $\rho$ denote
a local arc-length coordinate along $\Gamma_\mathrm{c}$. For each mesh level
$k \in \{\mathrm{M1}, \mathrm{M2}, \mathrm{M3}, \mathrm{M4}\}$, the relative
velocity error and the relative viscosity error with respect to the reference solution on the finest mesh M5 is defined as
\begin{equation}
    e_{{u}}
    = \frac{
        \displaystyle\left(\int_{\Gamma_\mathrm{c}}
        \bigl\|\vec{u}_h - \vec{u}_h^{\,\mathrm{ref}}\bigr\|^2 \,\mathrm{d}\rho
        \right)^{1/2}
      }{
        \displaystyle\left(\int_{\Gamma_\mathrm{c}}
        \bigl\|\vec{u}_h^{\,\mathrm{ref}}\bigr\|^2 \,\mathrm{d}\rho
        \right)^{1/2}
      },
    \qquad
    e_\eta
    = \frac{
        \displaystyle\left(\int_{\Gamma_\mathrm{c}}
        \bigl(\eta_h - \eta_h^{\,\mathrm{ref}}\bigr)^2 \,\mathrm{d}\rho
        \right)^{1/2}
      }{
        \displaystyle\left(\int_{\Gamma_\mathrm{c}}
        \bigl(\eta_h^{\,\mathrm{ref}}\bigr)^2 \,\mathrm{d}\rho
        \right)^{1/2}
      },
    \label{eq:eu_eeta}
\end{equation}
where $\vec{u}_h$ and $\eta_h$ are the velocity field and the viscosity field,
respectively, on mesh $k$, and the superscript $\mathrm{ref}$ denotes the
solution computed on M5. The integrals are evaluated numerically by dividing
$\Gamma_\mathrm{c}$ into $10{,}000$ equal intervals and applying the midpoint
rule. The reference solution on M5 is computed independently for a fixed
parameter value $(\lambda, n) = (5, 0.3)$ (the most nonlinear case in the
parameter range), and the same reference is used for all mesh levels. The
abscissa in the convergence plots is $\sqrt{N_\mathrm{elem}}$, where
$N_\mathrm{elem}$ is the number of elements on the respective mesh.

\subsection{Lid-Driven Cavity}

\begin{figure}[!ht]
    \centering
    \includegraphics[width=0.3\linewidth]{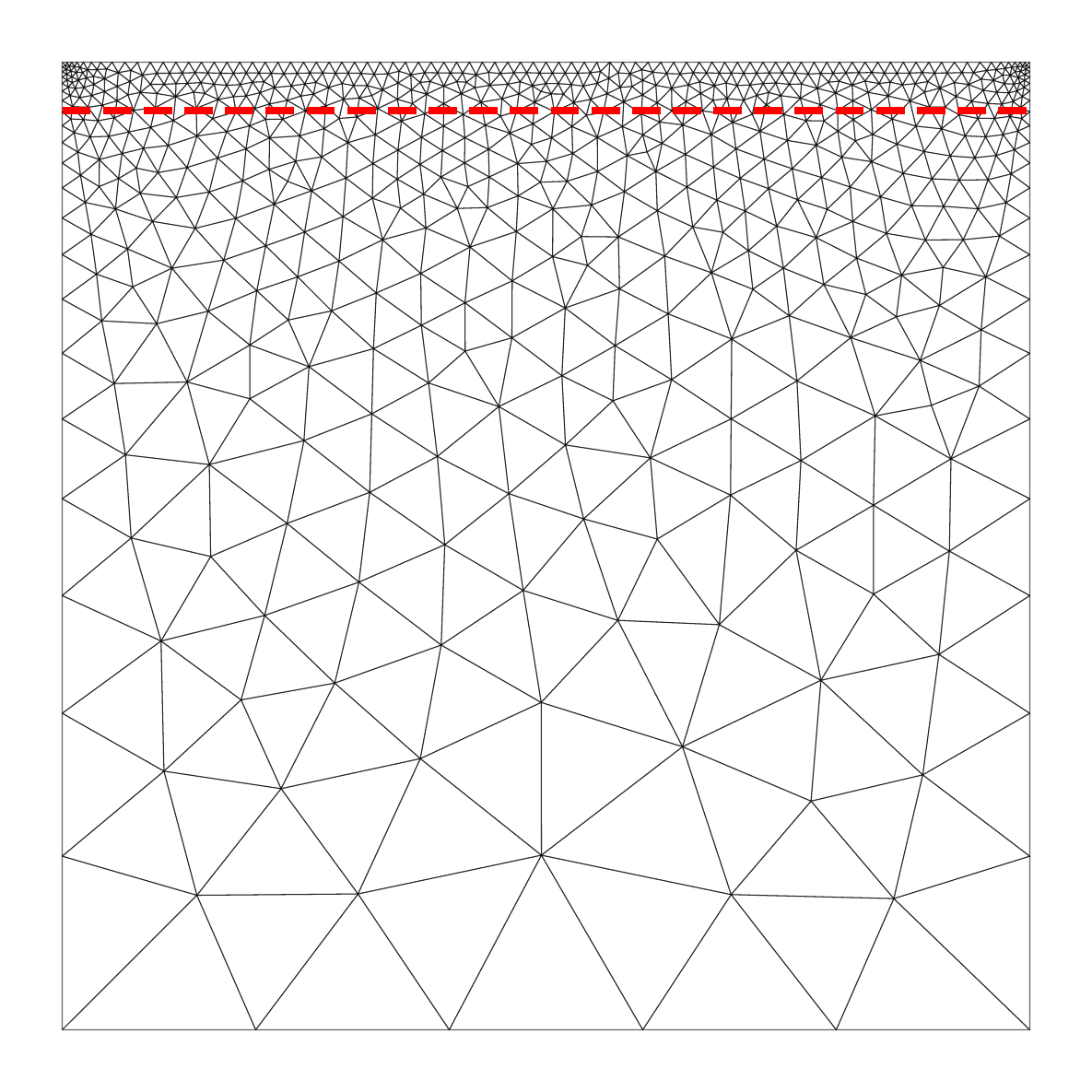}
\caption{Computational mesh M2 for the lid-driven cavity, with the sampling
line $\Gamma_\mathrm{c}$ (dashed, red) along which the convergence errors
$e_{{u}}$ and $e_\eta$ are evaluated.}
    \label{fig:sampling_lid}
\end{figure}

The lid-driven cavity domain is a unit square. Large velocity gradients occur
near the moving lid, and stress concentrations occur at the two upper corners.
The mesh is therefore graded: element sizes are smallest at the top corners
and along the top edge (where the lid velocity is prescribed), and largest
along the bottom edge. Five mesh levels, M1 to M5, are constructed by uniformly
halving all characteristic element sizes between successive levels. The three
characteristic element sizes $h_\mathrm{corners}$ (top corners),
$h_\mathrm{top}$ (top edge), and $h_\mathrm{bottom}$ (bottom edge) for each
mesh level are reported in Table~\ref{tab:mesh_characteristics}. The sampling
line $\Gamma_\mathrm{c}$ is the horizontal line at $y = 0.9$, as shown in
Figure~\ref{fig:sampling_lid}. The total mesh sizes are listed in
Table~\ref{tab:mesh_lid_info}.

\begin{table}[!ht]
\centering
\begin{tabular}{|c|c|c|c|}
\hline
Mesh & $h_\mathrm{corners}$ & $h_\mathrm{top}$ & $h_\mathrm{bottom}$ \\
\hline
M1 & $0.010$    & $0.027$   & $0.4$   \\
M2 & $0.005$    & $0.013$   & $0.2$   \\
M3 & $0.0025$   & $0.0067$  & $0.1$   \\
M4 & $0.00125$  & $0.0033$  & $0.05$  \\
M5 & $0.000625$ & $0.00167$ & $0.025$ \\
\hline
\end{tabular}
\caption{Characteristic element sizes for the lid-driven cavity convergence
study. $h_\mathrm{corners}$ is the element size at the top corners,
$h_\mathrm{top}$ is the element size along the moving lid, and
$h_\mathrm{bottom}$ is the element size along the stationary bottom edge.
Each successive mesh level halves all characteristic sizes.}
\label{tab:mesh_characteristics}
\end{table}

\begin{table}[!ht]
\centering
\begin{tabular}{|c|c|c|}
\hline
Mesh & Number of nodes & Number of elements \\
\hline
M1 & 789      & 360     \\
M2 & 2,588   & 1,229  \\
M3 & 9,916   & 4,829  \\
M4 & 38,842  & 19,165 \\
M5 & 152,784 & 75,881 \\
\hline
\end{tabular}
\caption{Mesh sizes used in the lid-driven cavity convergence study.}
\label{tab:mesh_lid_info}
\end{table}

\begin{figure}[!ht]
    \centering
    \begin{subfigure}[t]{0.33\linewidth}
        \centering
        \includegraphics[width=\linewidth]{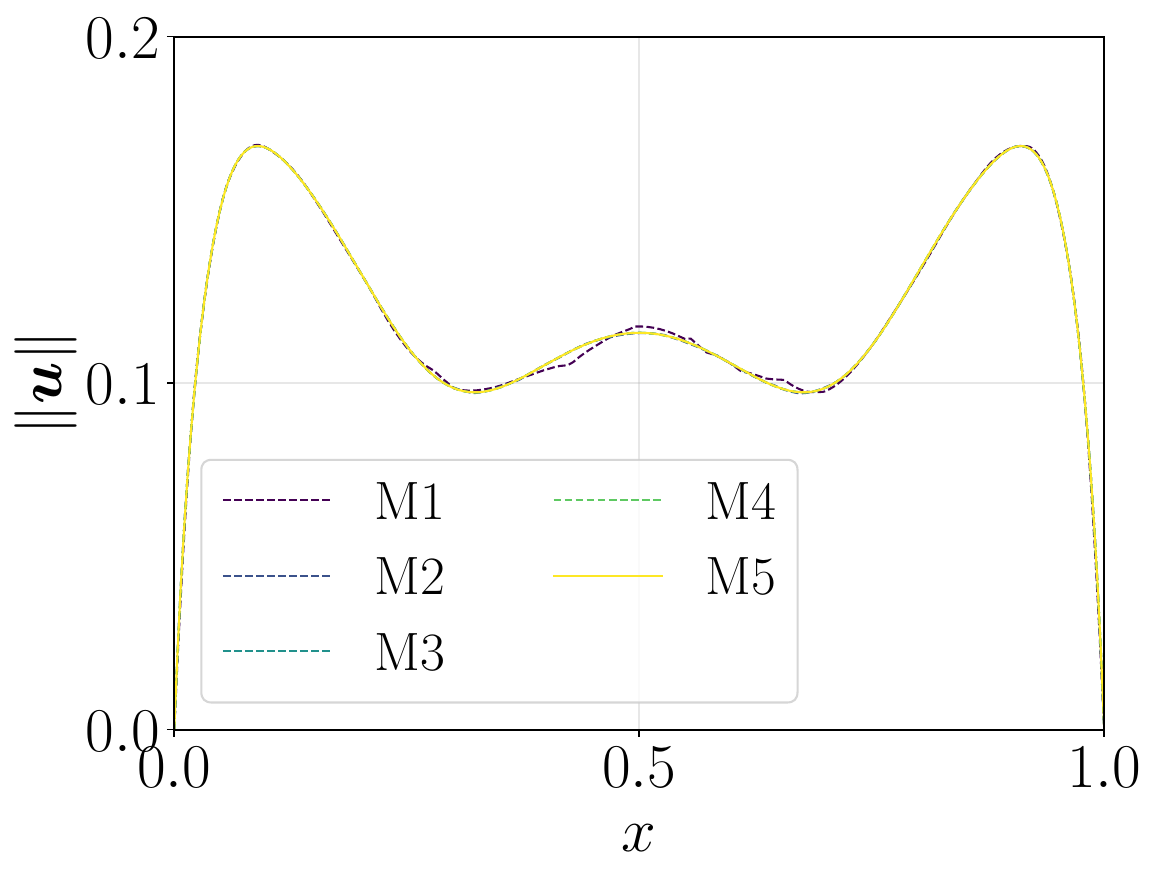}
        \caption{}
    \end{subfigure}
    \begin{subfigure}[t]{0.33\linewidth}
        \centering
        \includegraphics[width=\linewidth]{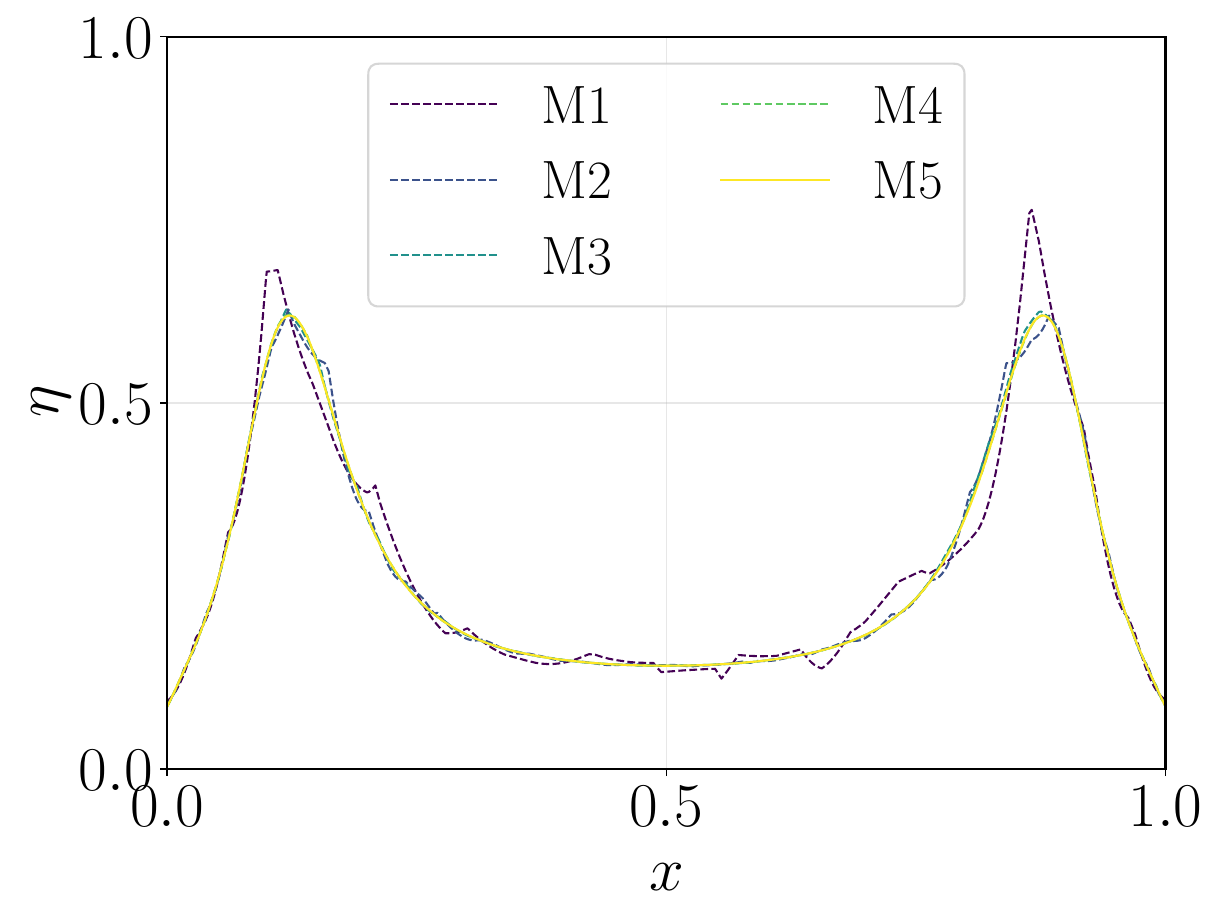}
        \caption{}
    \end{subfigure}

    \vspace{0.4cm}

    \begin{subfigure}[t]{0.33\linewidth}
        \centering
        \includegraphics[width=\linewidth]{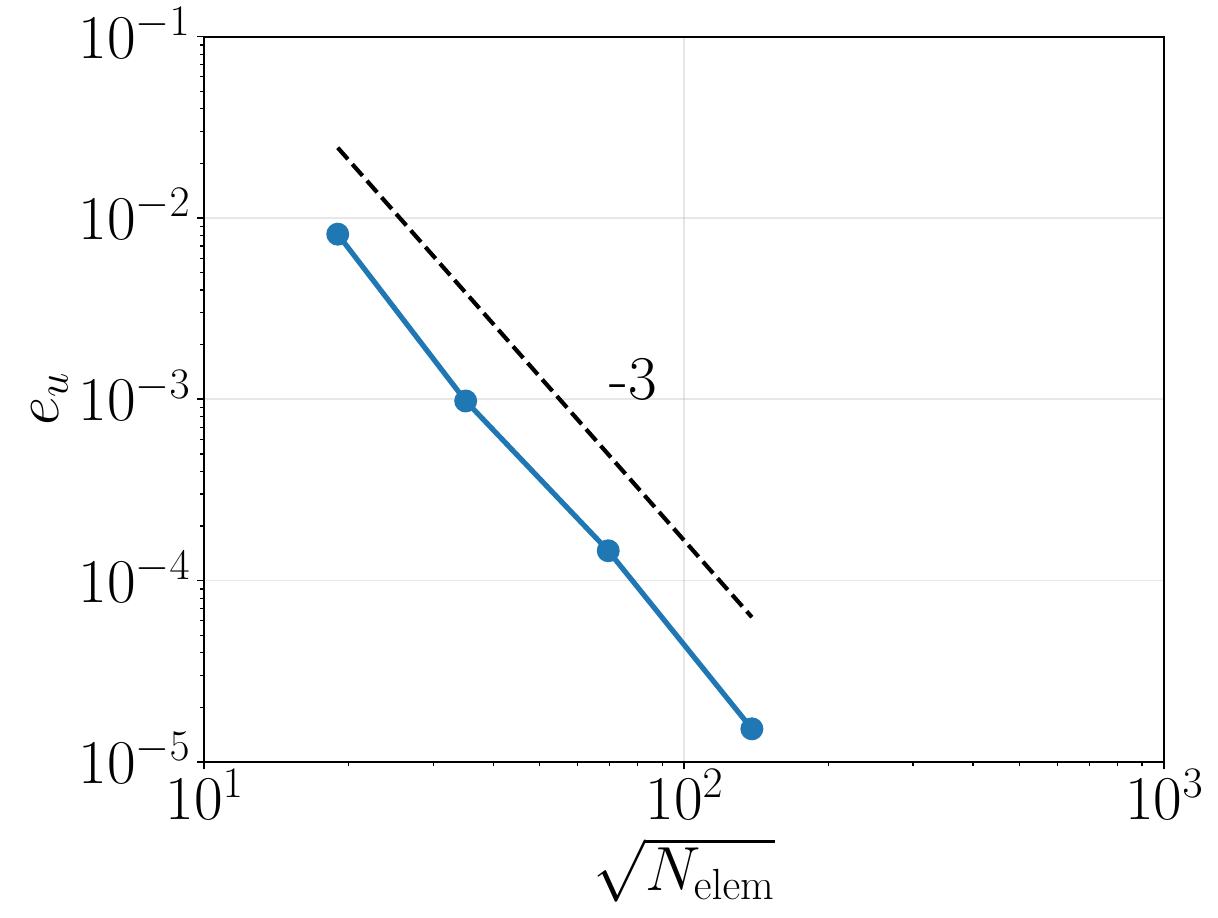}
        \caption{}
    \end{subfigure}
    \begin{subfigure}[t]{0.33\linewidth}
        \centering
        \includegraphics[width=\linewidth]{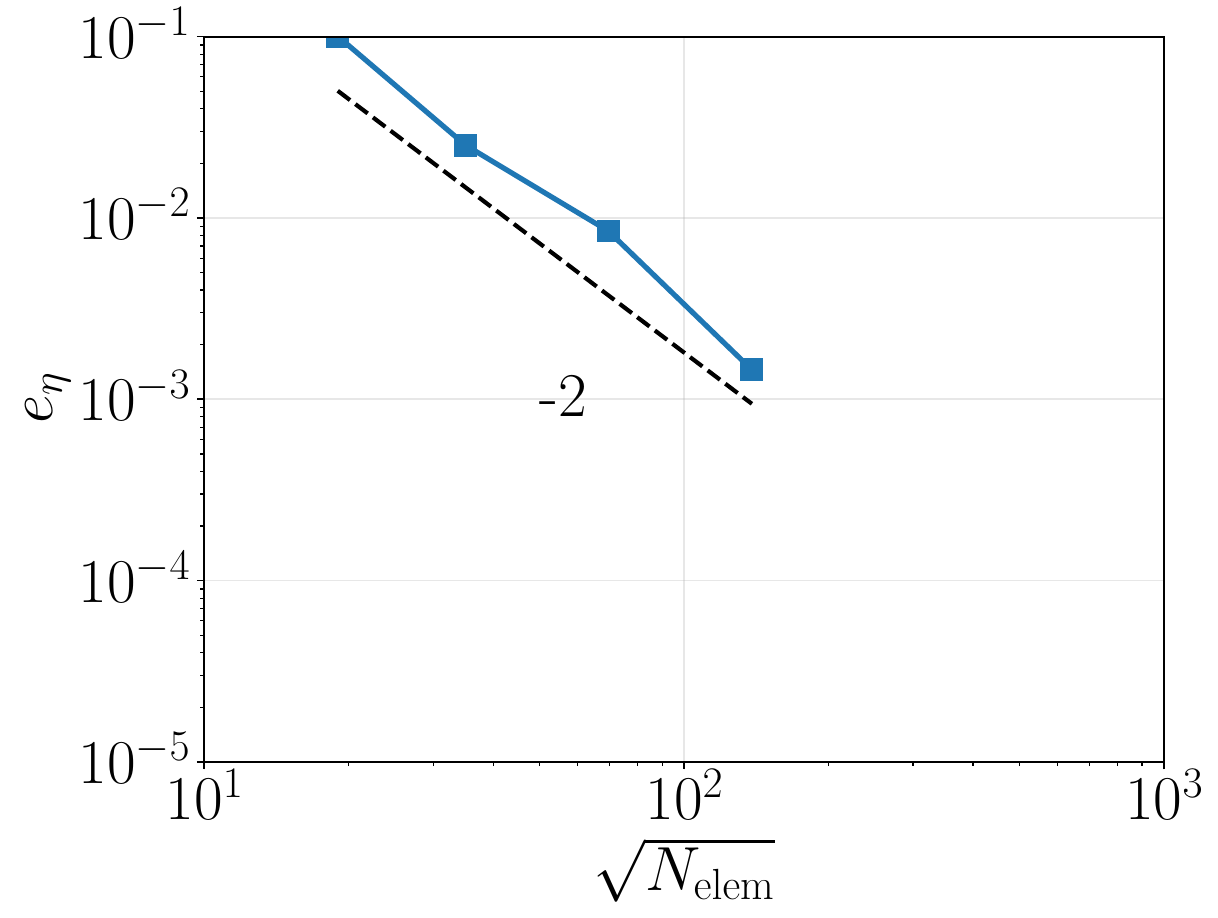}
        \caption{}
    \end{subfigure}
\caption{Mesh convergence study for the lid-driven cavity. Panels (a) and (b)
show the velocity magnitude $\|\vec{u}\|$ and viscosity $\eta$ profiles along
the sampling line $\Gamma_\mathrm{c}$ for the five meshes M1 to M5. Panels (c)
and (d) report the relative $L^2$ errors $e_{{u}}$ and $e_\eta$ as
functions of $\sqrt{N_\mathrm{elem}}$, with M5 taken as the reference solution.
The dashed lines indicate the theoretical convergence slopes of $-3$ and $-2$,
respectively.}
    \label{fig:mesh_conv_lid}
\end{figure}

The velocity and viscosity profiles along $\Gamma_\mathrm{c}$ for meshes M1 to
M5 are shown in the upper panels of Figure~\ref{fig:mesh_conv_lid}. The lower
panels display the corresponding $h$-convergence rates. Both $e_{{u}}$ and
$e_\eta$ decrease monotonically with mesh refinement and closely follow the
theoretical slopes of $-3$ and $-2$, respectively. Based on this study, mesh
M2 is selected for all high-fidelity lid-driven cavity simulations; it provides
sufficient accuracy at a manageable computational cost.

\subsection{Settling Sphere}

\begin{figure}[!ht]
    \centering
    \includegraphics[width=0.6\linewidth]{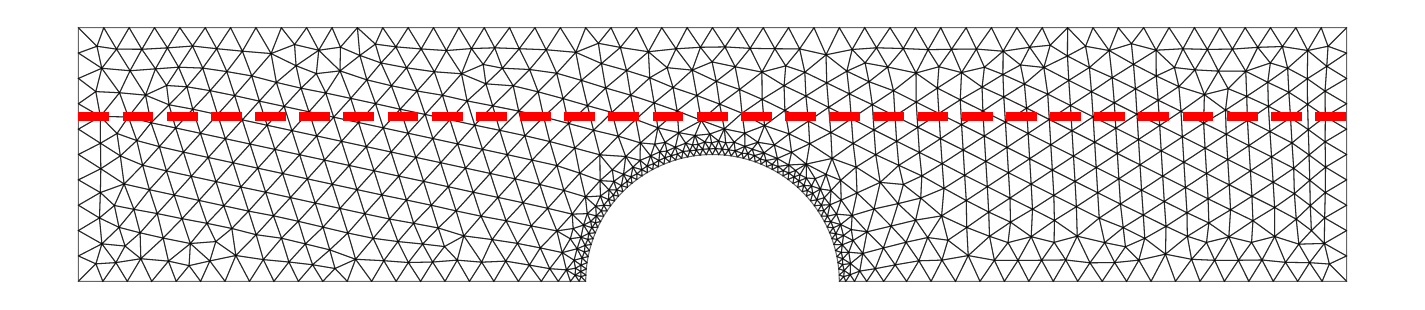}
\caption{Computational mesh M2 for the settling sphere problem, with the
sampling line $\Gamma_\mathrm{c}$ (dashed, red) along which the convergence
errors $e_{{u}}$ and $e_\eta$ are evaluated.}
    \label{fig:sampling_axi}
\end{figure}

The settling sphere problem is formulated in axisymmetric coordinates. The
flow domain is a closed cylinder enclosing the spherical particle, and the
primary source of error is the thin layer that develops around the particle
surface. To resolve this layer, a two-zone grading strategy is employed: the
characteristic element size on the particle surface $h_\mathrm{particle}$ is
set to one quarter of the far-field element size $h_\mathrm{far}$, \ie
$h_\mathrm{particle} = h_\mathrm{far}/4$. These two length scales for each
mesh level are reported in
Table~\ref{tab:mesh_characteristics_axi_simplified}. The sampling line
$\Gamma_\mathrm{c}$ is the axial mid-line of the domain, as shown in
Figure~\ref{fig:sampling_axi}. The total mesh sizes are listed in
Table~\ref{tab:mesh_axi_info}.

\begin{table}[!ht]
\centering
\begin{tabular}{|c|c|c|}
\hline
Mesh & $h_\mathrm{far}$ & $h_\mathrm{particle}$ \\
\hline
M1 & $0.4$    & $0.1$     \\
M2 & $0.2$    & $0.05$    \\
M3 & $0.1$    & $0.025$   \\
M4 & $0.05$   & $0.0125$  \\
M5 & $0.025$  & $0.00625$ \\
\hline
\end{tabular}
\caption{Characteristic element sizes for the settling sphere convergence
study. $h_\mathrm{far}$ is the element size in the far field and
$h_\mathrm{particle} = h_\mathrm{far}/4$ is the element size on the particle
surface. Each successive mesh level halves all characteristic sizes.}
\label{tab:mesh_characteristics_axi_simplified}
\end{table}

\begin{table}[!ht]
\centering
\begin{tabular}{|c|c|c|}
\hline
Mesh & Number of nodes & Number of elements \\
\hline
M1 & 904      & 405     \\
M2 & 3,041   & 1,430  \\
M3 & 11,105  & 5,376  \\
M4 & 42,714  & 21,004 \\
M5 & 168,109 & 83,350 \\
\hline
\end{tabular}
\caption{Mesh sizes used in the settling sphere convergence study.}
\label{tab:mesh_axi_info}
\end{table}

\begin{figure}[!ht]
    \centering
    \begin{subfigure}[t]{0.33\linewidth}
        \centering
        \includegraphics[width=\linewidth]{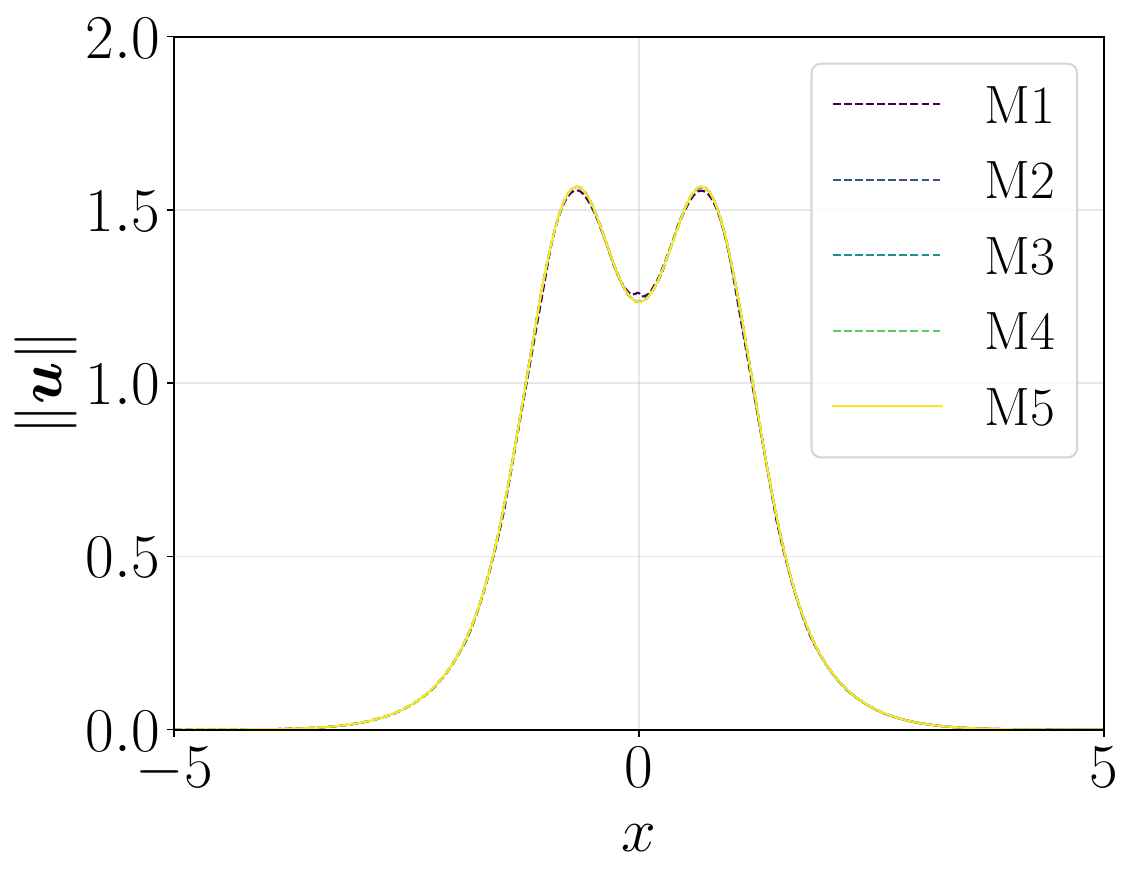}
        \caption{}
    \end{subfigure}
    \begin{subfigure}[t]{0.33\linewidth}
        \centering
        \includegraphics[width=\linewidth]{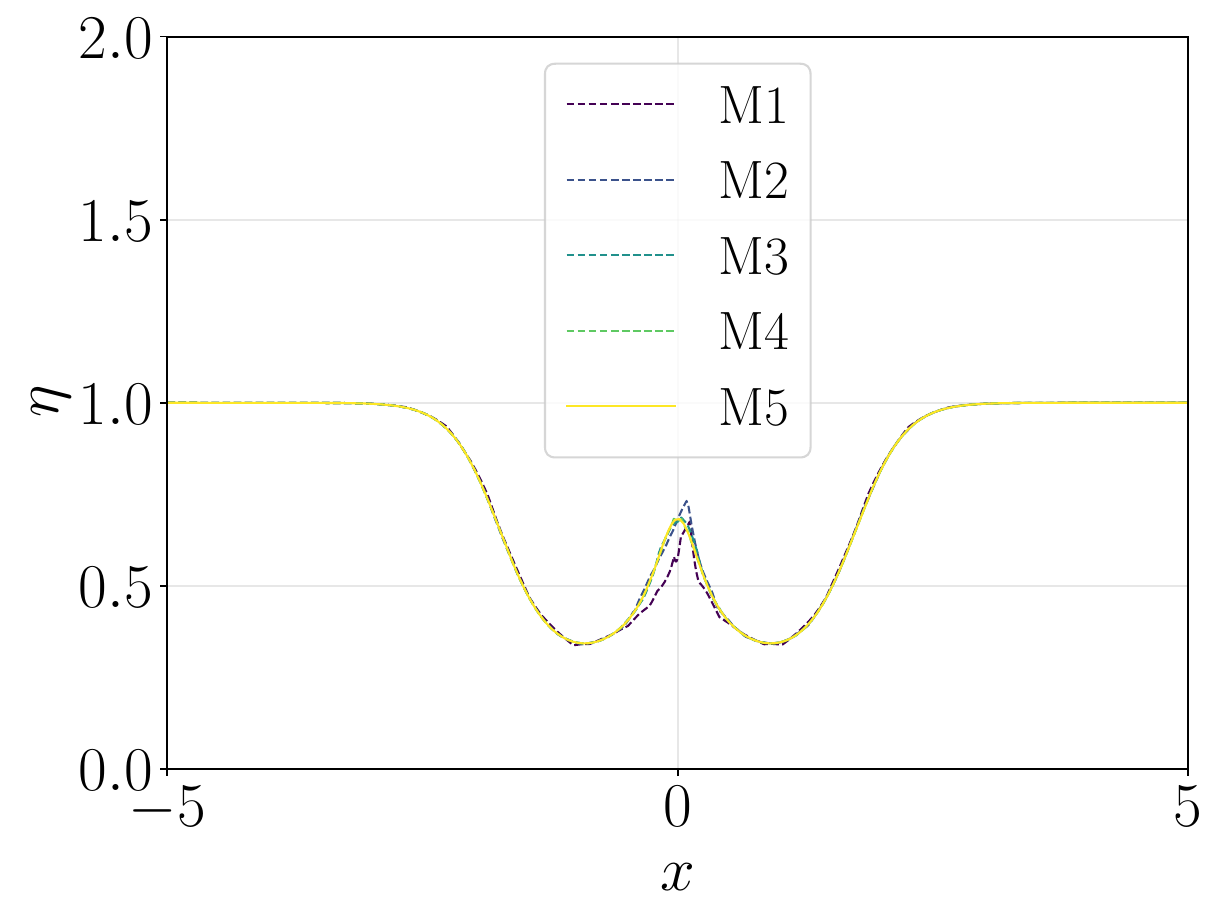}
        \caption{}
    \end{subfigure}

    \vspace{0.4cm}

    \begin{subfigure}[t]{0.33\linewidth}
        \centering
        \includegraphics[width=\linewidth]{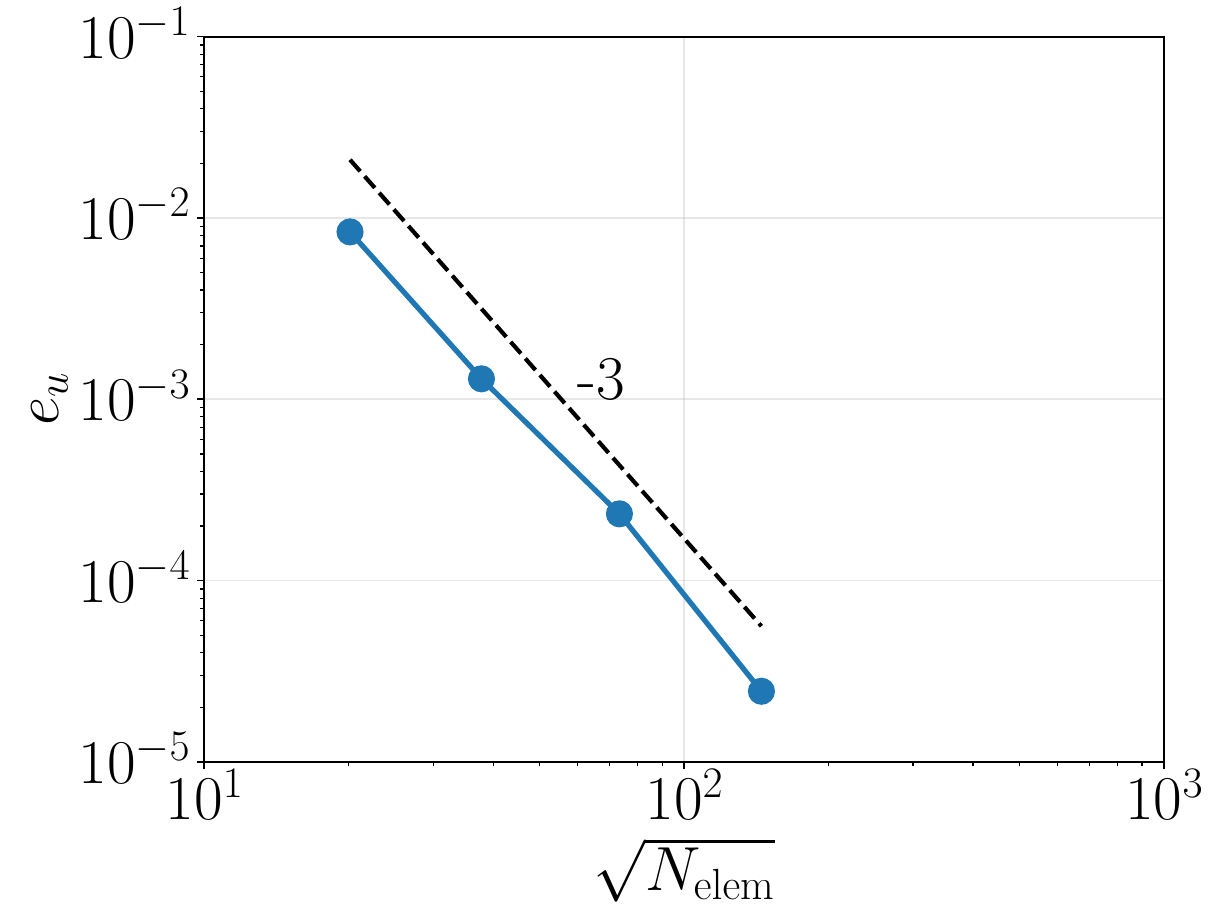}
        \caption{}
    \end{subfigure}
    \begin{subfigure}[t]{0.33\linewidth}
        \centering
        \includegraphics[width=\linewidth]{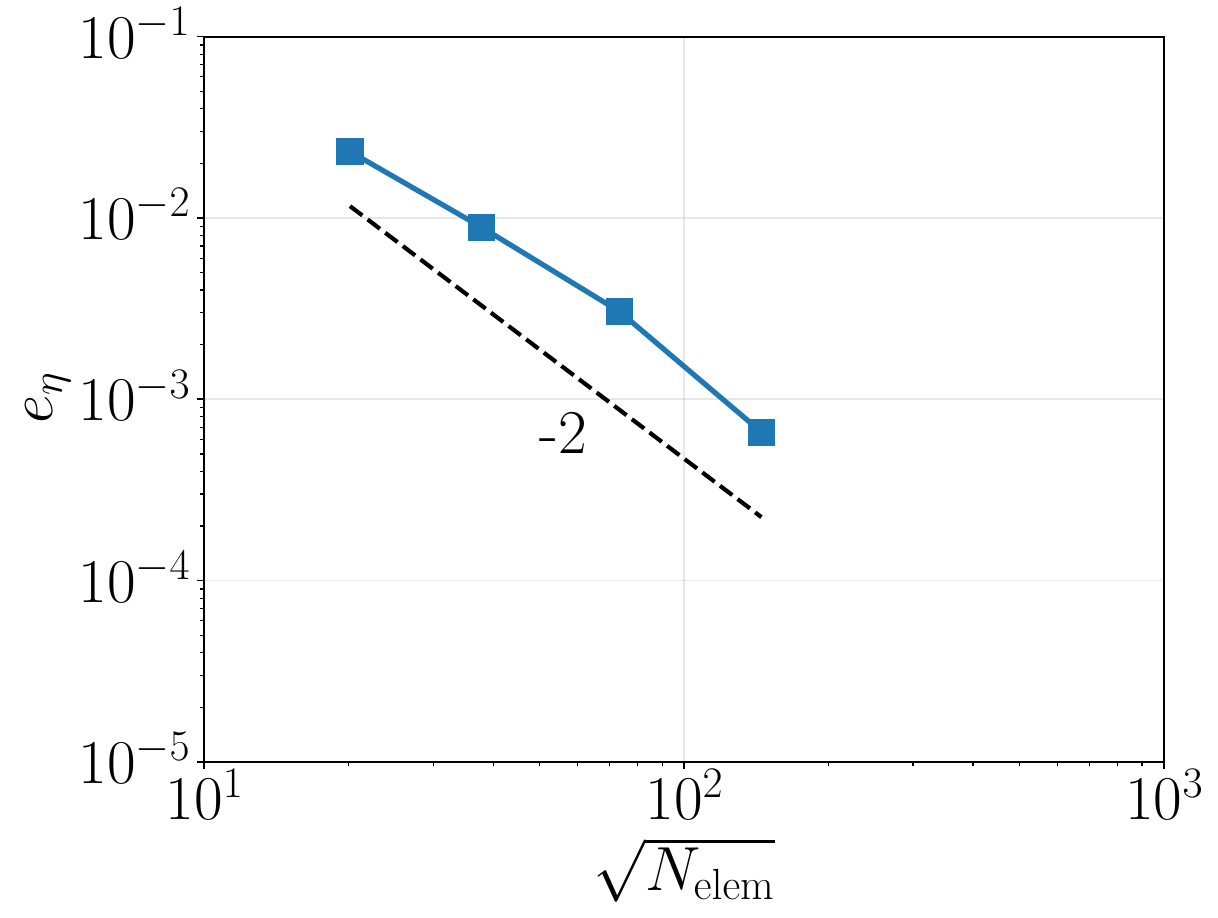}
        \caption{}
    \end{subfigure}
\caption{Mesh convergence study for the settling sphere problem. Panels (a)
and (b) show the velocity magnitude $\|\vec{u}\|$ and viscosity $\eta$ profiles
along the sampling line $\Gamma_\mathrm{c}$ for the five meshes M1 to M5.
Panels (c) and (d) report the relative $L^2$ errors $e_{{u}}$ and $e_\eta$
as functions of $\sqrt{N_\mathrm{elem}}$, with M5 taken as the reference
solution. The dashed lines indicate the theoretical convergence slopes of $-3$
and $-2$, respectively.}
    \label{mesh_conv_axi}
\end{figure}

The velocity and viscosity profiles along $\Gamma_\mathrm{c}$ for meshes M1 to
M5 are shown in the upper panels of Figure~\ref{mesh_conv_axi}, and the
$h$-convergence rates are shown in the lower panels. As for the lid-driven
cavity, both $e_{{u}}$ and $e_\eta$ follow the theoretical slopes of $-3$
and $-2$, confirming the expected convergence behavior and the suitability of
M5 as a reference. Based on this study, mesh M2 is selected for all
high-fidelity settling sphere simulations.

\section{ROM performance}
\label{sec:rom_time_accuracy_comparison}
We assess ROM performance in two complementary ways by measuring online timings and the number of Picard iterations. All computations were performed on an ASUS Vivobook laptop with an AMD Ryzen~7 7730U processor (8 cores, 16 threads, 2.0\,GHz) and 16\,GB of RAM, running
Windows~11. Similarly, all three ROMs were implemented in
Python~3.14.4, using NumPy~2.4.4 and SciPy~1.17.1 for dense and sparse linear
algebra.

\subsection{Measured timings}
Tables~\ref{tab:rom_time_accuracy_lid} and~\ref{tab:rom_time_accuracy_axi} summarise
the \emph{online} cost per query for all three ROM strategies on the lid-driven cavity
and settling-sphere benchmarks, respectively.
The reduced dimension $r$ denotes the number of retained POD velocity modes;
$s$ is the number of viscosity basis functions used in the DEIM approximation;
and $q = 2s$ is the number of (oversampled) DEIM interpolation points selected
by the GappyPOD$+$R procedure.
The column $n_{\mathrm{picard}}$ reports the number of Picard iterations required
for convergence of the \emph{online} nonlinear solve; ROM-RBF requires no iterative
solve and thus has no entry.

The results confirm that ROM-FULL, which reassembles the full-order operator at every
Picard iteration, incurs an \emph{online} cost on the order of $10$--$25$\,s per query.
ROM-DEIM reduces this cost by three to four orders of magnitude, to a few milliseconds,
by replacing the full-field viscosity assembly with $q$ point-wise evaluations.
ROM-RBF is the cheapest of the three, requiring only a single matrix-vector product
for the regression evaluation, with a cost below $5$\,ms independent of the rheological
regime.
The shear-thinning cases ($n = 0.35$) require more Picard iterations than the
shear-thickening cases ($n = 1.5$) for both ROM-FULL and ROM-DEIM, consistent
with the stronger nonlinearity of the viscosity field in that regime.

\begin{table}
\centering
\begin{tabular}{llcS[table-format=2.4]c}
\toprule
$\params = (n, \lambda)$ & ROM type & reduced sizes & {$t$ [s]}
       & {$n_\mathrm{picard}$} \\
\midrule
\multirow{3}{*}{\shortstack[l]{$(0.35,\,4.5)$}}
  & ROM-FULL & $r=50$              & 16.61  & {28}   \\
  & ROM-DEIM & $r=50,\ s=59,\ q=118$ & 0.013  & {36}   \\
  & ROM-RBF  & $r=50$              & 0.0033 & {-} \\
\midrule
\multirow{3}{*}{\shortstack[l]{$(1.5,\,4.5)$}}
  & ROM-FULL & $r=50$              & 11.16  & {19}   \\
  & ROM-DEIM & $r=50,\ s=59,\ q=118$ & 0.0046 & {21}   \\
  & ROM-RBF  & $r=50$              & 0.0033 & {-} \\
\bottomrule
\end{tabular}
\caption{Online computation time and number of Picard iterations of ROM-FULL,
ROM-DEIM and ROM-RBF for the lid-driven cavity in the shear-thinning
($n=0.35$, $\lambda=4.5$) and shear-thickening ($n=1.5$, $\lambda=4.5$) regimes.
ROM-RBF performs a single regression evaluation and involves no Picard iteration.}
\label{tab:rom_time_accuracy_lid}
\end{table}

\begin{table}
\centering
\begin{tabular}{llcS[table-format=2.4]c}
\toprule
$\params = (n, F)$ & ROM type & reduced sizes & {$t$ [s]}
       & {$n_\mathrm{picard}$} \\
\midrule
\multirow{3}{*}{\shortstack[l]{$(0.35,\,95)$}}
  & ROM-FULL & $r=50$               & 24.62  & {34}   \\
  & ROM-DEIM & $r=34,\ s=50,\ q=100$ & 0.0099 & {38}   \\
  & ROM-RBF  & $r=34$               & 0.0018 & {-} \\
\midrule
\multirow{3}{*}{\shortstack[l]{$(1.5,\,95)$}}
  & ROM-FULL & $r=50$               & 10.59  & {18}   \\
  & ROM-DEIM & $r=34,\ s=50,\ q=100$ & 0.0029 & {20}   \\
  & ROM-RBF  & $r=34$               & 0.0018 & {-} \\
\bottomrule
\end{tabular}
\caption{Online computation time and number of Picard iterations of ROM-FULL,
ROM-DEIM and ROM-RBF for the settling sphere in the shear-thinning
($n=0.35$, $F=95$) and shear-thickening ($n=1.5$, $F=95$) regimes.
ROM-RBF performs a single regression evaluation and involves no Picard iteration.}
\label{tab:rom_time_accuracy_axi}
\end{table}

\clearpage

\bibliographystyle{elsarticle-num}

\bibliography{references}